\documentclass[fleqn,usenatbib]{mnras}

\usepackage{newtxtext,newtxmath}

\usepackage[T1]{fontenc}

\DeclareRobustCommand{\VAN}[3]{#2}
\let\VANthebibliography\thebibliography
\def\thebibliography{\DeclareRobustCommand{\VAN}[3]{##3}\VANthebibliography}

\usepackage{graphicx}	
\usepackage{amsmath}	
\usepackage{amssymb}	
\usepackage{hyperref}
\usepackage{float}
\usepackage{subfig}
\usepackage[detect-weight=true, per-mode=reciprocal]{siunitx}

\newcommand{\figref}[2][Figure~]{#1\ref{#2}}
\newcommand{\tabref}[2][Table~]{#1\ref{#2}}
\newcommand{\secref}[2][Section~]{#1\ref{#2}}
\renewcommand{\eqref}[2][equation~]{#1\ref{#2}}

\DeclareSIUnit\Msol{M_{\odot}} 
\DeclareSIUnit\Lsol{L_{\odot}} 
\DeclareSIUnit\parsec{pc} 
\DeclareSIUnit\Jy{Jy} 
\DeclareSIUnit\beam{beam} 
\DeclareSIUnit\micron{\micro\metre} 
\DeclareSIUnit\pixel{px} 
\DeclareSIUnit\gauss{G} 

\newcommand{\aerr}{\substack}

\usepackage{xcolor} 

\title[ALMA study of hub-filament systems]{An ALMA study of hub-filament systems\\I. On the clump mass concentration within the most massive cores}

\author[M. Anderson et al.]{
Michael Anderson,$^{1}$\thanks{E-mail: michael.anderson@astro.cf.ac.uk (MA)}
Nicolas Peretto,$^{1}$
Sarah E. Ragan,$^{1}$
Andrew J. Rigby,$^{1}$
Adam Avison,$^{2,3}$
\newauthor{}
Ana Duarte-Cabral,$^{1}$
Gary A. Fuller,$^{2}$
Yancy L. Shirley,$^{4}$
Alessio Traficante$^{5}$
and Gwenllian M. Williams$^{6}$
\\
$^{1}$School of Physics and Astronomy, Cardiff University, Queens Buildings, The Parade, Cardiff, CF24 3AA, UK\\
$^{2}$Jodrell Bank Centre for Astrophysics, Department of Physics and Astronomy, School of Natural Sciences, The University of Manchester, Oxford Road, Manchester, M13 9PL, UK\\
$^{3}$UK Atacama Large Millimeter/submillimeter Array Regional Centre Node, Manchester, M13 9PL, UK\\
$^{4}$Steward Observatory, University of Arizona, 933 North Cherry Avenue, Tucson, AZ, 85721 USA\\
$^{5}$IAPS-INAF, Via Fosso del Cavaliere, 100, I-00133, Rome, Italy\\
$^{6}$Centre for Astrophysics Research, Department of Physics, Astronomy and Mathematics, University of Hertfordshire, College Lane, Hatfield, AL10 9AB, UK
}

\date{Accepted XXX. Received YYY; in original form ZZZ}

\pubyear{2021}

\begin{document}
\label{firstpage}
\pagerange{\pageref{firstpage}--\pageref{lastpage}}
\maketitle

\begin{abstract}
	The physical processes behind the transfer of mass from parsec-scale clumps to massive-star-forming cores remain elusive. We investigate the relation between the clump morphology and the mass fraction that ends up in its most massive core (MMC) as a function of infrared brightness, i.e. a clump evolutionary tracer. Using ALMA 12\si{~\m} and ACA we surveyed 6 infrared-dark hubs in 2.9\si{~\mm} continuum at $\sim3\si{\arcsecond}$ resolution. To put our sample into context, we also re-analysed published ALMA data from a sample of 29 high mass-surface density ATLASGAL sources. We characterise the size, mass, morphology, and infrared brightness of the clumps using \emph{Herschel} and \emph{Spitzer} data. Within the 6 newly observed hubs, we identify 67 cores, and find that the MMCs have masses between 15--911\si{~\Msol} within a radius of 0.018--0.156\si{~\parsec}. The MMC of each hub contains 3--24\% of the clump mass ($f_\mathrm{MMC}$), becoming 5--36\% once core masses are normalised to the median core radius. Across the 35 clumps, we find no significant difference in the median $f_\mathrm{MMC}$ values of hub and non-hub systems, likely the consequence of a sample bias. However, we find that $f_\mathrm{MMC}$ is $\sim7.9$ times larger for infrared-dark clumps compared to infrared-bright ones. This factor increases up to $\sim14.5$ when comparing our sample of 6 infrared-dark hubs to infrared-bright clumps. We speculate that hub-filament systems efficiently concentrate mass within their MMC early on during its evolution. As clumps evolve, they grow in mass, but such growth does not lead to the formation of more massive MMCs.
\end{abstract}




\section{Introduction}
\label{sec:introduction}
Understanding what physical processes determine the mass of stars is an active area of astrophysics research. The similarity between the shape of the mass distribution of prestellar cores identified in nearby star-forming regions and that of the initial mass function of stars suggests that the latter may be inherited from the former, with a one to one correlation between core and stellar masses, and a uniform core to star formation efficiency across all core masses of $\sim30\%$ \citep[e.g.][]{Motte1998,Johnstone2001,Nutter2007,Konyves2010,Konyves2015}. The determination of stellar masses via core accretion is often referred to as \emph{core-fed} accretion \citep[e.g.][]{Wang2010}. As a result of the analysis of \emph{Herschel} observations of Gould belt star-forming regions \citep{Andre2010}, it has been proposed that the mass of cores is, in turn, determined by the fragmentation of gravitationally unstable filaments whose local Jeans mass is $\sim1\si{~\Msol}$, i.e. the peak of the core mass function in these regions \citep{Andre2010,Andre2014,Andre2019,Roy2015}. 

While the scenario described above might be relevant for determining the masses of low-mass cores and stars, it seems rather inappropriate when it comes to the formation of the most massive stars ($M_{\star}>8\si{~\Msol}$). The most massive prestellar cores identified in the far-infrared and sub-millimetre surveys of Gould Belt regions are typically about 10\si{~\Msol} \citep{Konyves2015,Konyves2020}, implying a stellar mass of about 3\si{~\Msol} when accounting for the core to star formation efficiency derived by the same authors. Much more massive prestellar cores, typically 30\si{~\Msol} and above, would need to be found in order to form massive stars in a core-fed-type scenario. Searches for such massive cores have now failed to find a significant population \citep[e.g.][]{Motte2007,Svoboda2019,Sanhueza2019}, and as of today only a few exceptional cases are known \citep[e.g.][]{Cyganowski2014,Nony2018}, despite an ever increasing database of high-angular resolution observations of cold and compact sources. 

Recently, \cite{Peretto2020} have used (sub-)millimetre dust continuum observations of Galactic plane star-forming regions to show that the evolution of massive compact sources ($m_\mathrm{gas} > 30\si{~\Msol}$) in mass vs. temperature diagrams is better explained by an accretion scenario in which cores gain mass while simultaneously collapsing to form protostars. In a similar manner, \citep{Rigby2021} find evidence for the mass growth of clumps, suggesting that same accretion processes may occur over a wider range of scales. The mass growth of the core is believed to be the result of the collapse of the surrounding parsec-scale mass reservoir called \emph{clump}, hence the accretion scenario described above is referred to as \emph{clump-fed} \citep{Wang2010}. 

The results from \cite{Rigby2021} suggest that there must be a link between the properties of a clump and the stars that form within it. Such a link has been searched for in the past. For instance, \cite{Palau2014,Palau2021} found a correlation between the fragmentation level within massive 0.1\si{~\parsec}-size cores and their average volume density, as expected from Jeans instability. On larger scales, \cite{Barnes2021} found a similar result, larger parsec-size clouds having lager number of cores embedded within them. They also find a correlation between the cloud mass and the mass of its most massive core. The existence of such a relation has also been explored by \cite{Lin2019}, who found a tight correlation between the mass of a sub-sample of massive ATLASGAL clumps and the mass of the most massive fragment they identify on SABOCA 350\si{~\micron} continuum images. However, the small difference in angular resolution between LABOCA ($18\si{\arcsecond}$) and SABOCA ($8.5\si{\arcsecond}$) might play a significant part in driving the observed correlation. On the other hand, \citet{Urquhart2014} argued that clumps with signposts of active massive star formation are more spherical than those which do not have such associated tracers, while \citet{Rigby2018} suggested that more spherical clumps are more efficient at concentrating their mass within their most massive core. These studies suggest that a combination of clump mass and morphology might be important parameters for the formation of massive stars. 

Here, we focus on a specific morphological category of clumps: hub-filamentary systems (HFS) \citep{Myers2009}. Hubs are small networks of converging interstellar filaments, at the centre of which active star formation is often observed \citep[e.g.][]{Kirk2013,Peretto2013,Peretto2014,Liu2012,Trevino-Morales2019}. They are found in all types of region, from low-mass star-forming clouds \citep[e.g.][]{Myers2009, Kirk2013}, to high-mass star-forming regions \citep[e.g.][]{Peretto2013, Schworer2019}, and have even been observed in our closest neighbouring galaxy \citep{Fukui2019,Tokuda2019}. The formation mechanism of such hubs are not yet fully understood \citep[see][for a description of possible mechanisms]{Myers2009}. However, hubs are naturally formed in simulations of collapsing clouds with non-isotropic density fields \citep{Kuznetsova2018, Vazquez-Semadeni2019}. Observationally, there is increasing evidence that hubs are indeed in a state of global collapse \citep[e.g.][]{Peretto2013,Peretto2014,Kirk2013,Hacar2018,Schworer2019,Trevino-Morales2019}. \cite{Williams2018} argued that the centres of hub filament systems, where the filaments converge, are privileged locations of massive core formation as they correspond to the locations of maximum gradient of gravitational acceleration, as opposed to individual uniform density filaments where these are located at their ends \citep[e.g.][]{Hartmann2007,Clarke2015}. Clump global collapse, hub morphology, and formation of massive cores might therefore all be interconnected. 

In the present paper we aim at constraining the efficiency of hubs at concentrating their mass into their most massive core, and this for a large range of clump masses. The end goal is to disentangle the effects of clump mass to those related to clump morphology. We do this by analysing new ALMA observations of a sample of hubs. In \secref{sec:observations} we describe the observations and data used in this paper, in \secref{sec:mass_fragmentation} we discuss the method used to extract cores from the ALMA continuum data, present the extracted core properties, and describe how we obtain physical properties for the host clumps. In \secref{sec:clump_mmc_relations} we present our test sample of ALMA cores taken from the literature, describe our clump classification scheme, and discuss the clump efficiency at forming their most massive cores as a function of their morphology. Finally, we present our conclusions in \secref{sec:conclusions}.

\section{Observations}
\label{sec:observations}
\subsection{Sample selection}
\label{sub:sample_selection}
For the purpose of this study we selected 6 infrared dark clouds, all part of the \cite{Peretto2009} catalogue. One of these, SDC335, was already examined by our team in a series of studies \citep{Peretto2013, Avison2015, Avison2021}. These 6 clouds have been selected to exhibit a well defined hub morphology seen in extinction at 8\si{~\micron}, with an easily identified filament convergence point (see \secref{sub:clump_classification} for more details on the hub classification). They all have high extinction contrast against a relatively uniform mid-infrared background. They have been selected so that their distances lie within a narrow range, i.e. from 2\si{~\kilo\parsec} to 3.2\si{~\kilo\parsec}, so that their properties can easily be compared to each other. Finally, they have been chosen so that they cover a large range of masses, from a few hundred to a few thousand solar masses, to try to evaluate the impact of the hub morphology on core formation independently of the clump mass.

\subsection{ALMA observations}
\label{sub:alma_observations}
Five IRDCs (see \tabref{tab:obs_prop}) were observed with the Atacama Large Millimeter/submillimeter Array (ALMA) 12\si{~\m} array between 20\textsuperscript{th}--23\textsuperscript{rd} January 2016 with a total of 41--46 antennas (C36-1 configuration), and with the Atacama Compact Array (ACA) between 17\textsuperscript{th} April and 25\textsuperscript{th} July 2016 (during Cycle 3) with 11 antennas (Project ID: 2015.1.01014.S; PI: Peretto). The number of 12\si{~\m} (7\si{~\m}) pointings was 61 (23), with a total on-source observing duration of 3.81\si{~\hour} (10.76\si{~\hour}). 

An additional IRDC, SDC335, was observed with the ALMA 12\si{~\m} array between 27\textsuperscript{th} September--19\textsuperscript{th} November 2011 with the 16 available antennas during Cycle 0 in the compact configuration (Project ID: 2011.0.00474.S; PI: Peretto). A complete description of the observations are presented in \cite{Peretto2013}. Follow-up observations of SDC335 were performed with the ACA between 6\textsuperscript{th}--8\textsuperscript{th} November 2016 (during Cycle 4) with 10 antennas (Project ID: 2016.1.00810.S; PI: Peretto). The total number of 12\si{~\m} (7\si{~\m}) pointings was 11 (6), with a total on-source observing duration of 4.11\si{~\hour} (1.33\si{~\hour}).

We achieve an angular resolution of $\sim$2.8\si{\arcsecond}--4.7\si{\arcsecond}, which at the distance of the targets corresponds to a linear resolution of 0.029--0.073\si{~\parsec}. This is at least a factor of two smaller than the Jeans length (which ranges between 0.10--0.21\si{~\parsec}) computed from the clump's average density, assuming a sound speed of $0.2\si{~\kilo\meter\per\second}$.

\clearpage
\begin{figure*}
	\centering
	\includegraphics[width=0.95\textwidth]{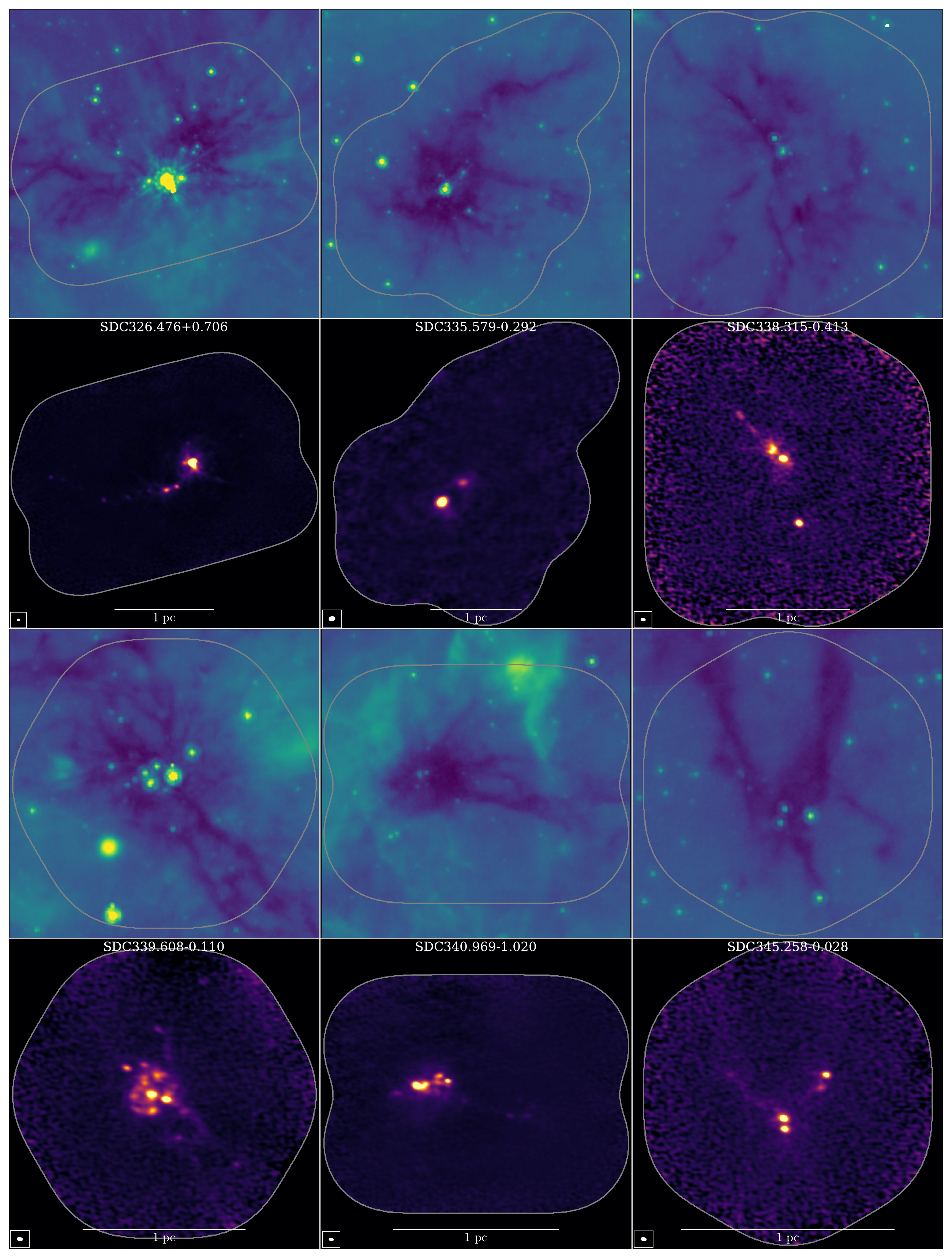}
		\caption{\emph{(First and third row)} \emph{Spitzer} 8\si{~\micron} images of the six IRDCs we observed with ALMA, showing prominent extinction features in a hub-filament system configuration. Below each Spitzer image is the corresponding ALMA combined 7\si{~\m}+12\si{~\m} continuum images at 2.9\si{~\milli\m} of each IRDC. The synthesised beam size of each image is shown in the lower left hand corner, the grey contour shows the extent of our ALMA fields. 
		}
	\label{fig:ALMA-Spitzer_fields}
\end{figure*}
\clearpage

The data were reduced and calibrated using the same CASA\footnote{\url{https://casa.nrao.edu}} \citep{McMullin2007} versions as used by the ALMA pipeline, using the standard pipeline scripts. The quasars J1531-5108, J1604-4441, J1617-5848, J1706-4600, J1650-5044 were used for phase calibration. Flux and bandpass calibration were performed using Mars, Ganymede, Neptune, J1427-4206, J1617-5848, J1733-1304, J1924-2914. The uncertainty in absolute flux calibration is $\sim5\%$ in Band 3, according to the ALMA Cycle 3 Technical Handbook\footnote{\url{https://almascience.eso.org/documents-and-tools/cycle3/alma-technical-handbook}}. The weights of the 12\si{~\m} SDC335 data were adjusted using \texttt{statwt()} on line-free channels prior to combination with the Cycle 4 ACA data. 

The calibrated ACA and 12\si{~m} visibilities were then concatenated and imaged using CASA version 5.5.0, utilising its implementation of the Multi-Scale CLEAN \citep{Cornwell2008} within the \texttt{tclean()} task. This was to better recover extended emission features that are larger than the beam. The data were imaged with Briggs weighting \citep{Briggs1995} with a robust parameter of 0.5. All of the images were primary beam corrected. For the five hubs observed in Cycle 3 the central frequency was 103.6\si{~\giga\Hz}, and the total continuum bandwidth used was 3.735\si{~\giga\Hz}. For SDC335 only 0.2\si{~\giga\Hz} of continuum bandwidth was used for imaging (split over two bands 104.0–-104.1\si{~\giga\Hz}, 105.0–-105.1\si{~\giga\Hz}), as this was the frequency coverage of the original Cycle 0 dataset. The central frequency was 104.55\si{~\giga\Hz}. \tabref{tab:obs_prop} contains a summary of the observational parameters for the six fields.

\subsection{Spitzer, WISE and Herschel data}
\label{sub:spitzer_wise_and_herschel_data}
We use publicly available \emph{Spitzer} GLIMPSE 8\si{~\micron} data\footnote{\url{https://irsa.ipac.caltech.edu/data/SPITZER/GLIMPSE}} \citep{Churchwell2009} and \emph{WISE} 12\si{~\micron} data\footnote{\url{https://irsa.ipac.caltech.edu/applications/wise/}} \citep{Wright2010}, at an angular resolution of $\sim2.4\si{\arcsecond}$ and $\sim6.5\si{\arcsecond}$, respectively. We use temperature and column density maps presented in \cite{Peretto2016} at a resolution of $\sim18\si{\arcsecond}$, which were constructed from 160\si{~\micron} and 250\si{~\micron} data from the \emph{Herschel} Hi-GAL survey \citep{Molinari2010a}. Finally, we also make use of the \cite{Molinari2016} 70\si{~\micron} compact source catalogue. \figref{fig:ALMA-Spitzer_fields} shows the \emph{Spitzer} 8\si{~\micron} fields for all six IRDCs, along with the final imaged ALMA 2.9\si{~\milli\m} continuum fields.

\begin{table*}
	\caption{The 6 IRDCs we observed with ALMA, their estimated distances, and a summary of observational properties of of our combined 7\si{~\m}+12\si{~\m} observations at 2.9\si{~\mm}. The IRDCs will hereafter be referred to by their shorthand names highlighted in bold. The linear resolution corresponds to the physical size of the beam major axis at the distance of the clump. The mass sensitivities were calculated assuming a source temperature of 12\si{~\kelvin}.
	}
	\label{tab:obs_prop}
	\centering
	\begin{tabular}{c c c c c c c c}
		\hline\hline
		Clump Name & SDC Name & $d$ & Synthesised Beam & PA & RMS noise & Linear resolution & $M_\mathrm{min}$ \\
		 & \citep{Peretto2009} & (\si{\parsec}) & ($\si{\arcsecond} \times \si{\arcsecond}$) & (\si{\degree}) & (\si[per-mode=symbol]{\micro\Jy\per\beam}) & (\si{\parsec}) & (\si{\Msol}) \\
		\hline
		G326.4745+0.7027  & \textbf{SDC326}.476+0.706  & 2610 & $2.80 \times 2.16$ & $ 69.12$ & 78.68  & 0.035 & 0.27 \\
		G335.5857--0.2906 & \textbf{SDC335}.579--0.292 & 3230 & $4.69 \times 3.63$ & $-79.00$ & 379.72 & 0.073 & 1.98 \\
   		G338.3150--0.4130 & \textbf{SDC338}.315--0.413 & 2940 & $2.91 \times 1.92$ & $ 80.19$ & 49.00  & 0.041 & 0.21 \\
   		G339.6080--0.1130 & \textbf{SDC339}.608--0.113 & 2740 & $2.88 \times 1.90$ & $ 80.42$ & 66.06  & 0.038 & 0.24 \\
		G340.9698--1.0212 & \textbf{SDC340}.969--1.020 & 2210 & $2.88 \times 1.92$ & $ 79.92$ & 100.10 & 0.031 & 0.24 \\
		G345.2580--0.0280 & \textbf{SDC345}.258--0.028 & 2090 & $2.84 \times 1.87$ & $ 80.65$ & 50.88  & 0.029 & 0.12 \\
		\hline
	\end{tabular}
\end{table*}

\section{Mass fragmentation}
\label{sec:mass_fragmentation}
\subsection{Core extraction}
\label{sub:core_extraction}
To extract the structures from our ALMA continuum images we use a dendrogram-based method using \texttt{astrodendro}, a Python package based on the \cite{Rosolowsky2008} implementation of dendrograms to analyse astronomical data. Our extraction method required that the minimum structure size $n_{\mathrm{pix,min}}$ must be greater than half the synthesised beam size (typically 18 pixels), the faintest included emission structure to be at a specific intensity of $I_{\mathrm{min}}=5\times\sigma_{\mathrm{global}}$, and minimum structure significance $\Delta I_{\mathrm{min}}=1\times\sigma_{\mathrm{global}}$, where $\sigma_{\mathrm{global}}$ is the RMS specific intensity calculated from the residual image of the field produced after imaging (see \tabref{tab:obs_prop}). This ensured that all of the extracted emission structures were at least detected five times above the global RMS in an image, with a peak at least 6 times the RMS.

For our analysis, we consider the leaves of the dendrogram (i.e. emission structures that do not have any detectable substructure) to be ``cores''. We are aware that these sources may well be sub-fragmented at higher resolution. We impose the constraint that only cores that are contained within the contour at $>$50\% of the primary beam power are included in the dendrogram. This is to avoid noise peaks that appear brighter and occur more frequently toward the edges of the fields, caused by the non-uniform response of the primary beam. After applying this constraint, the constructed dendrograms contained 71 candidate cores across the six fields. 

We produced error maps by performing a windowed RMS calculation on our residuals, with a window size of 4x4 beam major axis lengths. We then construct signal-to-noise (SNR) maps to better estimate the strength of the detections, given that the noise varies across the fields, and to help remove any spurious detections. Core candidates with at least $n_{\mathrm{pix,min}}$ pixels with a $\mathrm{SNR}\geq3$ are classed as detections. Extracted structures that do not satisfy this condition are discarded. After applying this criteria we obtain a set of 67 cores. \figref{fig:ALMA_cutouts} shows a zoomed in view of these 67 cores, along with their extent as defined by the dendrogram.

\subsection{Core sizes and masses}
\label{sub:core_properties}
Assuming that the cores are in local thermodynamic equilibrium (LTE) and that the dust emission is optically thin, the core masses can then be calculated using
\begin{equation}
	\label{eq:core_mass}
	M_\mathrm{core} = \frac{d^2 S_\nu}{\kappa_\nu B_\nu(T)}
\end{equation}
where $d$ is the distance to the IRDC, $S_\nu$ is the integrated flux density of the source, $\kappa_\nu$ is the specific dust opacity, and $B_\nu(T)$ is the Planck function at a given dust temperature $T$ \citep{Kauffmann2008}. We assume the same specific dust opacity relation as \cite{Marsh2015,Marsh2017}:
\begin{equation}
	\label{eq:kappa_dust}
	\kappa_\nu(\lambda) = 0.1\si{~\centi\metre\squared\per\gram} \left(\frac{\lambda}{300\si{~\micron}}\right)^{-\beta}
\end{equation}
with a dust opacity index $\beta=2$, for a given wavelength $\lambda$, and accounting for a gas-to-dust mass ratio of 100. The uncertainty in the dust opacity is around $\pm50\%$ \citep{Ossenkopf1994,Roy2013,Roy2015}. 

To estimate core temperatures we use a combination of two methods. Our primary method is to use dust temperature maps derived from \emph{Herschel} $160\si{~\micron}/250\si{~\micron}$ ratio maps as presented in \cite{Peretto2016}. We simply take the temperature ($T_\mathrm{col}$) at the position of each core's intensity-weighted centroid. These maps cover a temperature range of around 12--30\si{~\kelvin} for our set of fields. Note that because we assume a unique temperature along the line of sight and that the typical background temperature of the Galactic Plane is $\sim18\si{~\kelvin}$, we may overestimate the temperatures of dense clumps colder than this background value \citep{Peretto2010a, Battersby2011, Marsh2015}.

For warmer sources (such as massive protostellar cores), this may be significantly underestimating their temperature, and hence overestimating their mass. To try and counter this effect, we use the Hi-GAL 70\si{~\micron} Compact Source Catalogue \citep{Molinari2016} to see which cores in our sample have an associated 70\si{~\micron} source, as the 70\si{~\micron} flux density is known to be a good tracer of the luminosity of embedded sources \citep{Dunham2008, Ragan2012}. If a 70\si{~\micron} source is present within the equivalent radius $R_{\mathrm{eq}}$ of a core, which is the radius of a circle with equal area to the core's corresponding dendrogram mask, we say they are associated. We then convert the 70\si{~\micron} flux densities to bolometric (internal) luminosities using the following relation \citep{Elia2017}: 
\begin{equation}
	\label{eq:L_int}
	L_\mathrm{int} = 25.6 \left(\frac{S_{70\si{~\micron}}}{10\si{~\Jy}}\right) \left(\frac{d}{1\si{~\kilo\parsec}}\right)^2 \si{~\Lsol}
\end{equation}
Where $S_{70\si{~\micron}}$ is the integrated 70\si{~\micron} flux density of the source, and $d$ is the distance to the clump. Assuming that the dust emission from a protostellar core is optically thin and is predominantly in the far-infrared, we calculate the mean mass-weighted temperature of the core, $\overline{T_\mathrm{d}}$ \citep{Emerson1988, Terebey1993}:
\begin{equation}
	\label{eq:T_70}
	\overline{T_\mathrm{d}} = \frac{3}{2} T_0 \left(\frac{L_\mathrm{int}}{L_0}\right)^{1/6} \left(\frac{r}{r_0}\right)^{-1/3}
\end{equation}
Where $L_\mathrm{int}$ is the source's internal luminosity, $r$ is the core's radius, and reference values $T_0=25\si{~\kelvin}$, $L_0=520\si{~\Lsol}$, $r_0=0.032\si{~\parsec}$. This form of the equation assumes $\beta=2$, and that the density profile of the core follows $\rho(r)\propto r^{-2}$, \cite[as used by e.g.][]{Bontemps2010, Svoboda2019}.

We use \eqref{eq:T_70} to calculate the mean temperature within core equivalent radius $r=R_{\mathrm{eq}}$. For our set of sources, $\overline{T_\mathrm{d}}$ ranges between 18--76\si{~\kelvin}. 

If a 70\si{~\micron} flux density derived temperature can be obtained for a core, we assign the core $T_\mathrm{core} = \overline{T_\mathrm{d}}$, and otherwise assign $T_\mathrm{core} = T_\mathrm{col}$. We assume that the temperature of the gas and dust are coupled as the cores have a density at least $\sim10^6\si{~\cm^{-3}}$, the threshold at which \cite{Goldsmith2001} states that the dust and gas temperatures become essentially equal.

We use the Revised Kinematic Distance Calculator\footnote{\url{http://bessel.vlbi-astrometry.org/revised_kd_2014}} \citep{Reid2009,Reid2014} to estimate the distances to the IRDCs, using the LSR velocities for each clump. We assume that the IRDCs are located at the near distance as they are IR-dark at 8\si{~\micron}, but do not assume whether the clump is located within a spiral arm or in an inter-arm region. The typical distance uncertainty is between 10--20\%.

The integrated flux density of the cores comes from our dendrogram extraction, following the ``clipped'' paradigm \citep[see][]{Rosolowsky2008}. Since we care about the cores as being overdensities, by using a clipped method we minimise the contribution from the background on the mass estimates, which could be particularly large for the crowded areas at the centre of the hub-filamentary systems. This way, we are being conservative in the mass estimates, and are possibly underestimating the mass of some of these cores at the centre of the hubs. The error in integrated flux calculated from the quadrature sum within the core mask of our error maps, multiplied by pixel area.

\begin{figure}
	\centering
	\includegraphics[width=\columnwidth]{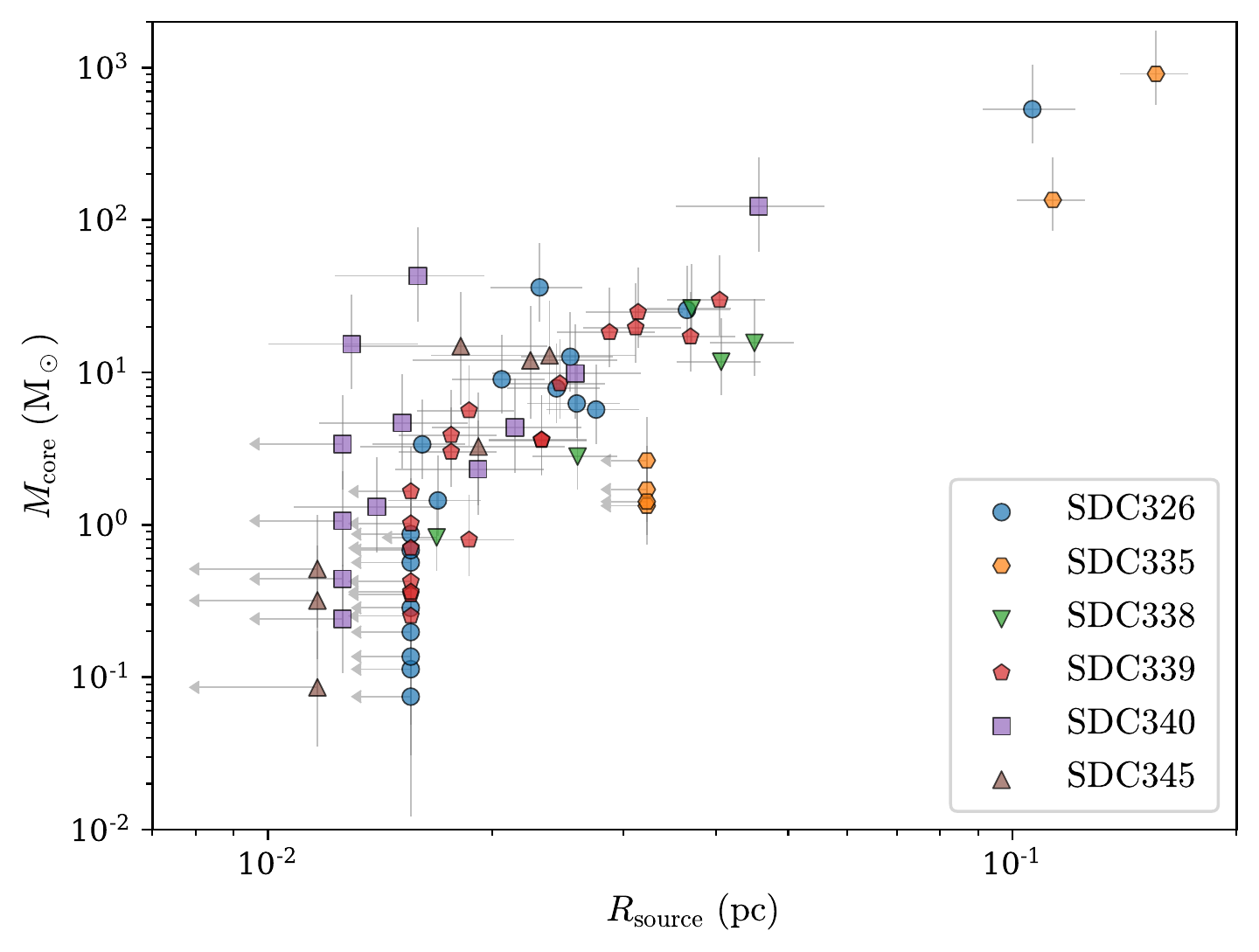}
		\caption{Core mass against deconvolved radius ($R_\mathrm{source}$) plot for all extracted cores from all 6 IRDCs. Upper limits for the radii of unresolved sources are indicated with arrows pointing towards the left. 
		}
	\label{fig:core_mass-radius}
\end{figure}

By substituting all of these values into \eqref{eq:core_mass}, we obtain masses for all cores. The error in the core masses was calculated using Monte-Carlo methods, by randomly sampling over each variable in \eqref{eq:core_mass}, assuming Gaussian errors. We also calculate a deconvolved source radius, $R_\mathrm{source}$, which is given by 
\begin{equation}
	\label{eq:deconvolved}
	R_\mathrm{source} = \sqrt{R_\mathrm{eq}^2 - \frac{\theta_\mathrm{maj} \theta_\mathrm{min}}{4}}
\end{equation}
where $\theta_\mathrm{maj}$ and $\theta_\mathrm{min}$ are the major and minor beam axes, respectively. A table of derived core properties is supplied as online supplementary material, and a plot of core mass against deconvolved radius is shown in \figref{fig:core_mass-radius}.

We see that our sample of extracted cores cover a broad mass range of 0.07--911\si{~\Msol}, with a mean mass of 32\si{~\Msol}. The core masses we present here follow the ``clipped'' paradigm, which subtracts all of the flux below the core's contour in the dendrogram (similar to a background subtraction). Our mass sensitivity ranges between 0.12--0.27\si{~\Msol}, depending on the field, with the exception of SDC335 for which the mass sensitivity is 1.98\si{~\Msol}. Note that these mass sensitivities were calculating assuming a source temperature of 12\si{~\kelvin}, an hence when cores are assigned a temperature warmer than 12\si{~\kelvin} they can have a lower calculated mass than our listed sensitivity. Also note that these mass sensitivities correspond to a clipped mass sensitivity, whereas often in literature the ``bijective'' mass sensitivity is quoted. Given the dendrogram parameters we have used for our extraction, a corresponding bijective mass sensitivity would be $\sim6$ times higher than the sensitivities quoted above.

Contrary to \cite{Csengeri2017}, we do find intermediate-mass cores in the sample, likely due to not using a single $T_\mathrm{core}=25\si{~\kelvin}$ for all cores, the assumption made in their core mass calculations. Two cores (SDC335-MM1 and SDC326-MM1) are exceptionally high mass, at 911\si{~\Msol} and 534\si{~\Msol} contained within a deconvolved radius of 0.156\si{~\parsec} and 0.106\si{~\parsec}, respectively. They also correspond to the two of the largest sources identified. They are therefore excellent candidates for the formation of very high-mass stars. 

Four of the HFS contain at least one core with $M_\mathrm{core}\geq30\si{~\Msol}$ and so, assuming a core to star formation efficiency of 30\%, could form at least one high-mass star with $M_{\star}>8\si{~\Msol}$.

\subsection{Core formation efficiencies}
\label{sub:core_formation_efficiencies}
\begin{table*}
	\caption{Core properties of the most-massive cores within each IRDC in our sample, ordered by clump mass. $R_{\mathrm{source}}$ is the deconvolved equivalent radius of the core, and $R_{\mathrm{clump}}$ is the equivalent radius of the clump. The full table can be found online, along with a table of properties for all of the extracted cores.
	}
	\label{tab:MM_cores}
	\centering
	\begin{tabular}{c c c c c c c c}
		\hline\hline
		Core ID & $R_{\mathrm{source}}$ (\si{\parsec}) & $T_{\mathrm{core}}$ (\si{\kelvin}) & $M_{\mathrm{core}}$ (\si{\Msol}) & $R_{\mathrm{clump}}$ (\si{\parsec}) & $M_{\mathrm{clump}}$ (\si{\Msol}) & $f_{\mathrm{MMC}}$ (\%) & CFE (\%) \\
		\hline
		SDC345-MM2 & $0.018$ & 14.6 & $ 15\aerr{+19\\-9}$     & $0.27$ & $ 135$ & $11.0$ & $32.7$ \\
		SDC338-MM3 & $0.037$ & 15.9 & $ 26\aerr{+25\\-10}$    & $0.42$ & $ 213$ & $12.4$ & $27.0$ \\
		SDC339-MM3 & $0.040$ & 15.9 & $ 30\aerr{+29\\-12}$    & $0.54$ & $ 942$ & $3.2$ & $15.3$ \\
		SDC340-MM1 & $0.046$ & 46.3 & $123\aerr{+134\\-61}$   & $0.53$ & $1768$ & $7.0$ & $11.8$ \\
		SDC326-MM1 & $0.106$ & 41.5 & $534\aerr{+512\\-216}$  & $0.80$ & $2399$ & $22.2$ & $26.9$ \\
		SDC335-MM1 & $0.156$ & 41.2 & $911\aerr{+835\\-338}$  & $0.95$ & $3739$ & $24.4$ & $28.2$ \\
		\hline
	\end{tabular}
\end{table*}
As discussed in the introduction, the ability of a clump to concentrate its mass within cores is a fundamental, but poorly understood characteristic of star-forming regions. In this paper we will refer to parsec-scale dense molecular cloud structures as ``clumps'', within which stellar clusters and large systems can form \citep{Eden2012,Motte2018}. Here, we calculate the core formation efficiency (CFE), 
\begin{equation}
	\label{eq:CFE}
	\mathrm{CFE} = \frac{\sum_i M_{\mathrm{core,}i}}{M_{\mathrm{clump}}}
\end{equation}
which is the sum of core masses in a given clump, divided by the clump's mass. This tells us how much of a clump's mass is contained within compact sources. The clump masses are obtained from \emph{Herschel} column density maps \citep{Peretto2016}, where the clump boundary is defined by the $\mathrm{H}_2$ column density contour at $N_{\mathrm{H}_2}=3\times10^{22}\si{~{\centi\metre}^{-2}}$.

As far as massive star formation is concerned, another quantity of interest is the fraction of the clump mass contained within its most-massive core (MMC), 
\begin{equation}
	\label{eq:f_MMC}
	f_\mathrm{MMC} = \frac{M_\mathrm{MMC}}{M_\mathrm{clump}}. 
\end{equation}
\tabref{tab:MM_cores} shows a summary of the properties of the MMCs for each IRDC in our sample, and the CFE for each clump. We see that the CFE varies between 11--33\%, while $f_\mathrm{MMC}$ ranges between 3--24\%. Note that the CFE calculated here does not take into account the variation in sensitivity between each field, and hence are not directly comparable.

\section{The relationship between clump and core masses}
\label{sec:clump_mmc_relations}
\subsection{Broader sample of clumps and cores}
\label{sub:broader_sample_of_clouds_cores}
In order to get a sense of how $f_\mathrm{MMC}$ values from our hub sample compare to those from a less biased Galactic plane population of clumps, we use the \cite{Csengeri2017} sample of high-mass ATLASGAL sources observed with ALMA (Project ID: 2013.1.00960.S; PI: Csengeri). This sample contains 35 clumps that have been observed with ALMA ACA at 878\si{~\micron} (Band 7). These ACA data have similar angular resolution as ours, with a mean beam size of 3.8\si{\arcsecond}. Also, the distance of these clumps span a very similar range ($1.3\si{~\kilo\parsec} < d < 4.2\si{~\kilo\parsec}$) to our set of sources. Note that as our 7\si{~\m}+12\si{~\m} observations are Band 3, and hence the dust emission we are comparing between datasets may arise from slightly different layers of the cores.

For consistency we use the same procedure for source extraction as described in \secref{sub:core_extraction}. However, note that the \cite{Csengeri2017} observations are single-pointing only, and are somewhat less sensitive. We therefore cannot compare the CFE values from both samples, and instead focus on comparing $f_\mathrm{MMC}$. Core temperatures and clump masses for the \cite{Csengeri2017} sample are estimated in the same way as for our sample of clumps (see \secref{sec:mass_fragmentation}). 

Three of the clumps overlap between our samples, so we preferentially choose extracted fluxes from our data due to greater coverage, sensitivity and resolution. In the two instances where two clumps share the same $N_{\mathrm{H}_2}$ contour, we merge the clumps and assign it the name of the ``original'' clump containing the brightest source. Given that our method to measure clump mass is dependent on \emph{Herschel} coverage, one source from the \cite{Csengeri2017} sample has been discarded. The joined sample therefore contains 35 clumps in total, and within those clumps we detect 129 cores. \emph{Spitzer} 8\si{~\micron} cutout images of each clump are shown in \figref{fig:Spitzer_cutouts}, with the \emph{Herschel} column density contours (that define our clump boundaries) overlaid.

\subsection{Clump classification}
\label{sub:clump_classification}
In this paper, we use two distinct clump classification schemes, one that qualitatively identifies the amount of star formation activity within it, and another that determines whether or not a most massive core is at the centre of a hub filament system. Despite both schemes having their own limitations (see below), they can still provide insight into the time evolution of the clumps for the former, and the filamentary environment of the most massive cores for the latter.

\begin{figure}
	\centering
	\includegraphics[width=\columnwidth]{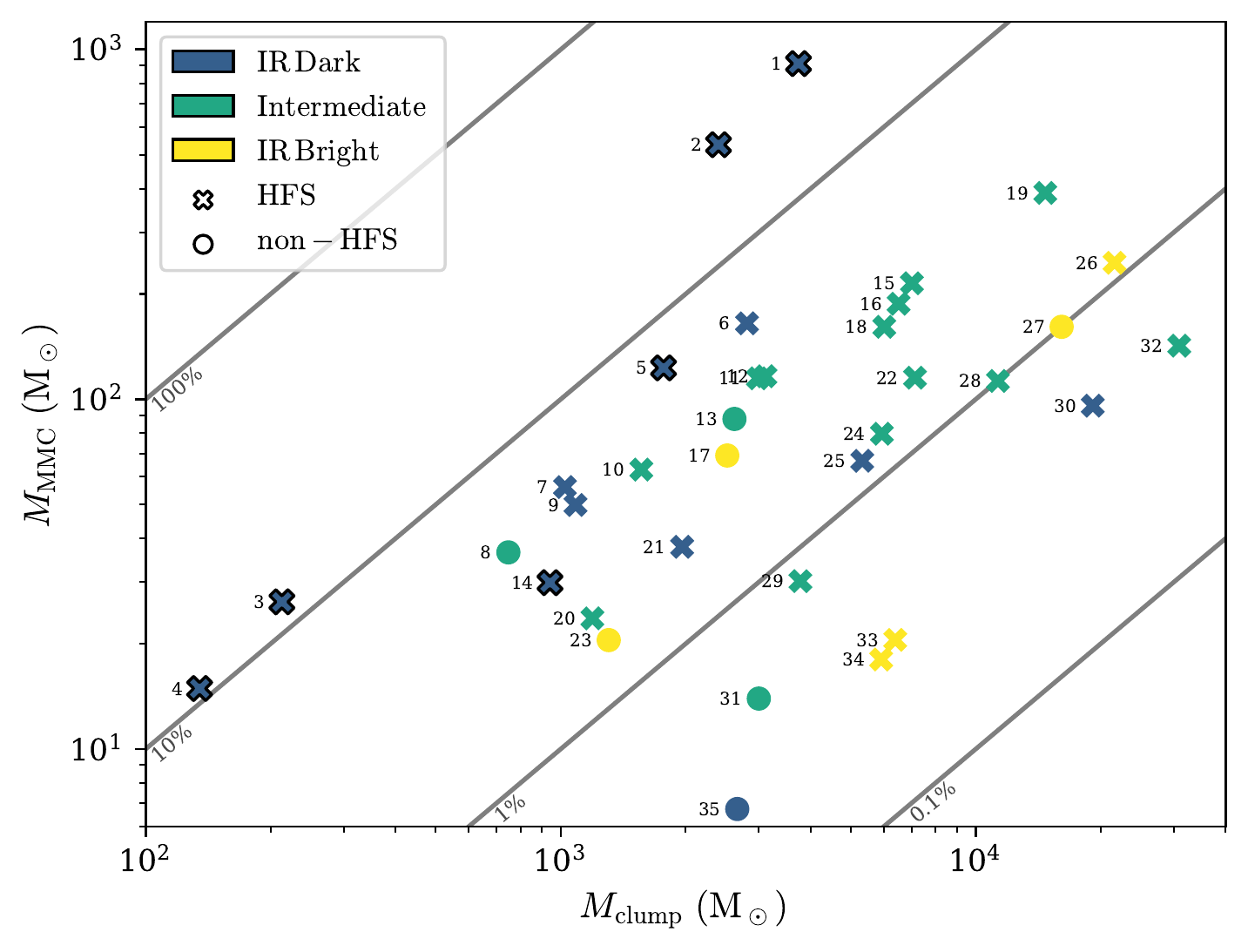}
		\caption{Mass of clumps against the mass of the most-massive core (MMC) within that clump. The crosses represent clumps that have been classified as HFS, and circular points are non-HFS clumps. The point fill colours represent the three IR-brightness classes. Points with a black outline are sources observed at 2.9\si{~\mm}, and points without outlines are sources observed at 878\si{~\micron}. Clump 20 (G339.6802-1.2090) had no \emph{Spitzer} 8\si{~\micron} coverage, so \emph{WISE} 12\si{~\micron} was used for IR-brightness classification. The diagonal grey lines represent lines of constant $f_\mathrm{MMC}$.
		}
	\label{fig:clump_mass-mmc_mass}
\end{figure}

We first classify the clumps based on the mid-infrared brightness within the $N_{\mathrm{H}_2}=3\times10^{22}\si{~{\centi\metre}^{-2}}$ contour used to define the clump boundaries (see \figref{fig:Spitzer_cutouts}). Infrared brightness has recently been shown to be a reliable time evolution tracer (\cite{Rigby2021}; Watkins et al. in prep). We classify clumps into three infrared brightness category, from the less evolved to the more evolved: ``IR-dark'', no 8\si{~\micron} extended emission within clump, prominent extinction features; ``IR-bright'', significant 8\si{~\micron} extended emission within the clump, without prominent extinction features; or ``Intermediate'', having both clear extinction and emission features within the clump. This classification is made by eye, and is therefore subject to some subjectivity, especially for borderline cases. However, it still provides a reasonable classification of the inner star formation activity of a clump. Out of the the 35 clumps, we classify 13 as IR-dark, 16 as Intermediate, and 6 as IR-bright.

Clumps are then further classified as either HFS or non-HFS according to the location of the most massive core with respect to its local network of filaments. For that purpose we utilise a Hessian-based method, similar to \cite{Schisano2014, Orkisz2019}, to extract filamentary structures from \emph{Herschel} 250\si{~\micron} images of the clumps. We then classify a clump as a HFS if there are at least three filaments pointing towards the location of the most massive core. One caveat of this method is the relatively low angular resolution of the \emph{Herschel} 250\si{~\micron} image compared to the ALMA data ($\sim18\si{\arcsecond}$ vs. $\sim3\si{\arcsecond}$) which prevents us from making a robust association between filaments and cores. Also, for the same reason, a lot of the filamentary structures within the clumps will not be resolved or even identified. We therefore use the \emph{Spitzer} 8\si{~\micron} images in conjunction with our extracted filaments to inform our final classification, by checking each one of the clumps for filamentary structures seen in extinction at 8\si{~\micron}. Instances where clumps were classified by \emph{Spitzer} 8\si{~\micron} are noted in our table of MMC properties as part of the online supplementary materials associated with this paper.

Out of the 35 clumps, 28 are classified as hubs and 7 as non-hubs, making our sample hub-dominated. This is likely to be a consequence of how the sample has been built: the merging of 6 infrared dark hubs with a sample of 29 massive clumps, which are known to often be associated to hubs \citep{Kumar2020}.

\subsection{Mass concentration within most massive cores}
\label{sub:mass_concentration_onto_mmcore}
One argument is that a clump's ability to form high-mass stars is directly linked to the amount of material within that clump \citep{Beuther2013}. Therefore we first investigate the relation between the clump mass ($M_\mathrm{clump}$) and the mass of their most massive cores ($M_\mathrm{MMC}$). \figref{fig:clump_mass-mmc_mass} shows that, when considering the entire clump sample, there is only a fairly moderate correlation between these two quantities, with a Spearman's rank correlation coefficient $r_\mathrm{S}=0.535$, and a $p$-value=0.0009. It is possible that this correlation may be influenced by the sparse sampling of the parameter space below a clump mass of $<1000\si{~\Msol}$. Above a clump mass of 1000\si{~\Msol}, the distribution of the most massive core mass is fairly uniform between 10\si{~\Msol} and 1000\si{~\Msol}, suggesting a wide range of $f_\mathrm{MMC}$ values. If we exclude all datapoints (4 clumps) with $M_\mathrm{clump}<1000\si{~\Msol}$, then we obtain a correlation coefficient of $r_\mathrm{S}=0.447$ ($p$-value=0.01), which is moderately weaker than for the full sample. However, if we now only consider the 6 new infrared-dark hubs we observed, we notice that the correlation, even though less statistically significant, is much stronger, with a correlation coefficient of 1 ($p$-value=0). We speculate that this could point towards a time-dependent correlation between clump and core mass. We will discuss that point further below.

\begin{figure*}
	\centering
	\includegraphics[width=0.7\textwidth]{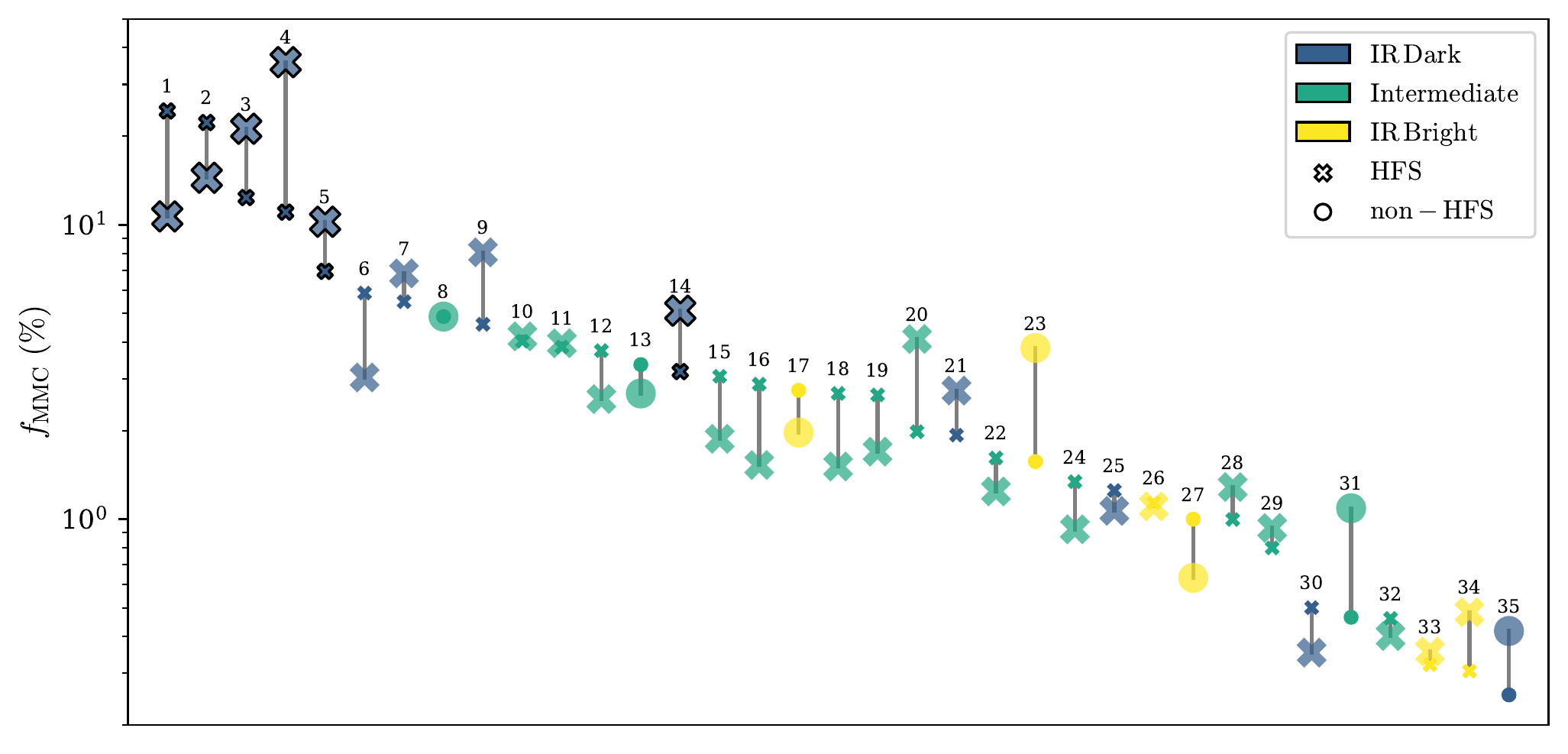}
		\caption{Fraction of each clump's total mass contained within its most-massive core ($f_\mathrm{MMC}$). The larger points are the $f_\mathrm{MMC}$ values multiplied by the median core radius of 0.07\si{~\parsec} over the radius of that MMC. The number above each pair of points represents the clump ID number.
		}
	\label{fig:f_MM}
\end{figure*}

A tight correlation between clump mass and mass of the most massive core was found by \cite{Lin2019} for a sample of ATLASGAL clumps covering a large range of evolutionary stages as traced by their luminosity to mass ratio. This is at odds with the results discussed above for the full sample. It is likely that the tight correlation observed by \cite{Lin2019} is artificially driven by the small range of scales they probe, typically 0.3\si{~\parsec} for what they call cores and 0.7\si{~\parsec} for their clumps (a factor of $\sim$2.3). In our study, the range of scales we probe between the median core size (0.08\si{~\parsec}) and the median clump size (1.5\si{~\parsec}) is a lot larger, a factor of $\sim$18.8, therefore probing clearly distinct structures.

The smaller set of symbols in \figref{fig:f_MM} shows the same information as presented in \figref{fig:clump_mass-mmc_mass} but in the form of $f_\mathrm{MMC}$ values, with each clump marked by their unique ID number. The points use the same colour scheme as used in \figref{fig:clump_mass-mmc_mass}. What is apparent is that some of the clump categories, such as IR-dark clumps, have on average larger $f_\mathrm{MMC}$ values than others. However, one possible bias that may affect such comparison is the difference in core radii, with some cores being more massive simply by being much larger. In order to remove that bias, the larger set of symbols in \figref{fig:f_MM} shows the same quantity as the small set of symbols but rescaled by the median $R_\mathrm{eq}$ of the MMCs (0.07\si{~\parsec}) over the the core's $R_\mathrm{eq}$. By doing this rescaling we effectively compare $f_\mathrm{MMC}$ at the same core radius, assuming that the density profiles of these cores scale as $\rho(r) \propto r^{-2}$ \citep{Bontemps2010, Svoboda2019}. We now see that, even though there has been a bit of reshuffling, the individual $f_\mathrm{MMC}$ have not drastically changed.

\begin{figure}
	\centering
	\includegraphics[width=\columnwidth]{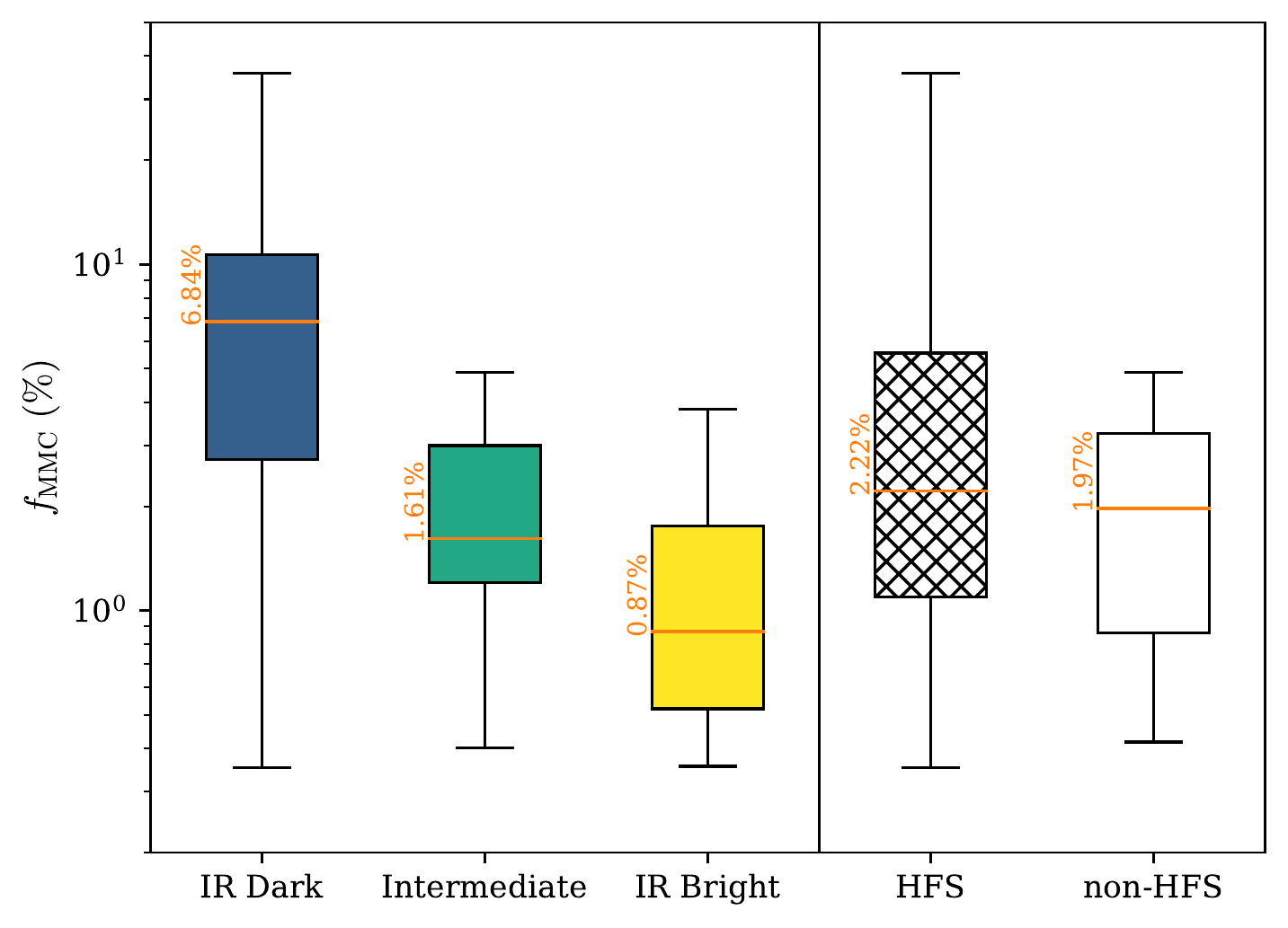}
		\caption{Distribution of rescaled $f_\mathrm{MMC}$ values for each of our clump categories. The orange lines represent the median (with the values also in orange), the boxes represent the interquartile range (IQR), and the ``whiskers'' represent the full extent (i.e. the 0th and 100th percentile) of the data. From left to right, there are 13, 16, 6, 28, and 7 clumps in each category.
		}
	\label{fig:f_MM_classes}
\end{figure}

\figref{fig:f_MM_classes} shows the distribution of rescaled $f_\mathrm{MMC}$ values for each category of clump, where the orange line represents the median, and the ``whiskers'' of the box-plot represent the full extent of the data. We can see that IR-dark clumps have a median $f_\mathrm{MMC}$ around 7.9 times higher than IR-bright clumps, while Intermediate clumps have a median value 2.4 times higher than their IR-bright counterparts. Even more striking, is our sample of 6 IR-dark hubs (see \tabref{tab:MM_cores}) that have a median value of 12.6\%, which is 14.5 times higher than IR-bright clumps. In contrast, the median rescaled $f_\mathrm{MMC}$ values for HFS and non-HFS are only separated by a factor of 1.1. Although the median values for these two clump categories are close, the distributions shown in \figref{fig:f_MM_classes} appear to be different. To test this we perform a two-sample Kolmogorov–Smirnov (K-S) test to check whether the two samples (HFS and non-HFS) come from the same distribution. We find a $p$-value of 0.705 for the test, and therefore we cannot reject the null hypothesis that the two samples were drawn from the same distribution (at a significance level of 5\%). Although this test is inconclusive, it is likely biased by the very small sample size of non-HFS clumps.

\section{Conclusions}
\label{sec:conclusions}

As shown in \secref{sec:clump_mmc_relations}, although the distributions of $f_\mathrm{MMC}$ values of the hub filament and non-hub systems appear different, this apparent difference is not statistically significant. Whether this is because there is a common mass concentration efficiency between the two types of clumps or due to the small size of the non-HFS sample and a bias in the sample construction is unclear. Distinguishing between these two possibilities requires observations of a larger, well selected sample of non-HFS sources. The source selection bias is such that we are, by construction, focusing on high-mass star-forming clumps. These have been shown to be preferentially associated to hubs \citep{Kumar2020}. As a result, we may be missing out on a large population of non-hub clumps that have much lower $f_\mathrm{MMC}$ values. The relatively low resolution of the data used to derive filament skeletons (compared to the ALMA data used for core characterisation) may cause us to mis-classify a large fraction of clumps  altogether, in either direction, which would lead to averaging out $f_\mathrm{MMC}$ values for both hub and non-hub clumps. As it is, we believe that we cannot provide any robust conclusions on the ability of hubs to concentrate more mass within their most massive cores compared to non-hub clumps.

Interestingly, \figref{fig:f_MM_classes} shows a clear trend of $f_\mathrm{MMC}$ values with our infrared brightness classification, $f_\mathrm{MMC}$ decreases by more than one order of magnitude when going from IR-dark to IR-bright clumps. If one takes this infrared brightness classification as a rough proxy for time evolution, then our results suggest that the clump efficiency in concentrating mass within their most massive cores decreases with time. When inspecting in details the origin of this decrease, we realise that this trend is due to an increase of median clump masses (IR-dark clump: 1961\si{~\Msol}; Intermediate clump: 4859\si{~\Msol}; IR-bright clump: 6155\si{~\Msol}) and not due to a decrease of median core masses (IR-dark clumps: 56\si{~\Msol}; Intermediate clumps: 114\si{~\Msol}; IR-bright clumps: 45\si{~\Msol}). Note as well that the sub-sample of 6 infrared dark hubs we observed displays the highest median $f_\mathrm{MMC}$ value (12.6\%) of all categories. While it is not completely clear what bias in the way we selected these 6 sources is responsible for driving such high $f_\mathrm{MMC}$ values, in the context of the trend discussed above, these sources represent some of the earliest stages of clump evolution (with a median clump mass of 1355\si{~\Msol} and a median MMC mass of 76\si{~\Msol}). It is possible that we overestimate the temperatures of our clumps, and hence underestimate their mass leading to a potentially artificially higher $f_\mathrm{MMC}$ for IR-dark clumps in particular. However Figure 3 in \cite{Peretto2016} shows that for a clump with a mean temperature of 12\si{~\kelvin} and $N_{\mathrm{H}_2}\geq3\times10^{22}\si{~{\centi\metre}^{-2}}$ the column density (and therefore mass) is at worst underestimated by $\sim$30\%, far from the factor of 3 required to bring the median mass of IR-dark and IR-bright clumps in line.

In light of these results, we propose a scenario in which HFS are formed very early on during the time evolution of a clump, efficiently funnelling mass into its most massive core. The early global collapse of the clump is likely to be driving force behind the early formation of these cores \citep{Peretto2013,Peretto2014,Williams2018}. During these early stages of clump evolution the mass of the MMC most likely correlates with the mass of the clump itself. As time goes on, clump mass grows \citep{Peretto2020, Rigby2021}, accreting matter from its surrounding environment (without increasing the mass of its MMC), resulting in a decreasing $f_\mathrm{MMC}$ over time. 

There are a couple of consequences to this scenario. First, the core mass function (CMF) at early stages is likely to be top-heavy, as observed by \cite{Zhang2015,Motte2018a}. Second, despite the subsequent mass growth of the clump, the most-massive cores that are formed within are those that are formed at early stages. This could be explained by radiative feedback disrupting the hubs after the first few massive stars have formed \citep{Geen2017}, or by mechanisms such as fragmentation induced starvation \citep{Peters2010}. By using observations of the optically thick HCO\textsuperscript{+}(1--0) line, \cite{Jackson2019} measured the level of the blue asymmetry — which signifies the presence of gravitational collapse — in a sample of $\sim1000$ MALT90 clumps. The significance of the asymmetry feature was found to decrease as a function of evolutionary stage in these clumps, which would seem to support our proposed reduction in the efficiency of mass concentration over time.

This scenario needs to be further tested by enlarging the sample to cover a wider range of masses, and selected to be representative of the population of Galactic clumps. Mapping the kinematics of these HFS would allow us to look for signatures of clump collapse and accretion, infer whether the filaments in these hubs are really converging, and investigate any link with various core properties. We will address the latter in a following study.

\section*{Acknowledgements}
	MA is supported by the Science and Technology Facilities Council (STFC). NP and AJR acknowledges the support of the STFC consolidated grant number ST/S00033X/1. ADC acknowledges the support from the Royal Society University Research Fellowship (URF/R1/191609). GMW acknowledges support from STFC under grant number ST/R000905/1. 
	
	This paper makes use of the following ALMA data: ADS/JAO.ALMA\#2011.0.00474.S, ADS/JAO.ALMA\#2015.1.01014.S, ADS/JAO.ALMA\#2016.1.00810.S, and ADS/JAO.ALMA\#2013.1.00960.S. ALMA is a partnership of ESO (representing its member states), NSF (USA) and NINS (Japan), together with NRC (Canada), MOST and ASIAA (Taiwan), and KASI (Republic of Korea), in cooperation with the Republic of Chile. The Joint ALMA Observatory is operated by ESO, AUI/NRAO and NAOJ.
	We would like to thank Timea Csengeri for supplying us with the additional ALMA data.
	
	This research made use of the Python packages Astropy\footnote{\url{https://astropy.org/}} \citep{AstropyCollaboration2013,AstropyCollaboration2018}, astrodendro\footnote{\url{http://dendrograms.org/}}, IPython\footnote{\url{https://ipython.org/}} \citep{Perez2007}, NumPy\footnote{\url{https://numpy.org/}} \citep{Harris2020}, SciPy\footnote{\url{https://scipy.org/}} \citep{Virtanen2020}, Matplotlib\footnote{\url{https://matplotlib.org/}} \citep{Hunter2007}, and scikit-image\footnote{\url{https://scikit-image.org}} \citep{vanderWalt2014}. 
	This research also made use of NASA's Astrophysics Data System Bibliographic Services, TOPCAT\footnote{\url{http://www.star.bris.ac.uk/~mbt/topcat/}} \citep{Taylor2005}, and SAOImageDS9\footnote{\url{http://ds9.si.edu/}} \citep{Joye2003}.

\section*{Data Availability}

\emph{The data underlying this article are available in the article and in its online supplementary material. Any additional data will be shared on reasonable request to the corresponding author.}



\bibliographystyle{mnras}
\bibliography{hfs-paper-1} 




\appendix

	\onecolumn
\section{ALMA 2.9\si{~\mm} continuum images}
\label{sec:ALMA_cutouts}
\begin{figure*}
	\centering
	\subfloat{
	\includegraphics[width=0.35\textwidth]{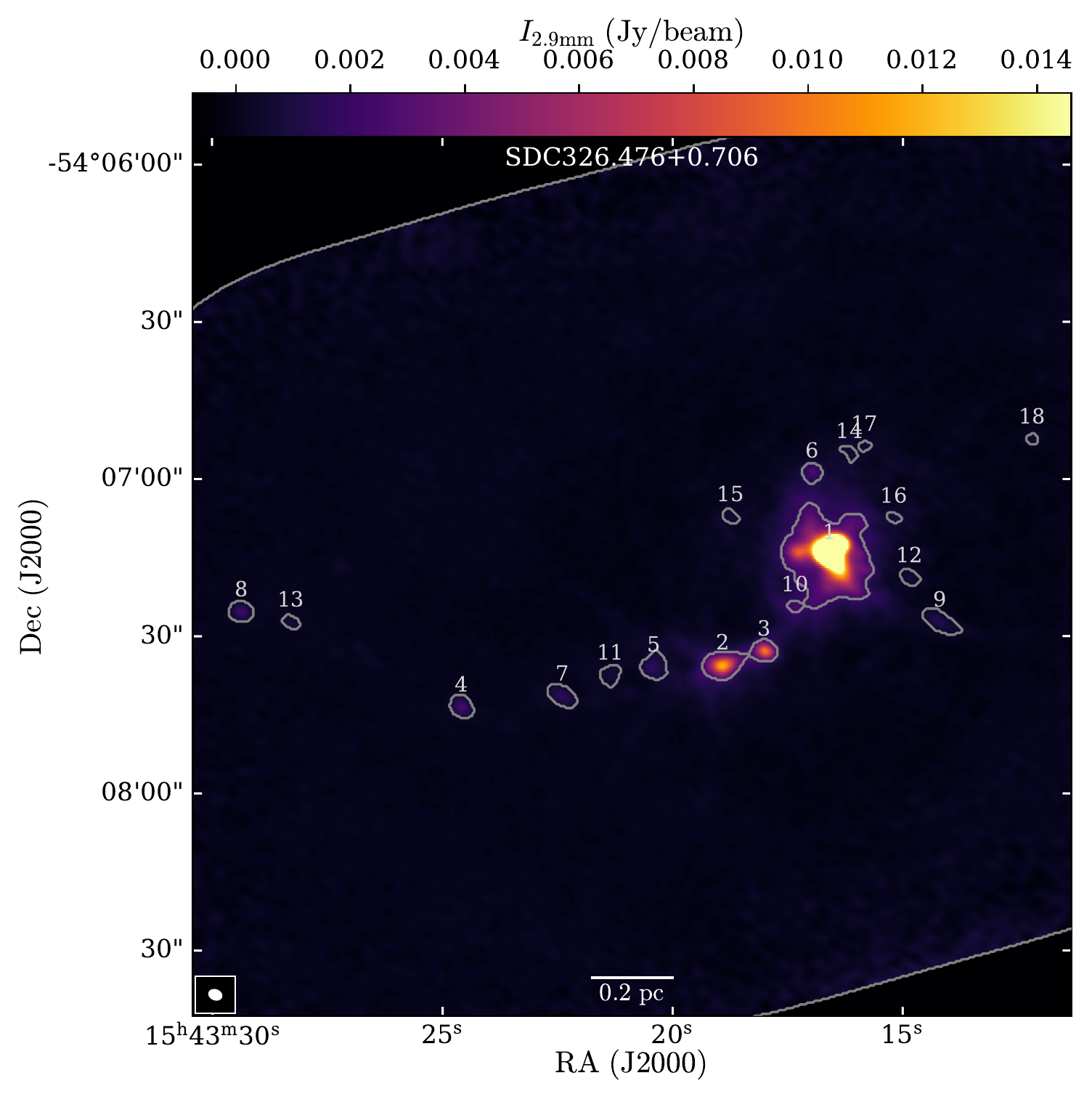}
	}
	\subfloat{
	\includegraphics[width=0.35\textwidth]{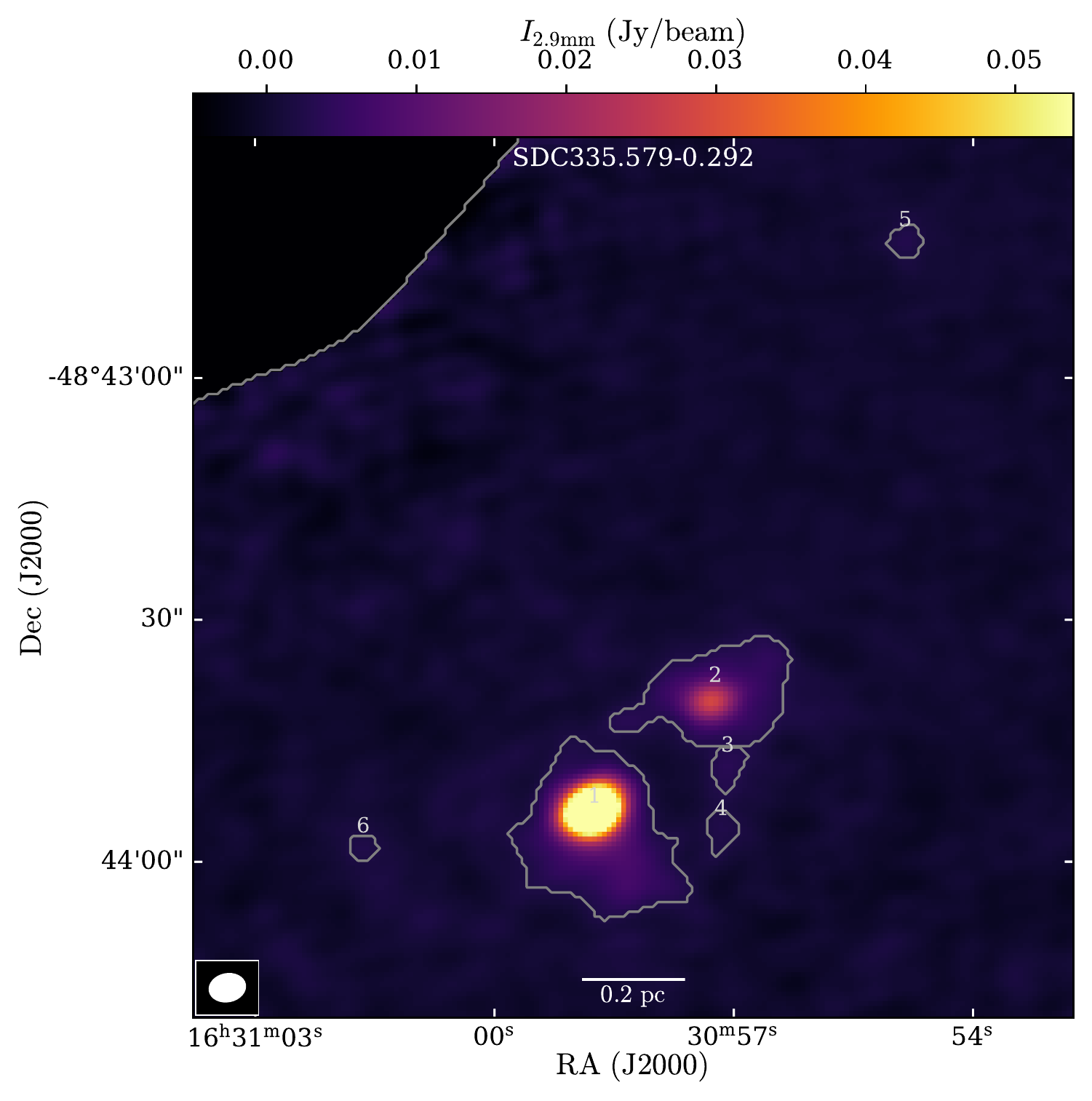}
	}\\[-5mm]
	\subfloat{
	\includegraphics[width=0.35\textwidth]{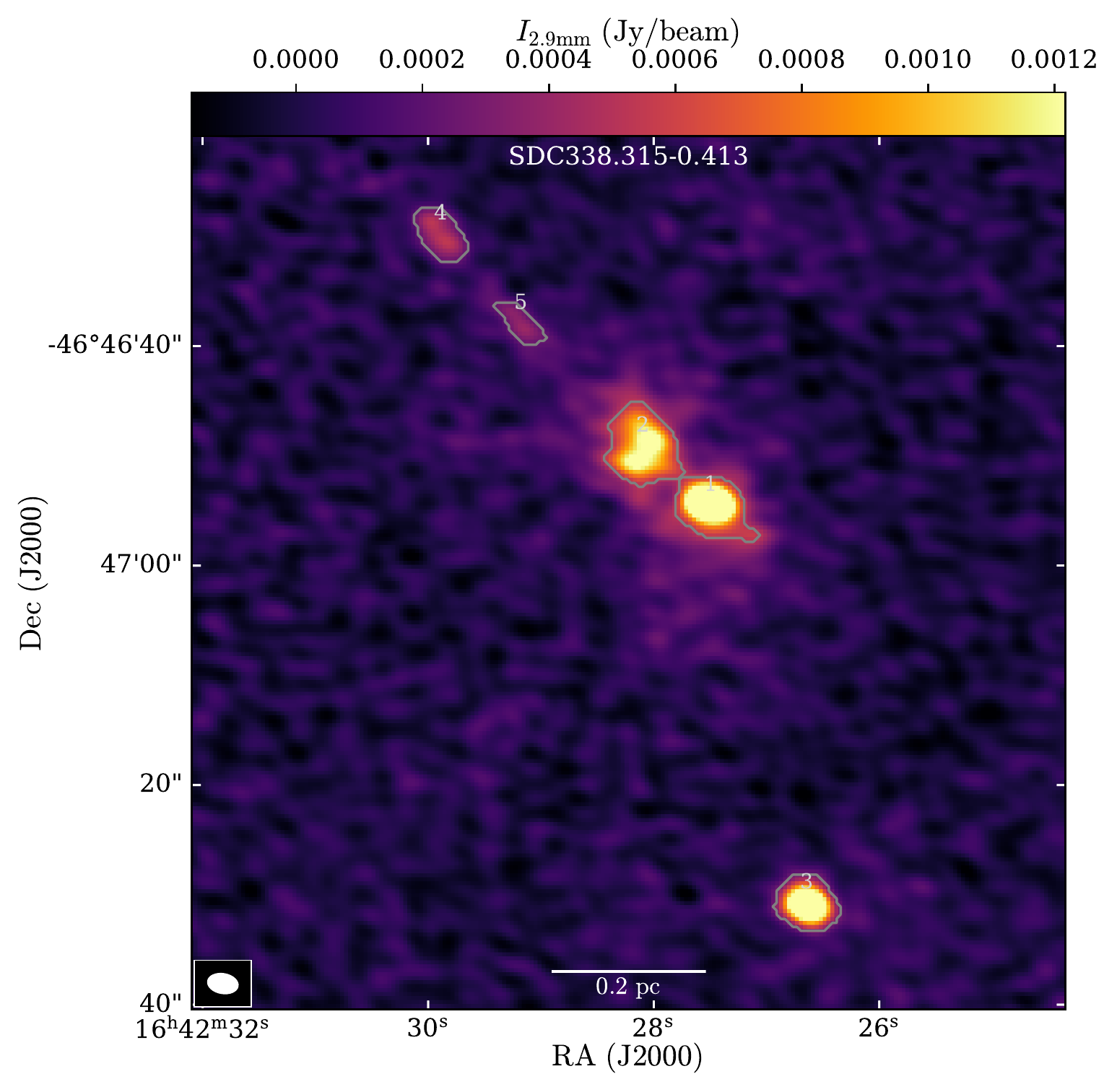}
	}
	\subfloat{
	\includegraphics[width=0.35\textwidth]{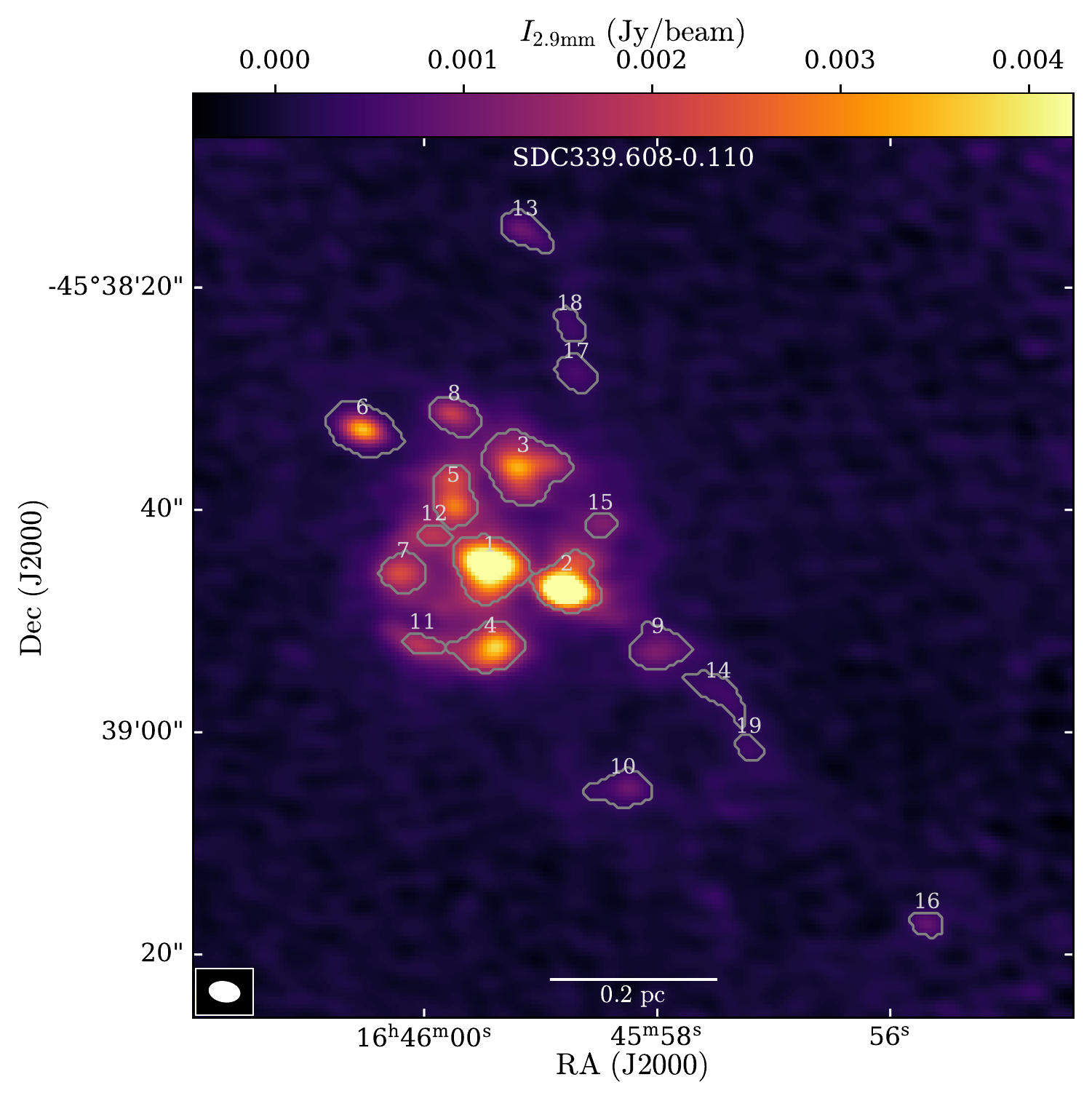}
	}\\[-5mm]
	\subfloat{
	\includegraphics[width=0.35\textwidth]{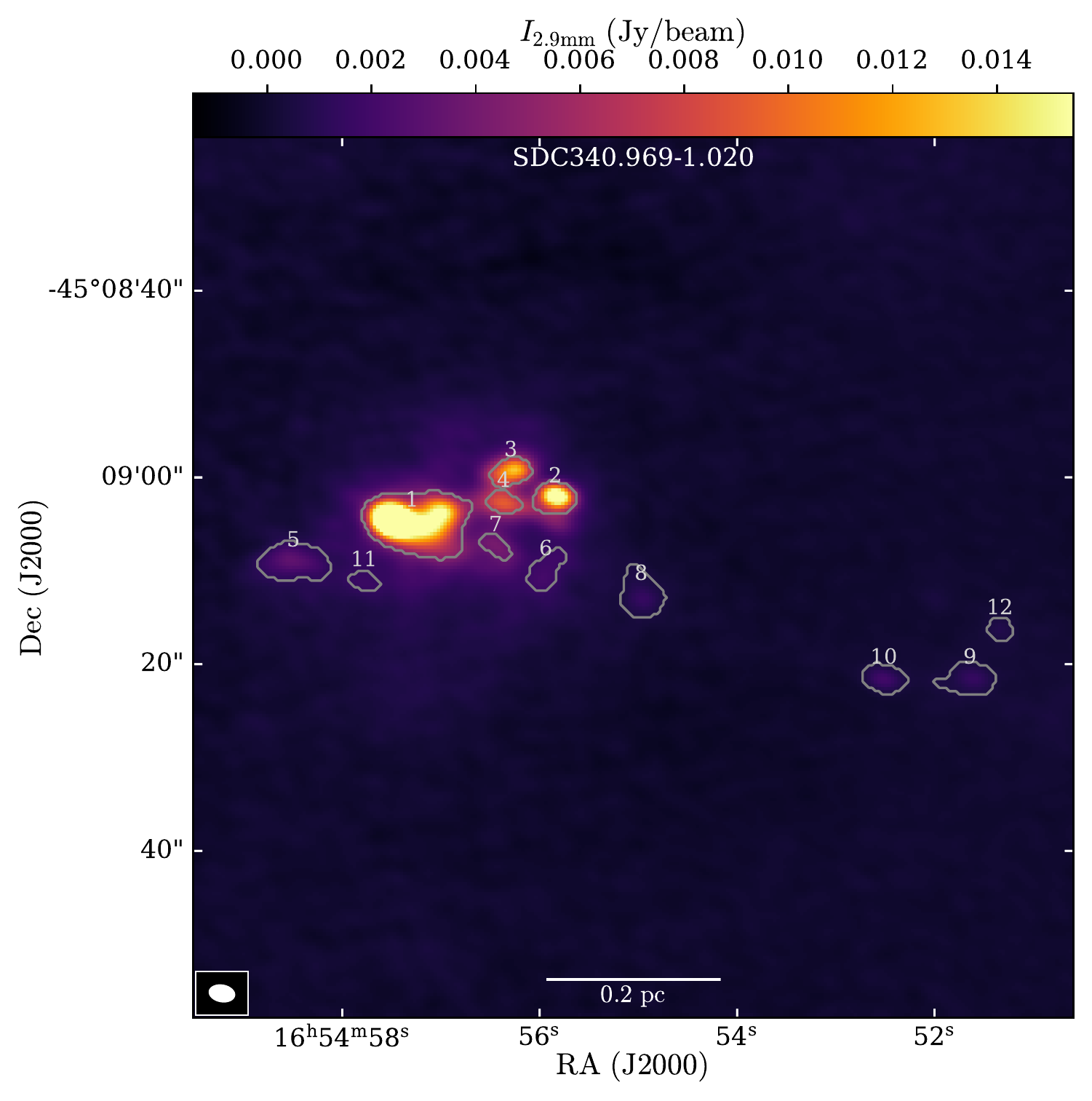}
	}
	\subfloat{
	\includegraphics[width=0.35\textwidth]{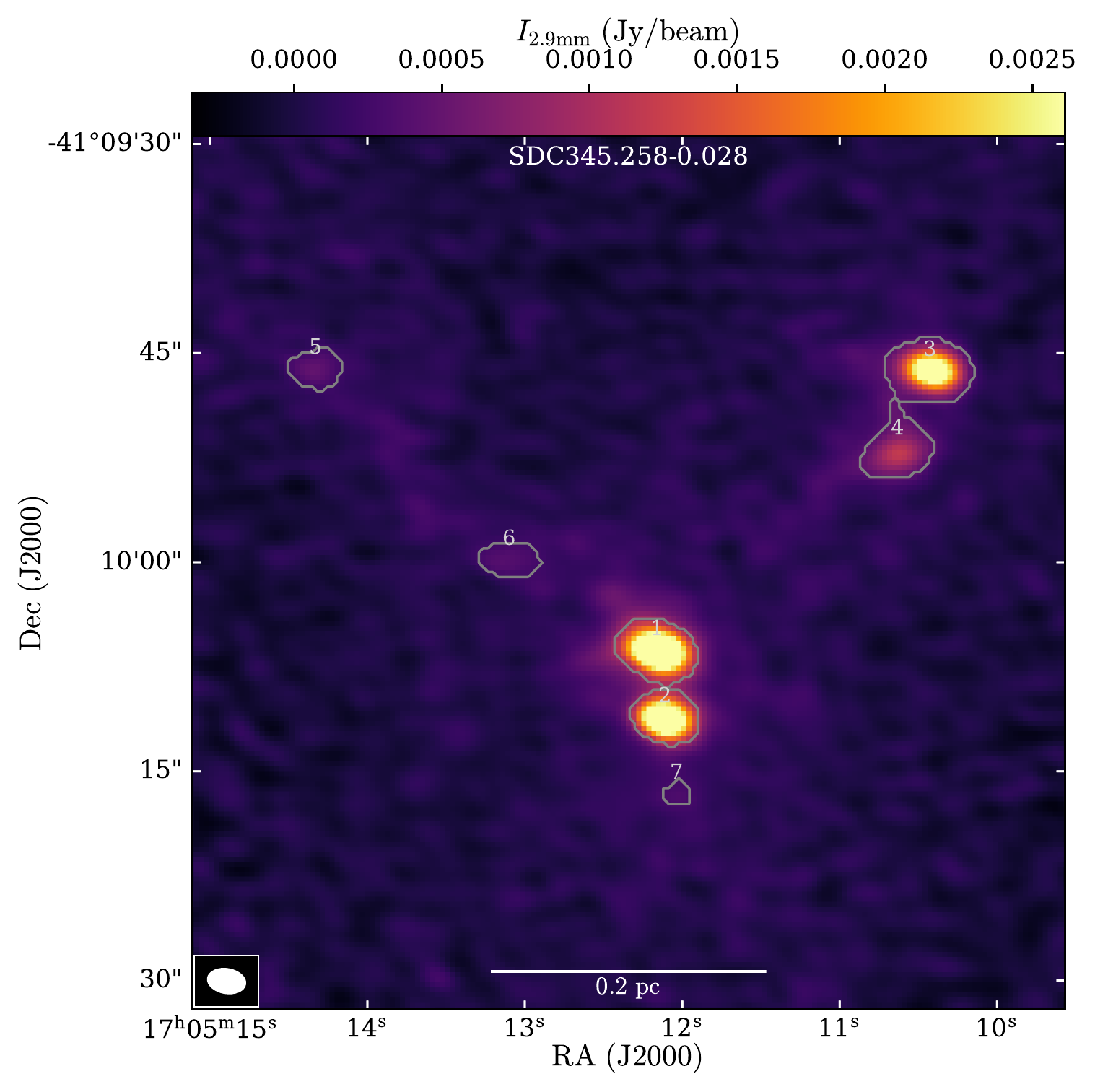}
	}
	\caption{Close up view of ALMA 2.9\si{~\mm} continuum images of our sample of 6 hub-filament systems, showing all of our extracted ``cores''. Each extracted core is labeled with their MM\#, with the grey contours showing each core's dendrogram structure footprint.
	}\label{fig:ALMA_cutouts}
\end{figure*}
\newpage
\section{\emph{Spitzer} 8\si{~\micron} Clump Cutouts}
\label{sec:Spitzer_cutouts}
\begin{figure*}
	\centering
	\subfloat{
	\includegraphics[width=43mm]{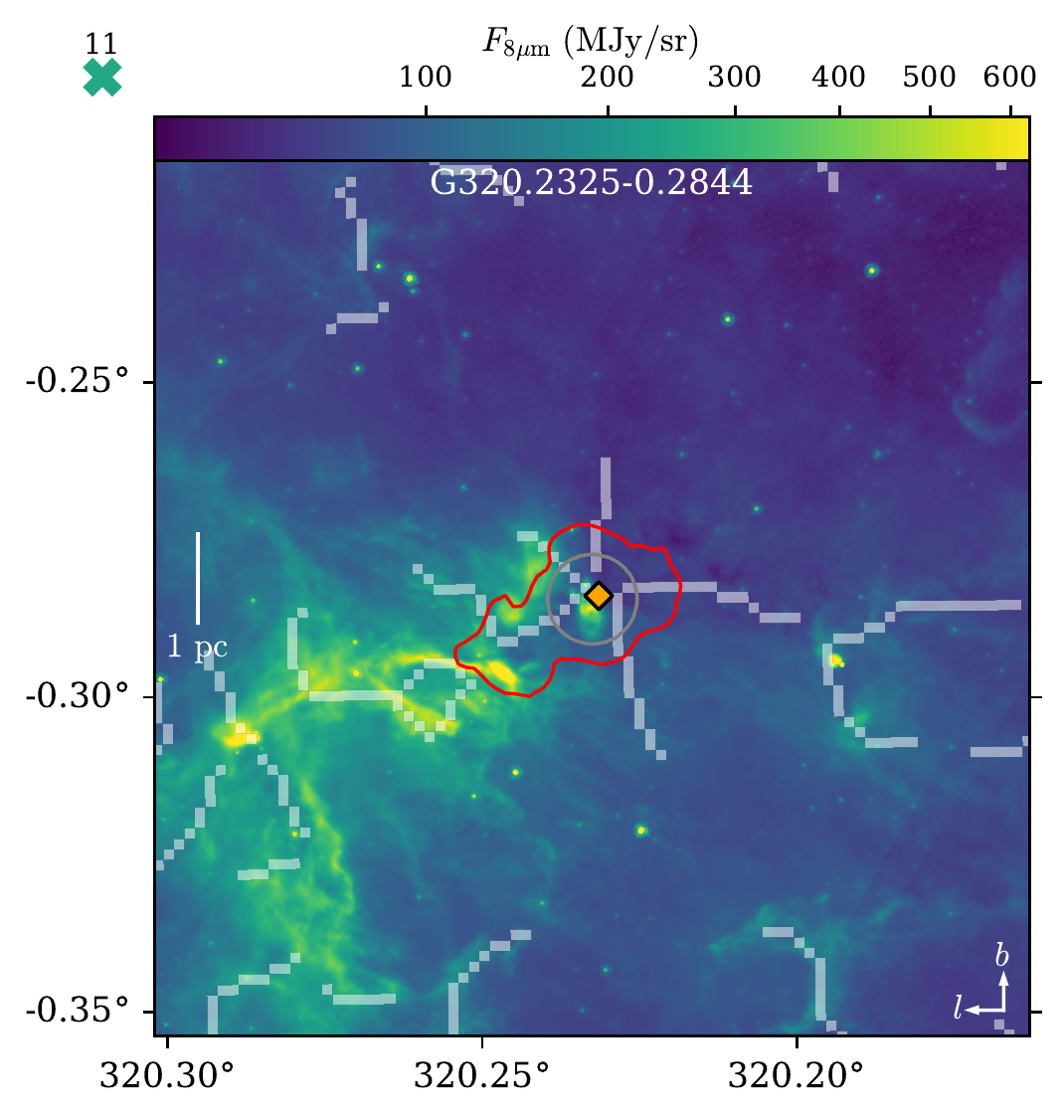}
	}
	\subfloat{
	\includegraphics[width=43mm]{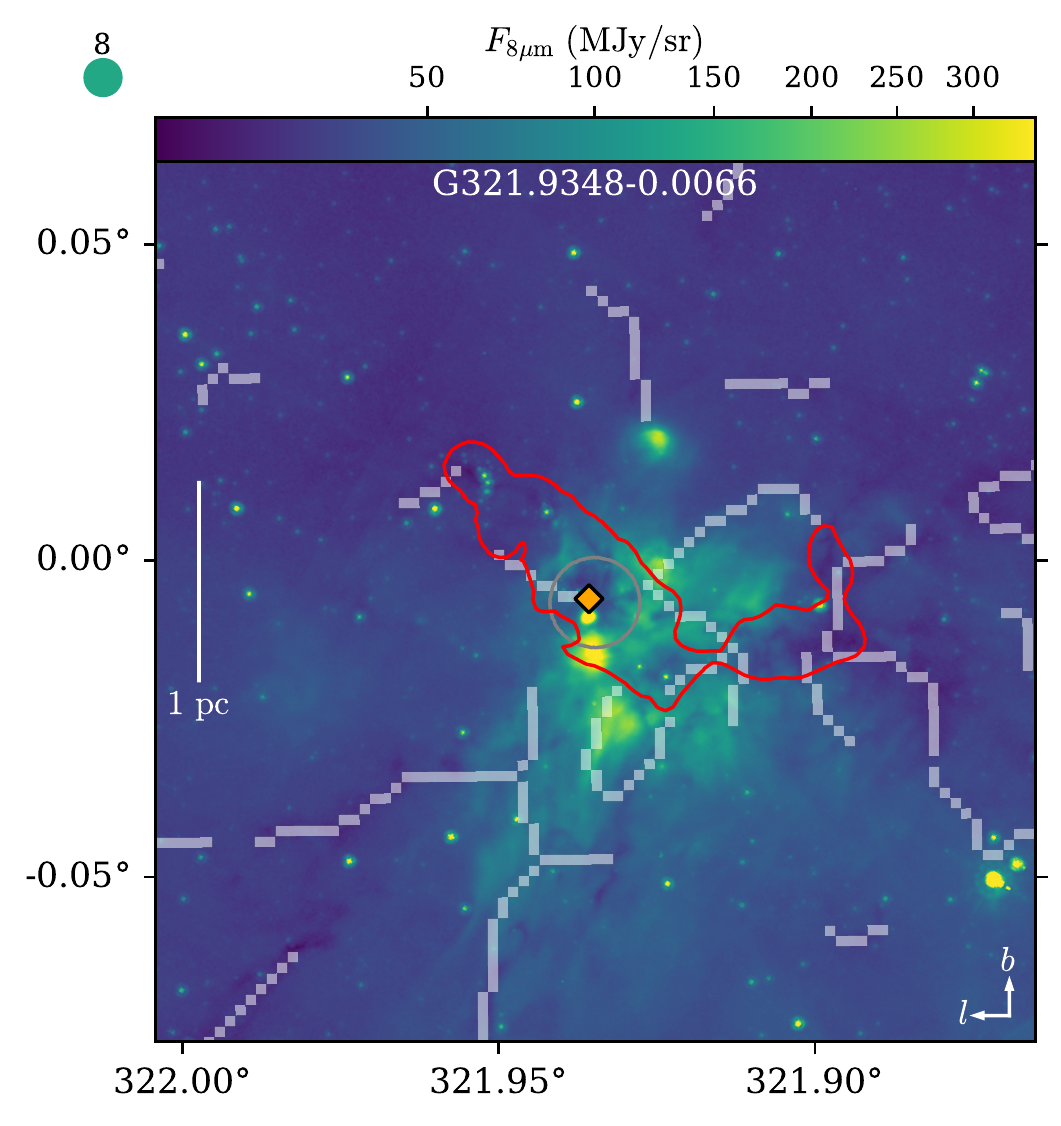}
	}
	\subfloat{
	\includegraphics[width=43mm]{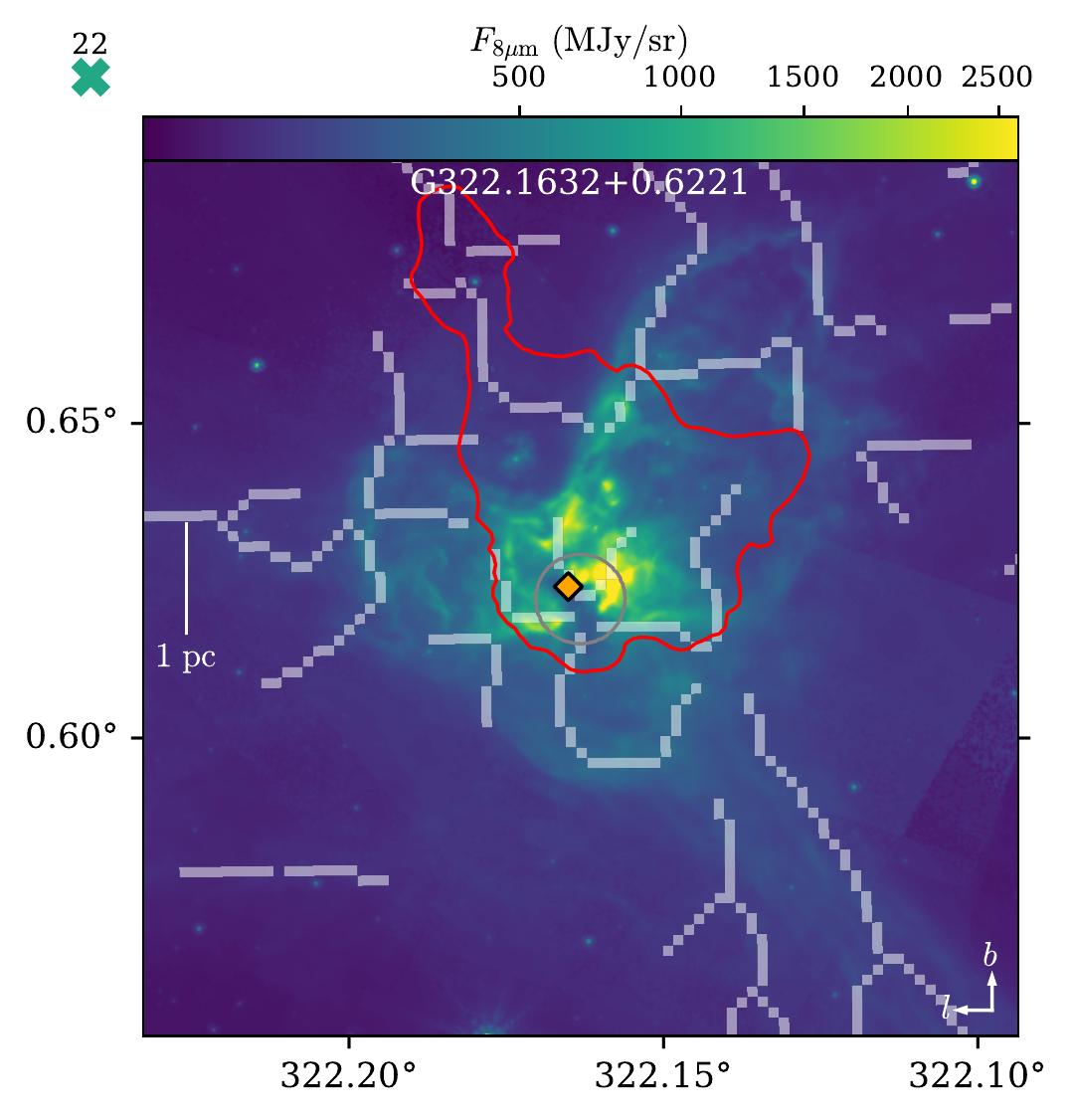}
	}\\[-5mm]
	\hspace{0mm}	
	\subfloat{
	\includegraphics[width=43mm]{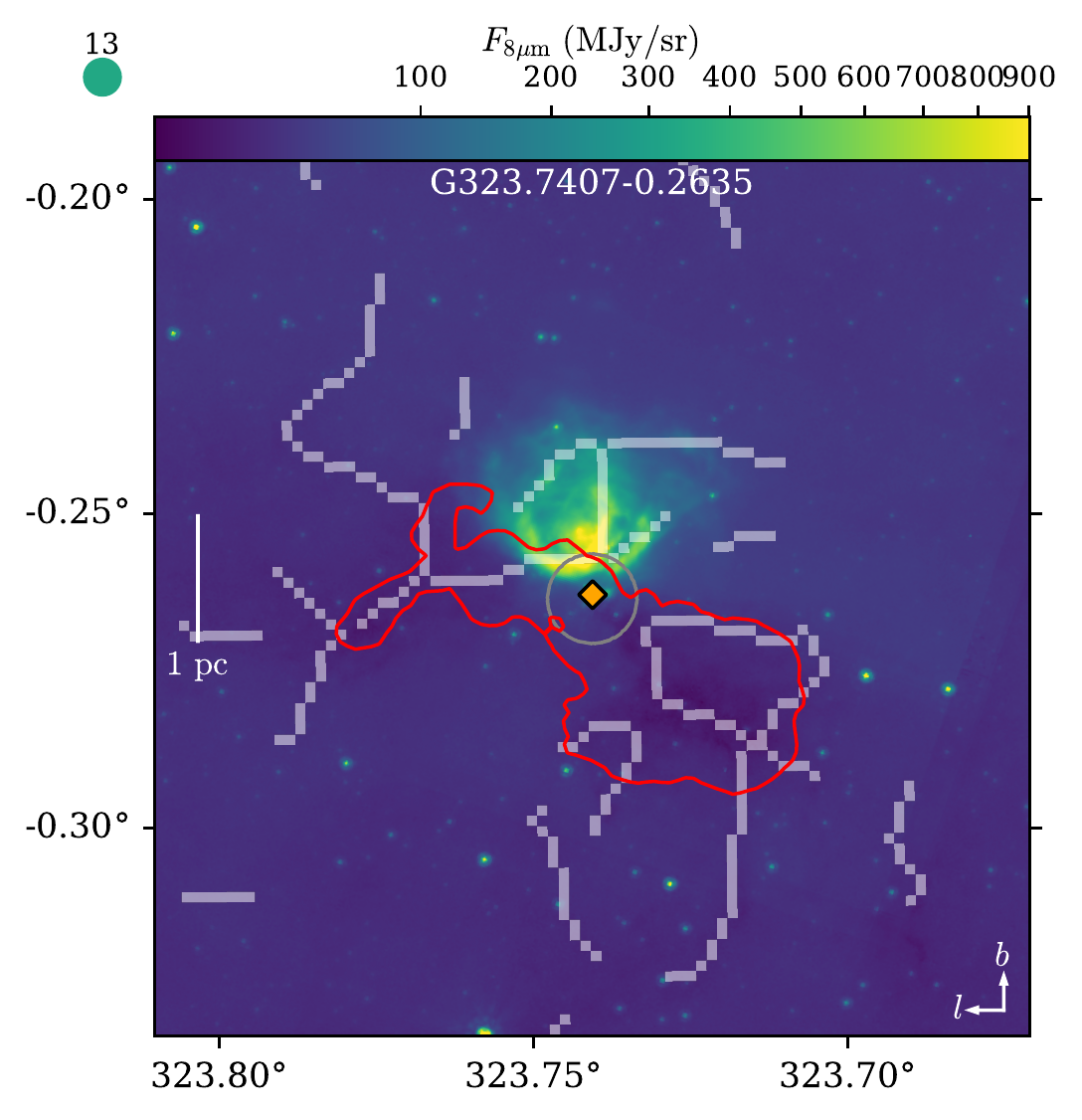}
	}
	\subfloat{
	\includegraphics[width=43mm]{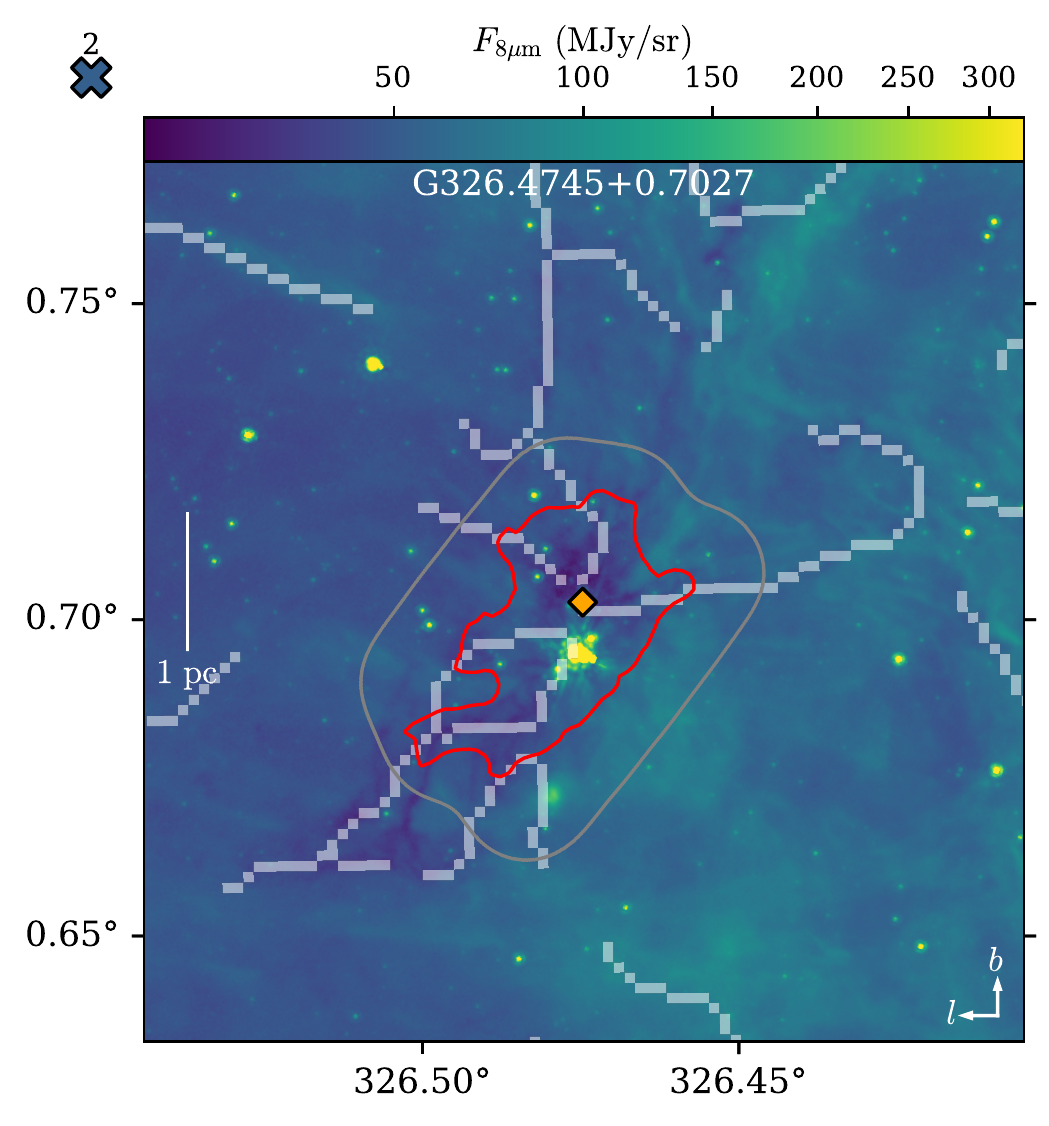}
	}
	\subfloat{
	\includegraphics[width=43mm]{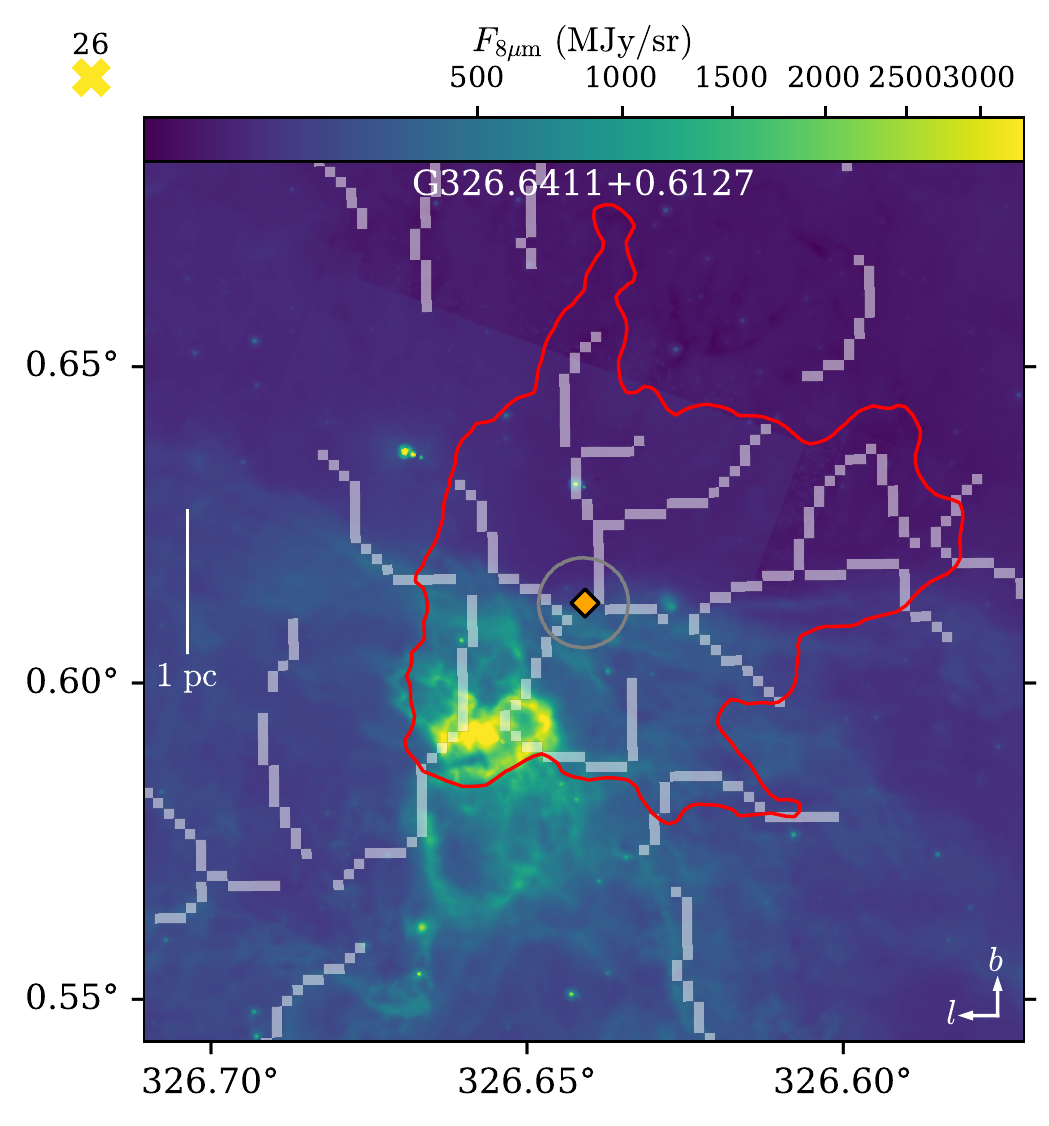}
	}\\[-5mm]
	\hspace{0mm}
	\subfloat{
	\includegraphics[width=43mm]{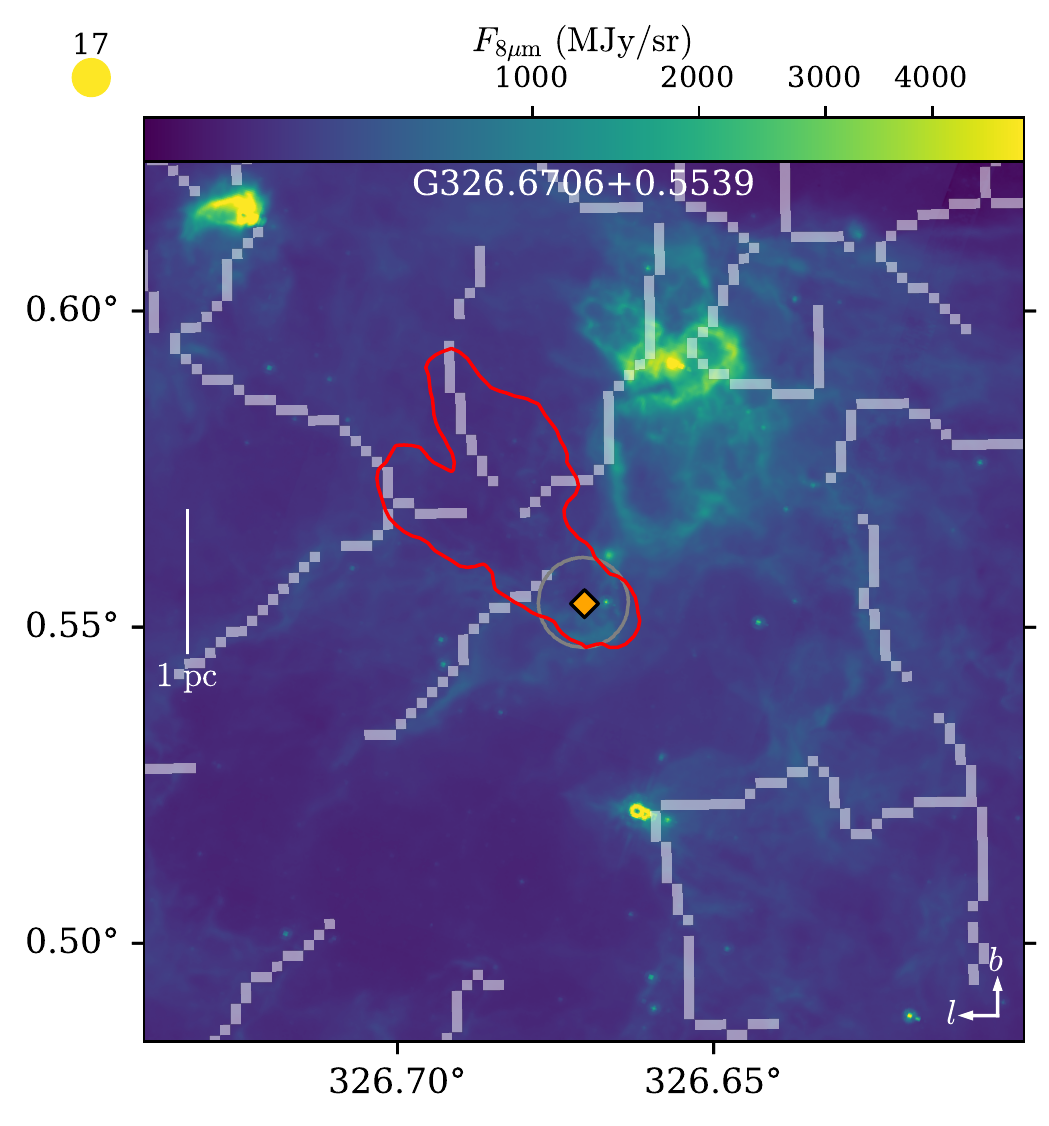}
	}
	\subfloat{
	\includegraphics[width=43mm]{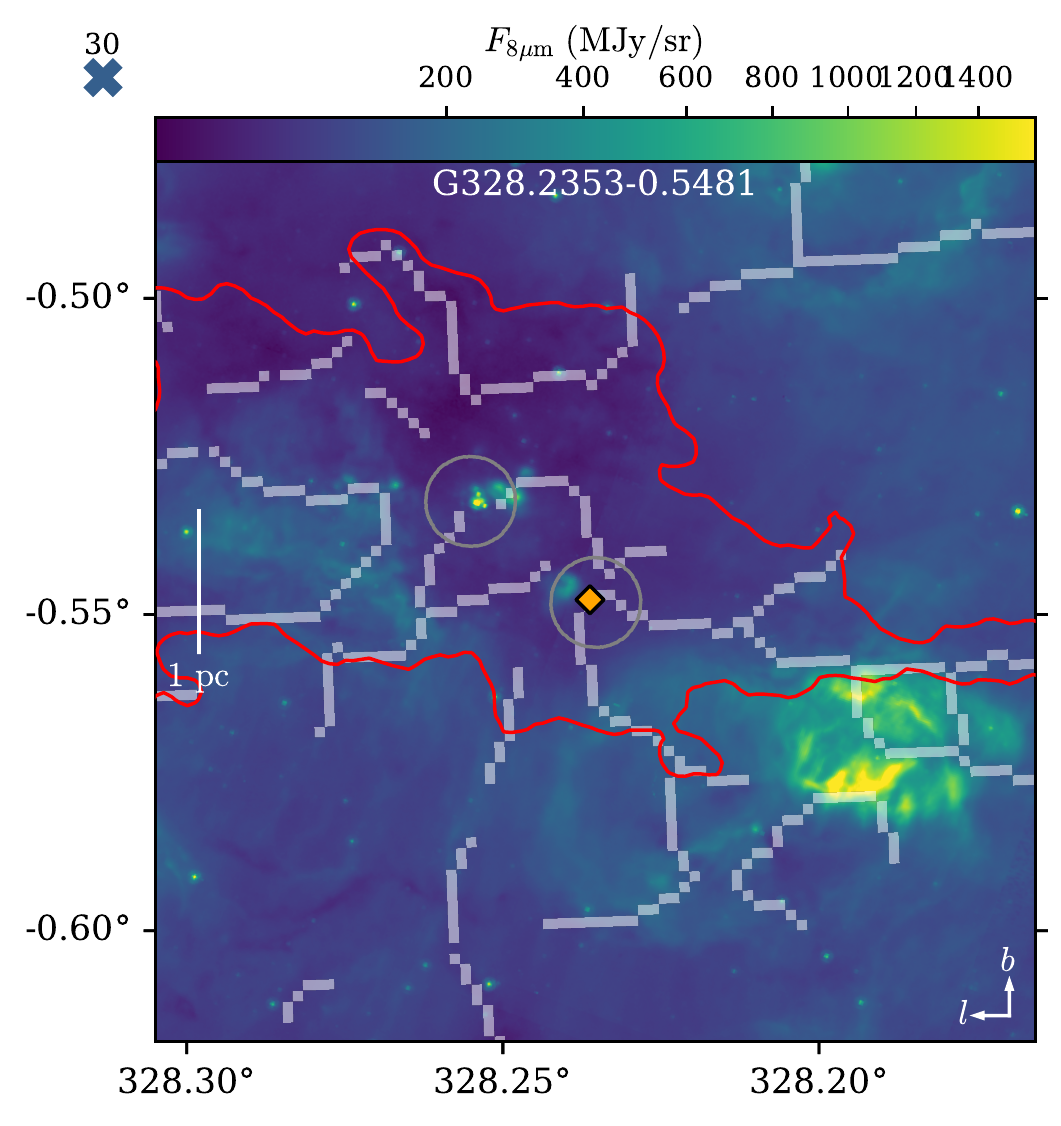}
	}
	\subfloat{
	\includegraphics[width=43mm]{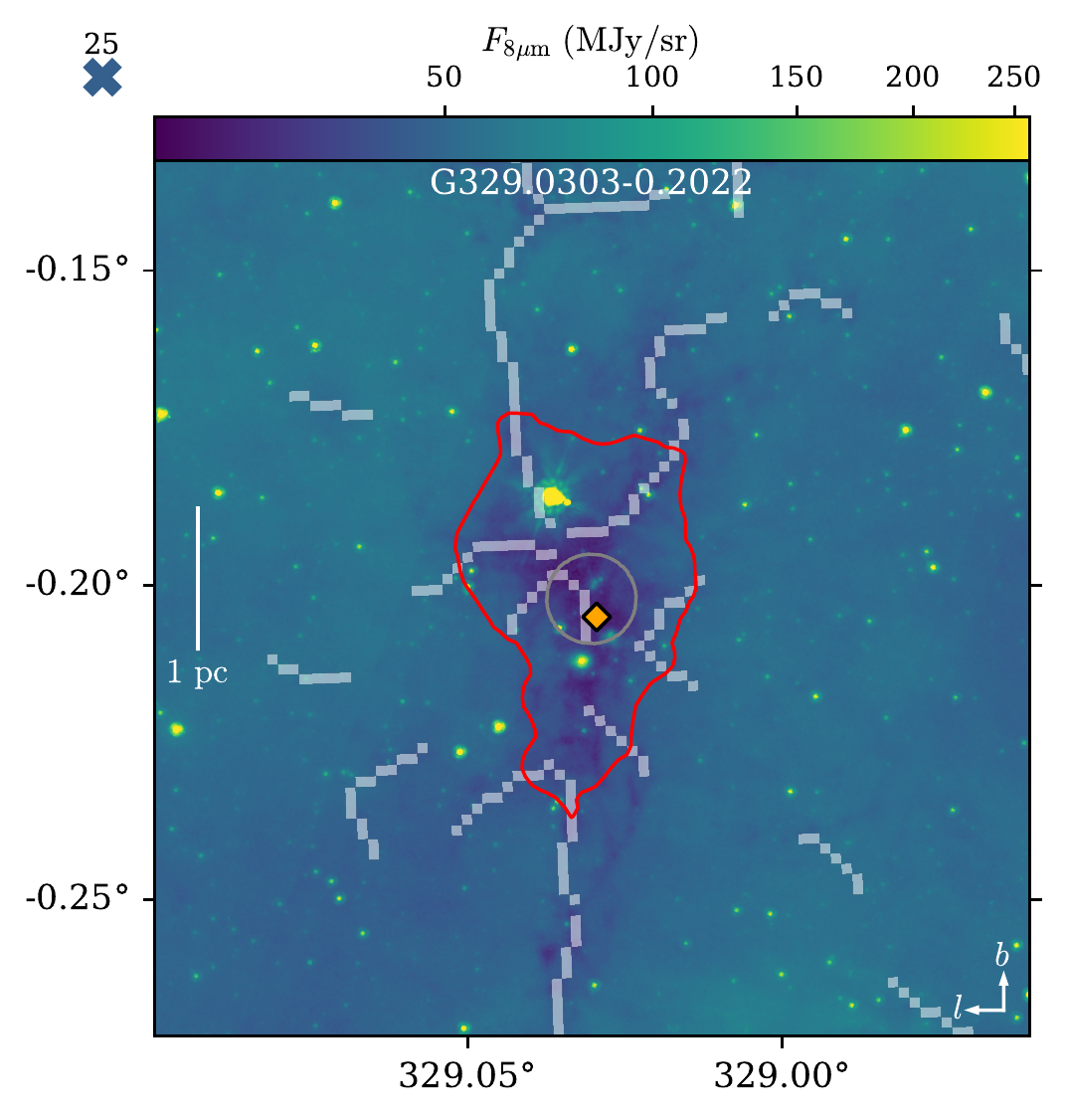}
	}\\[-5mm]
	\hspace{0mm}
	\subfloat{
	\includegraphics[width=43mm]{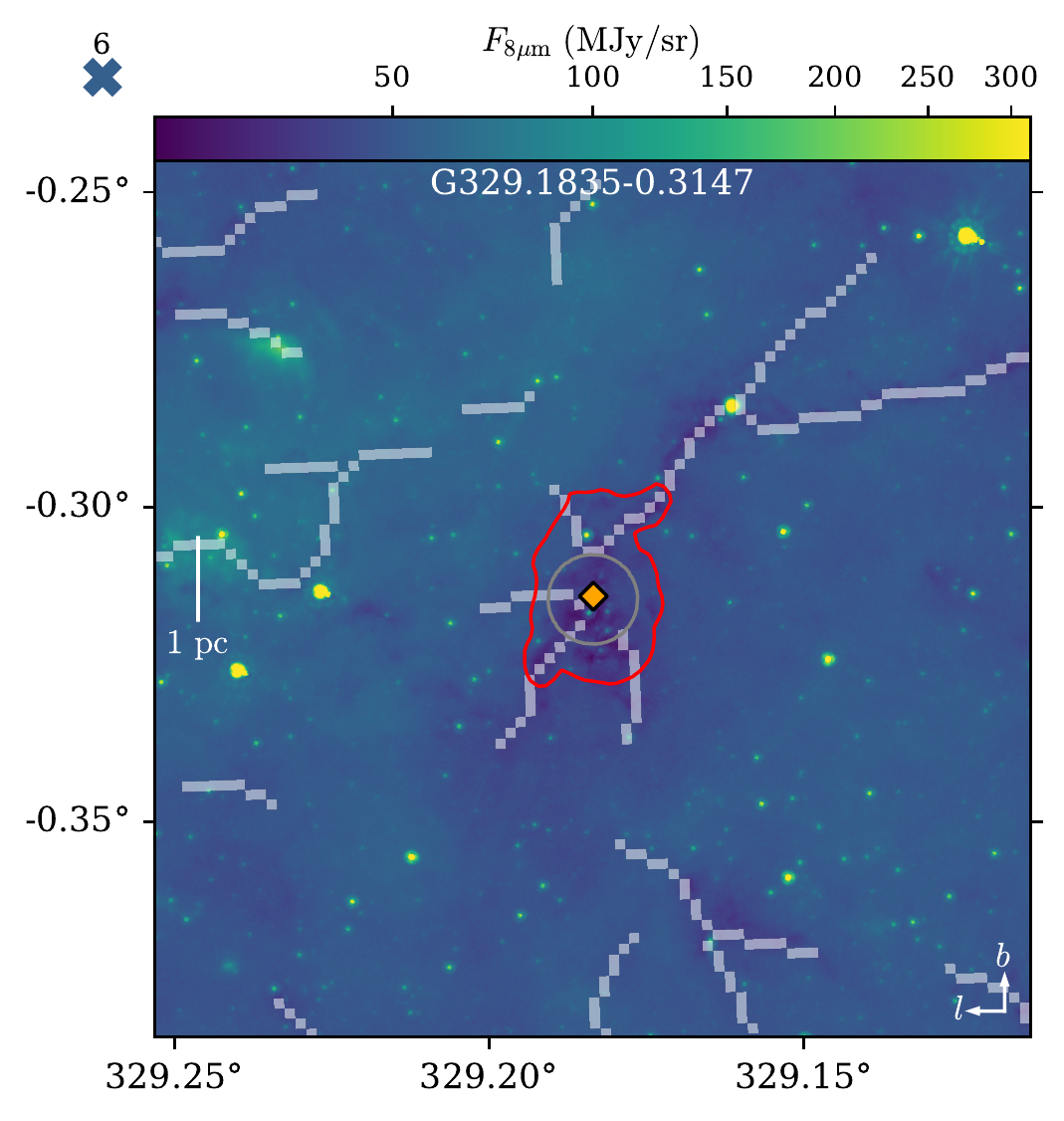}
	}
	\subfloat{
	\includegraphics[width=43mm]{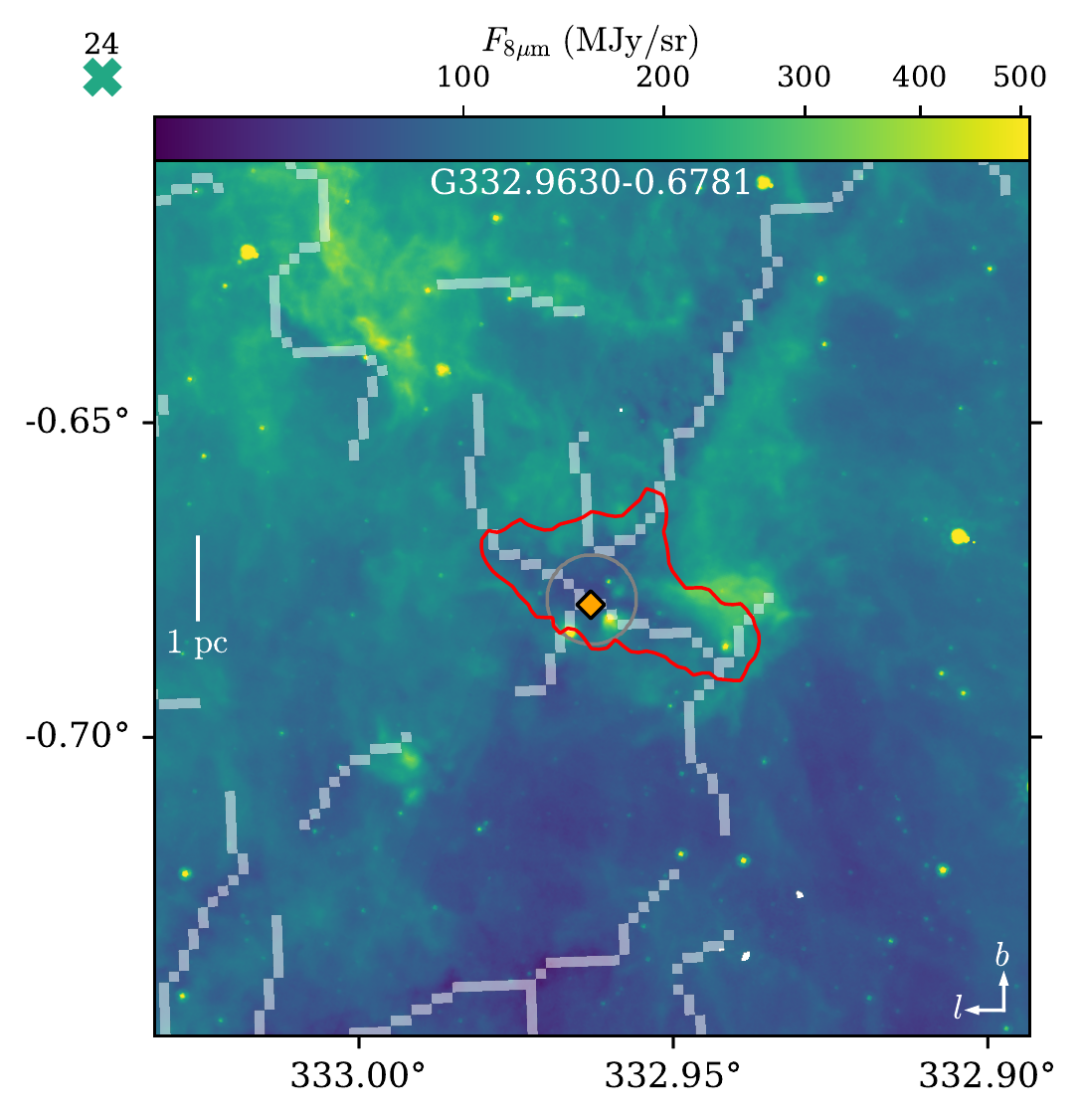}
	}
	\subfloat{
	\includegraphics[width=43mm]{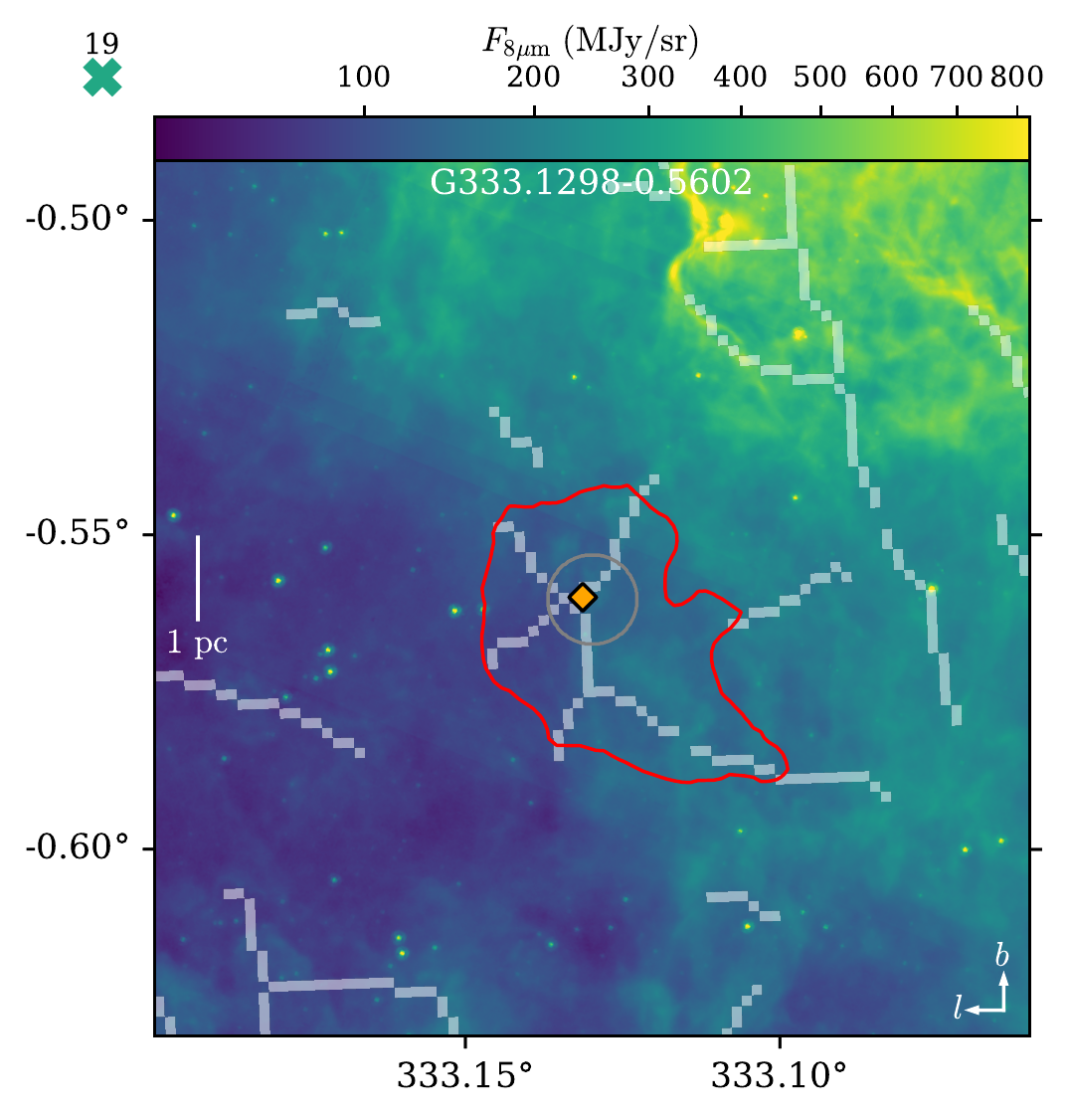}
	}
	\caption{\emph{Spitzer} 8\si{~\micron} images of each clump in our sample. The images are in Galactic longitude and latitude ($l,b$), with the axes marked in the lower right. The red contours represent the ``boundary'' of our clumps defined by the $N_{\mathrm{H}_2}=3\times10^{22}\si{~\cm^{-2}}$ level in our \emph{Herschel} column density maps. The orange diamond marks the location of the most-massive core (MMC) in each clump, and the grey contours show the ALMA field of view coverage for each object. The filamentary structures identified in the \emph{Herschel} 250\si{~\micron} maps are overlaid in white. The symbol to the upper left represents how we classified each clump, along with the clump ID number. Crosses represent clumps that have been classified as HFS, and circular points are non-HFS clumps. The point fill colours represent the three IR-brightness classes. Points with a black outline are sources observed at 2.9\si{~\milli\m}, and points without outlines are sources observed at 878\si{~\micron}. 
	}\label{fig:Spitzer_cutouts}
\end{figure*}
\begin{figure*}\ContinuedFloat
	\centering
	\subfloat{
	\includegraphics[width=43mm]{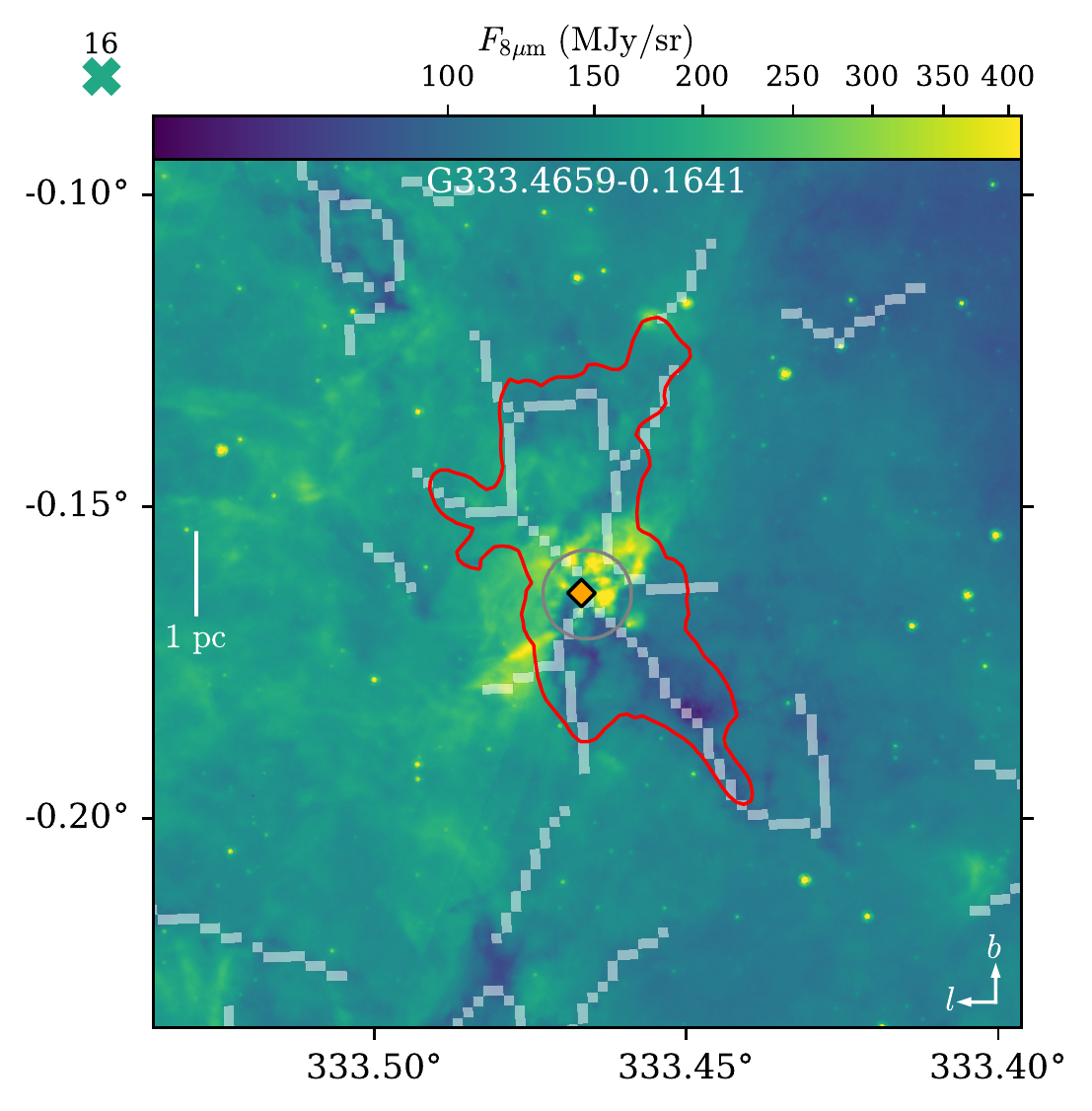}
	}
	\subfloat{
	\includegraphics[width=43mm]{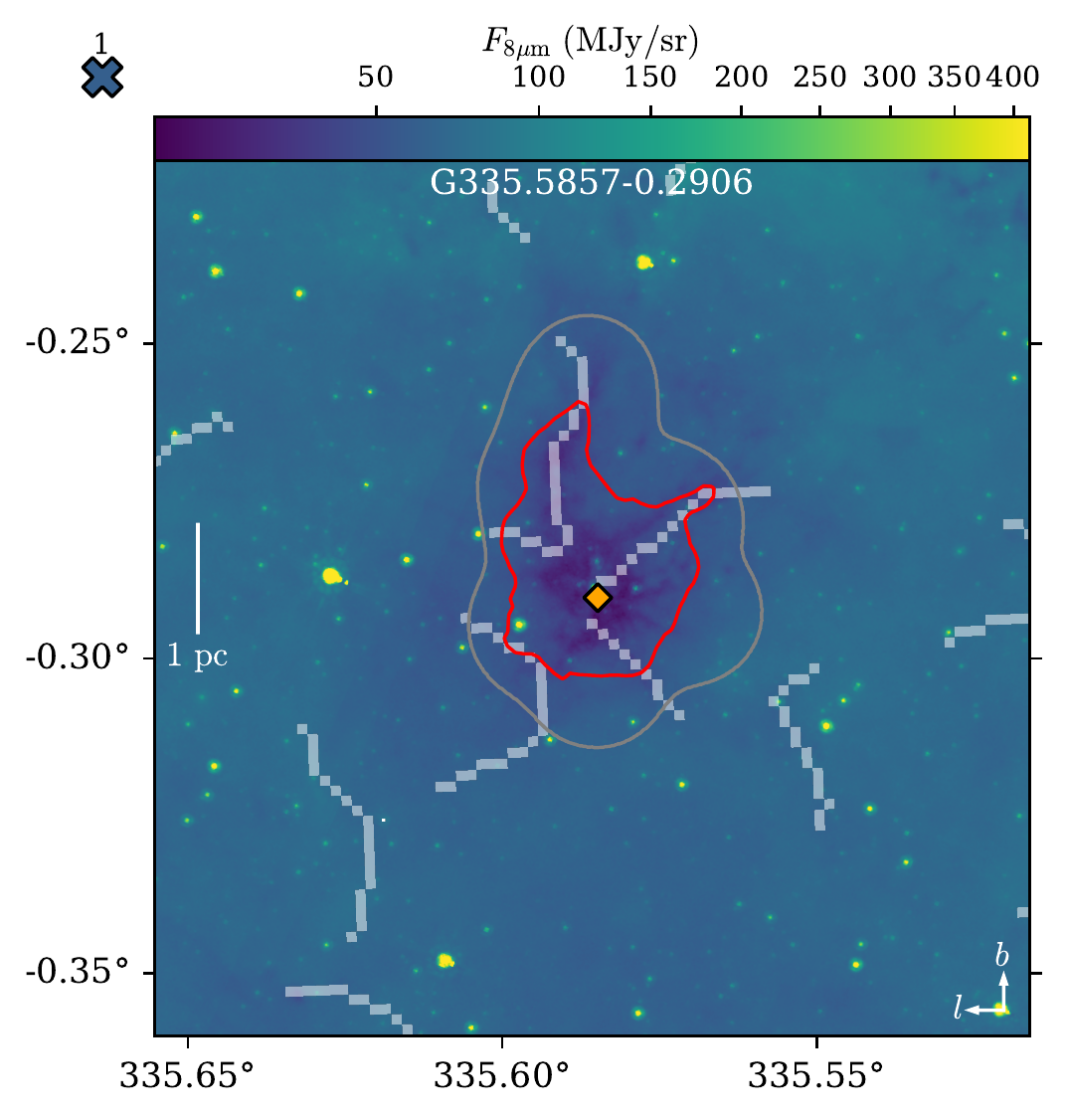}
	}
	\subfloat{
	\includegraphics[width=43mm]{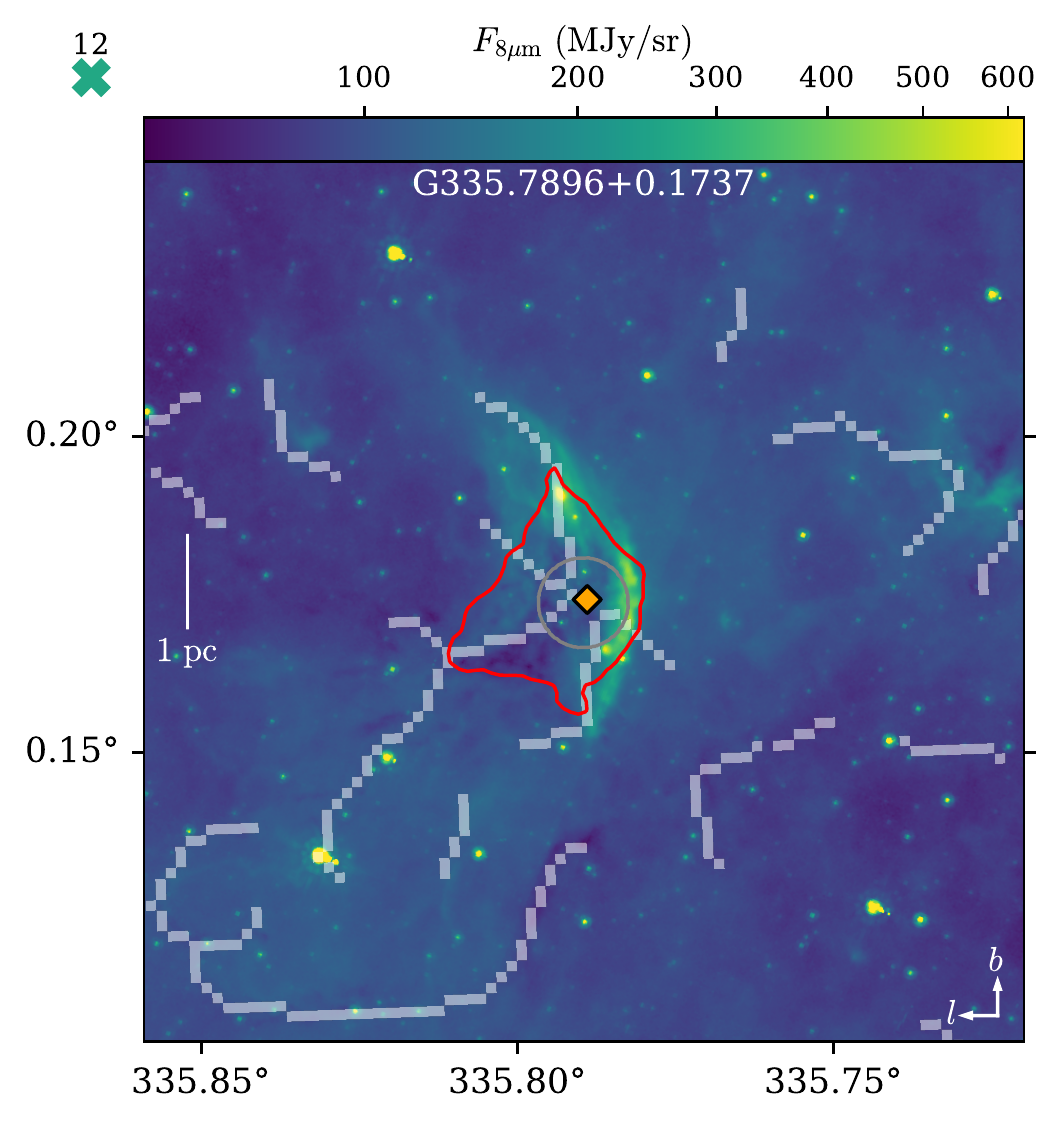}
	}\\[-5mm]
	\hspace{0mm}
	\subfloat{
	\includegraphics[width=43mm]{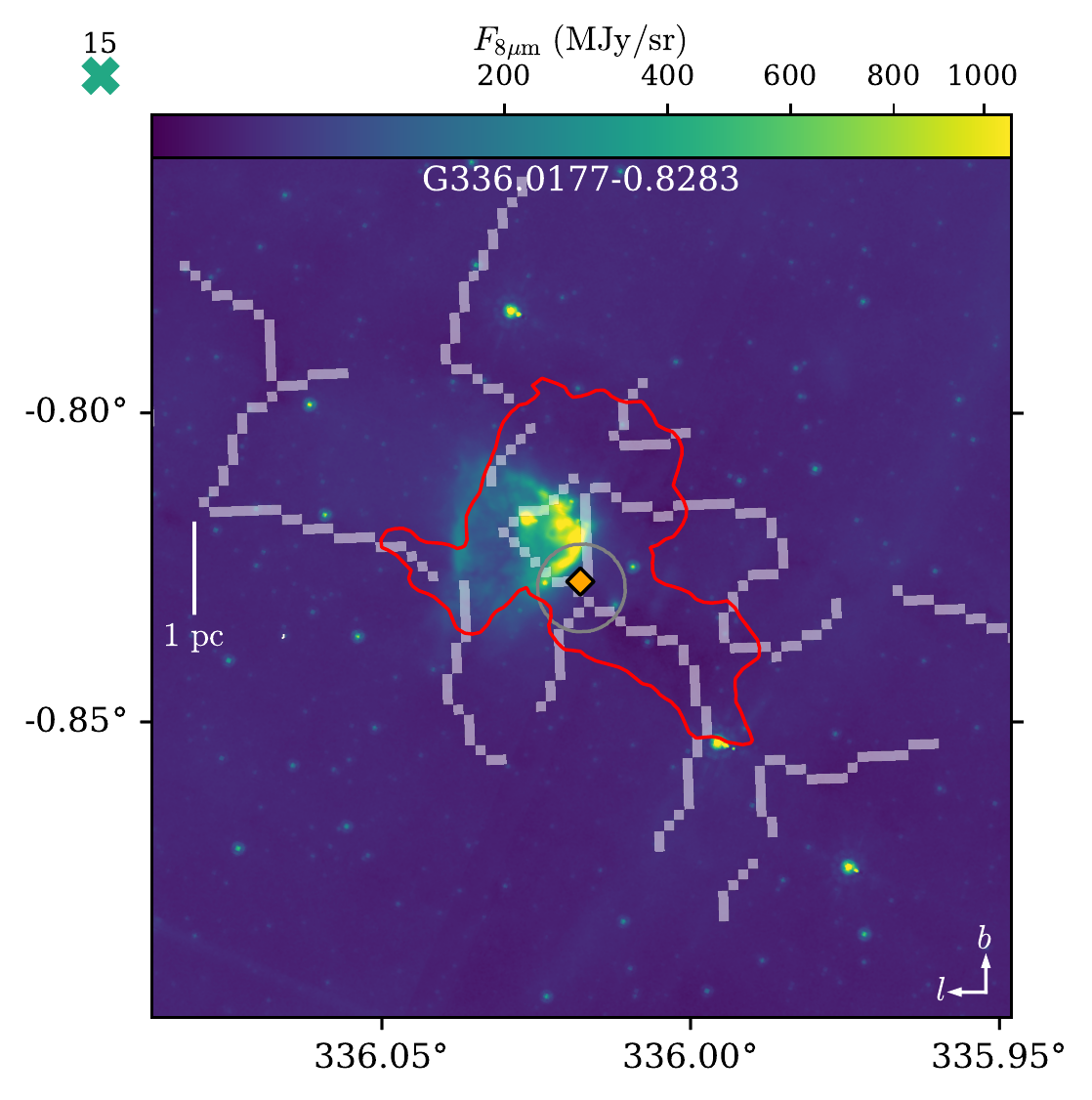}
	}
	\subfloat{
	\includegraphics[width=43mm]{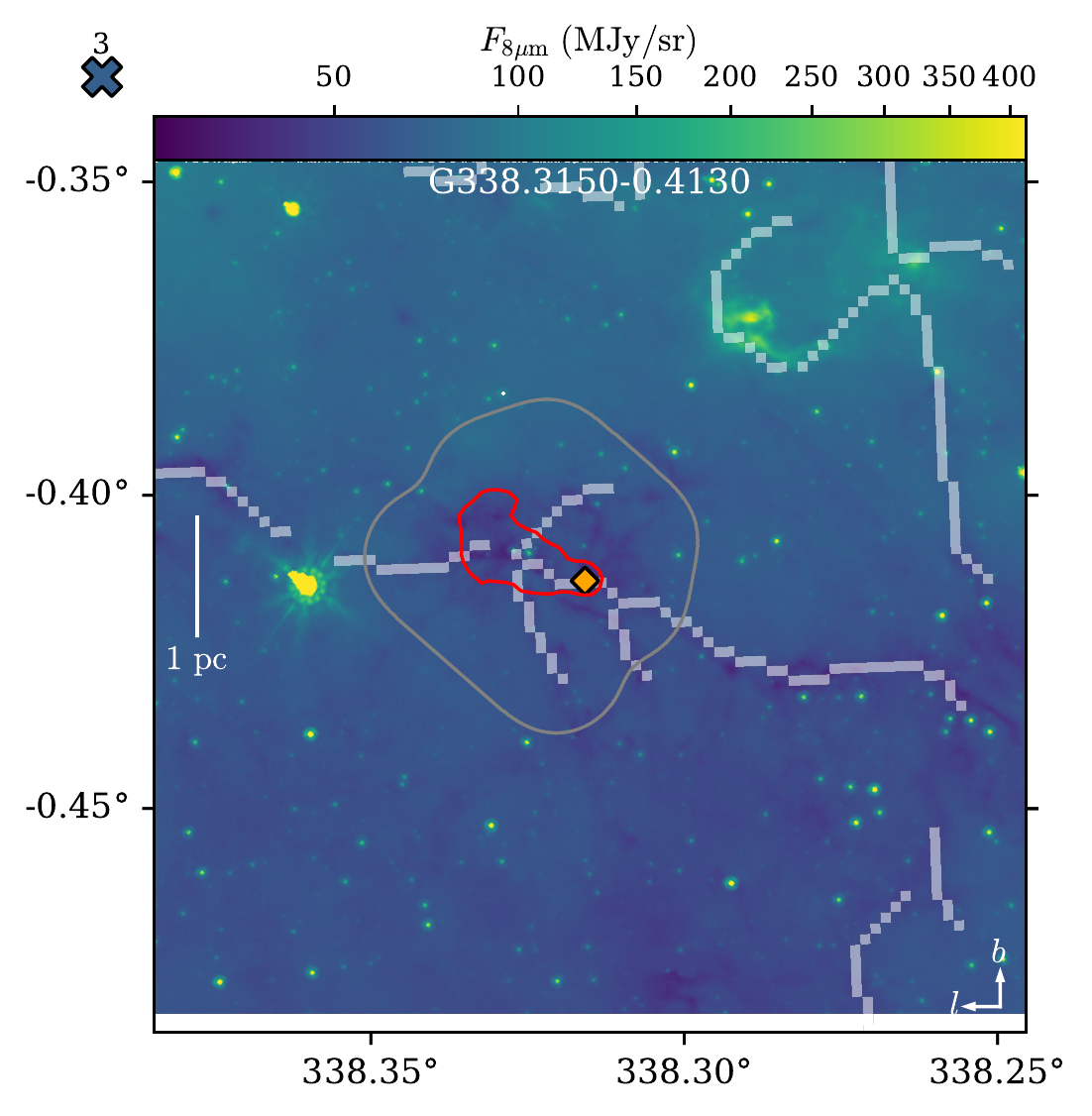}
	}
	\subfloat{
	\includegraphics[width=43mm]{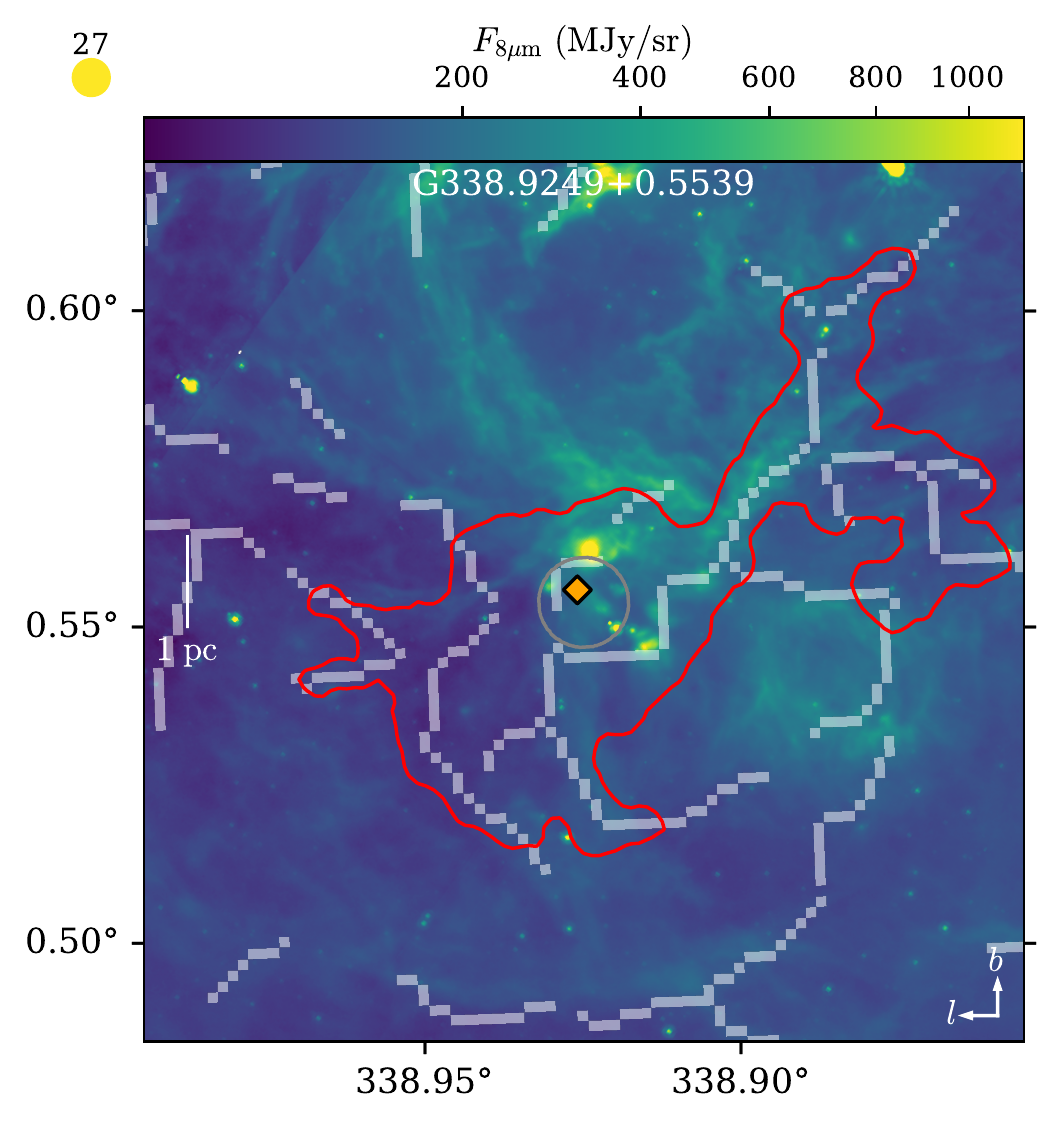}
	}\\[-5mm]
	\hspace{0mm}
	\subfloat{
	\includegraphics[width=43mm]{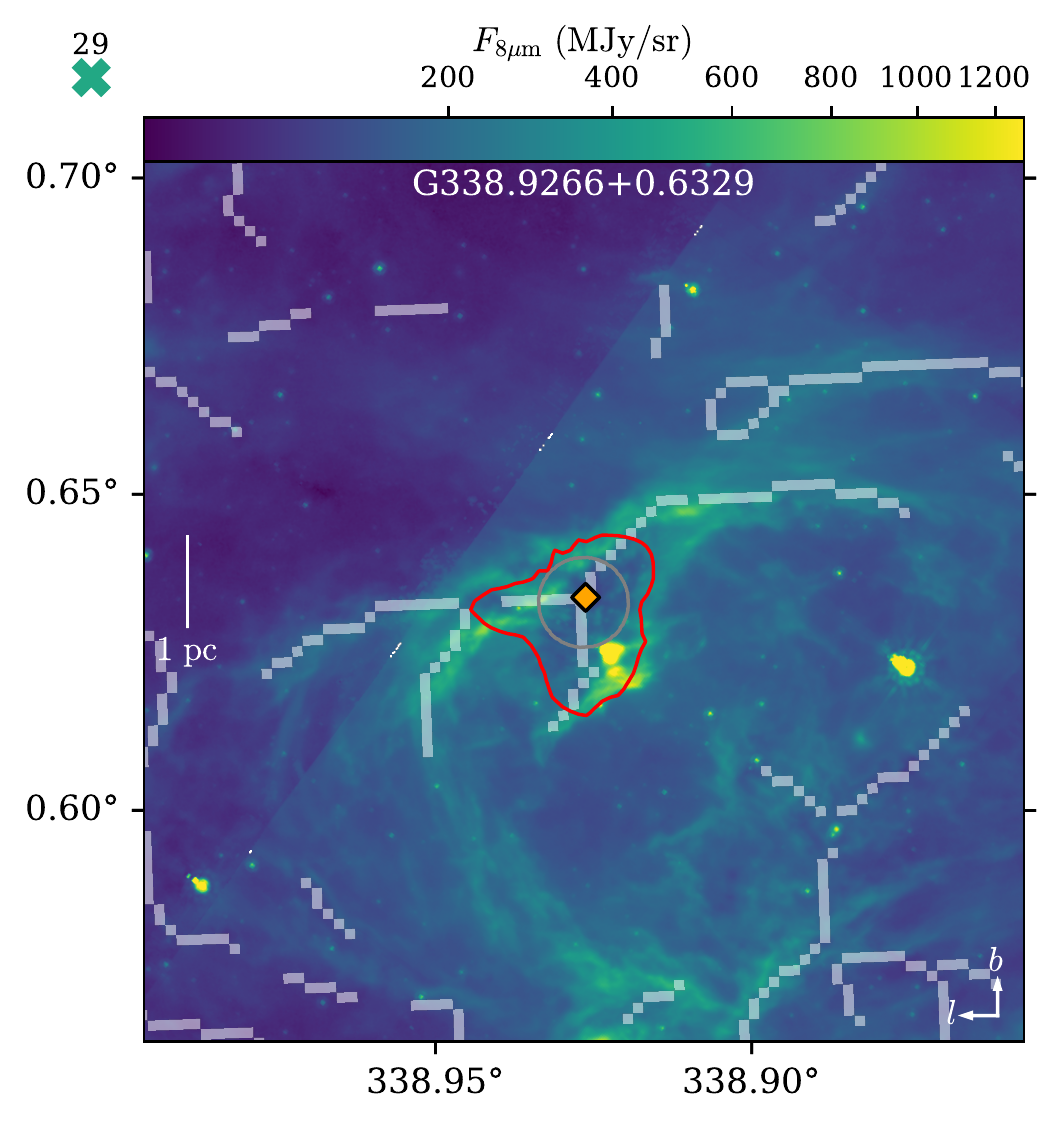}
	}
	\subfloat{
	\includegraphics[width=43mm]{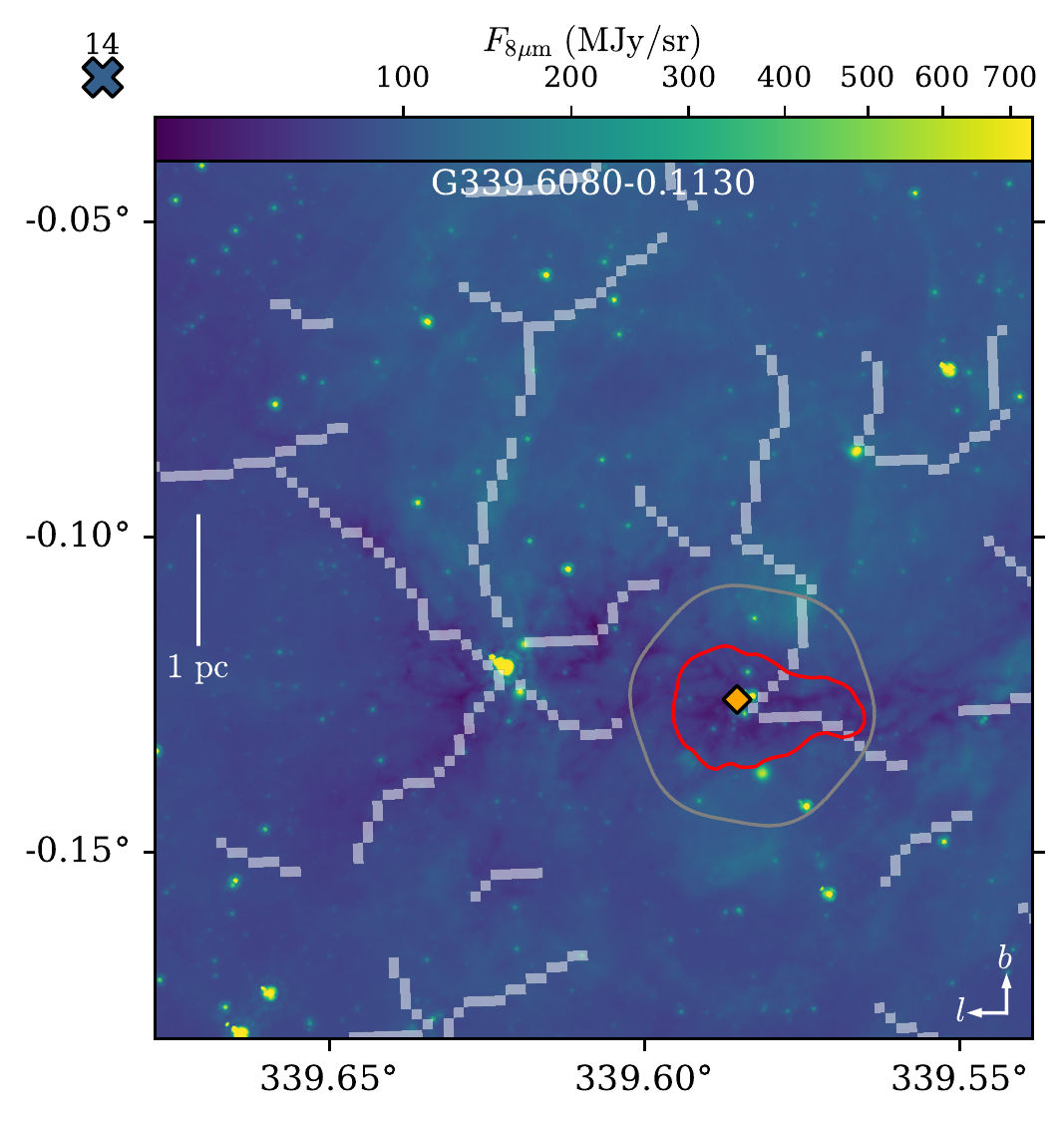}
	}
	\subfloat{
	\includegraphics[width=43mm]{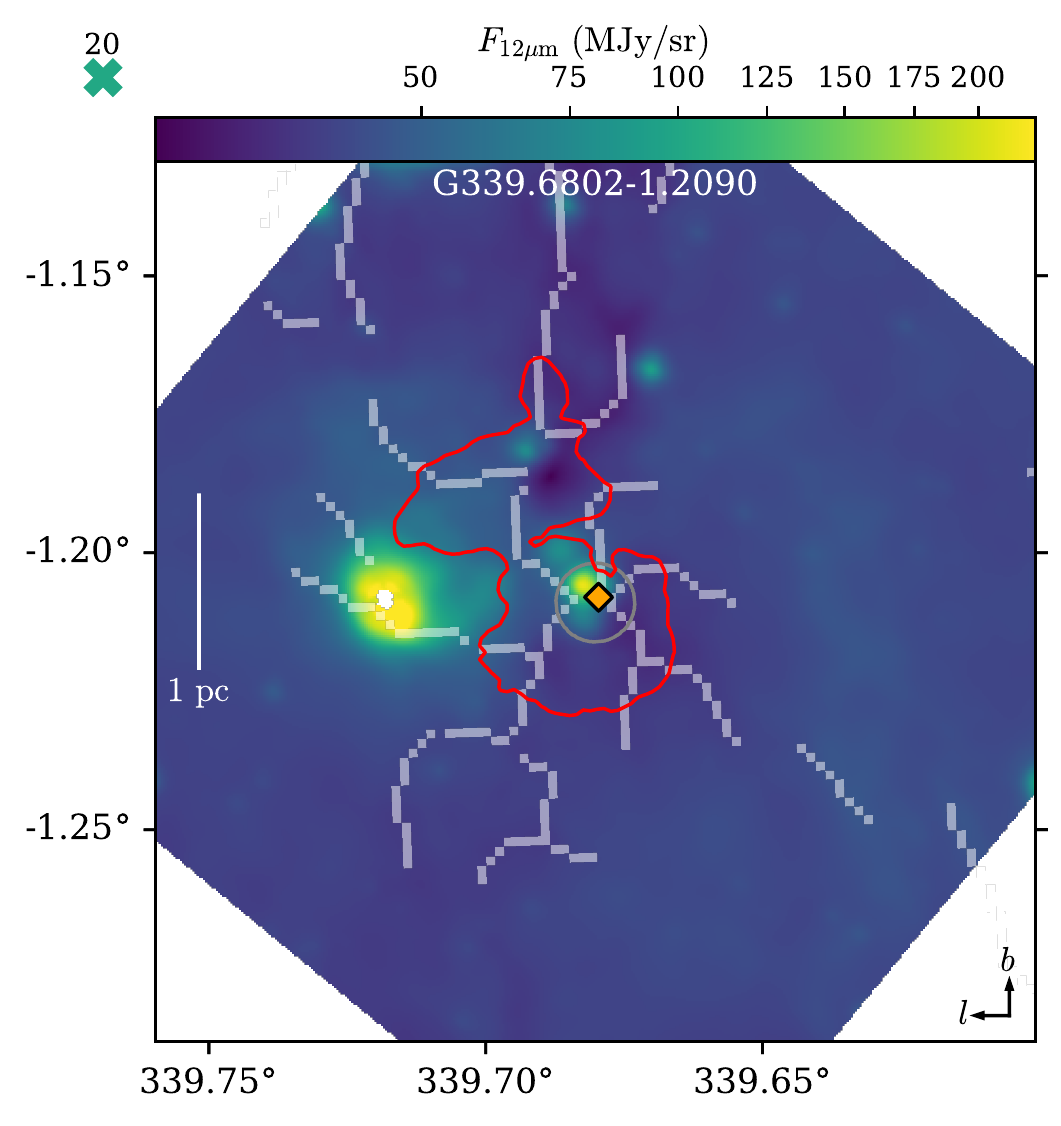}
	}\\[-5mm]
	\hspace{0mm}
	\subfloat{
	\includegraphics[width=43mm]{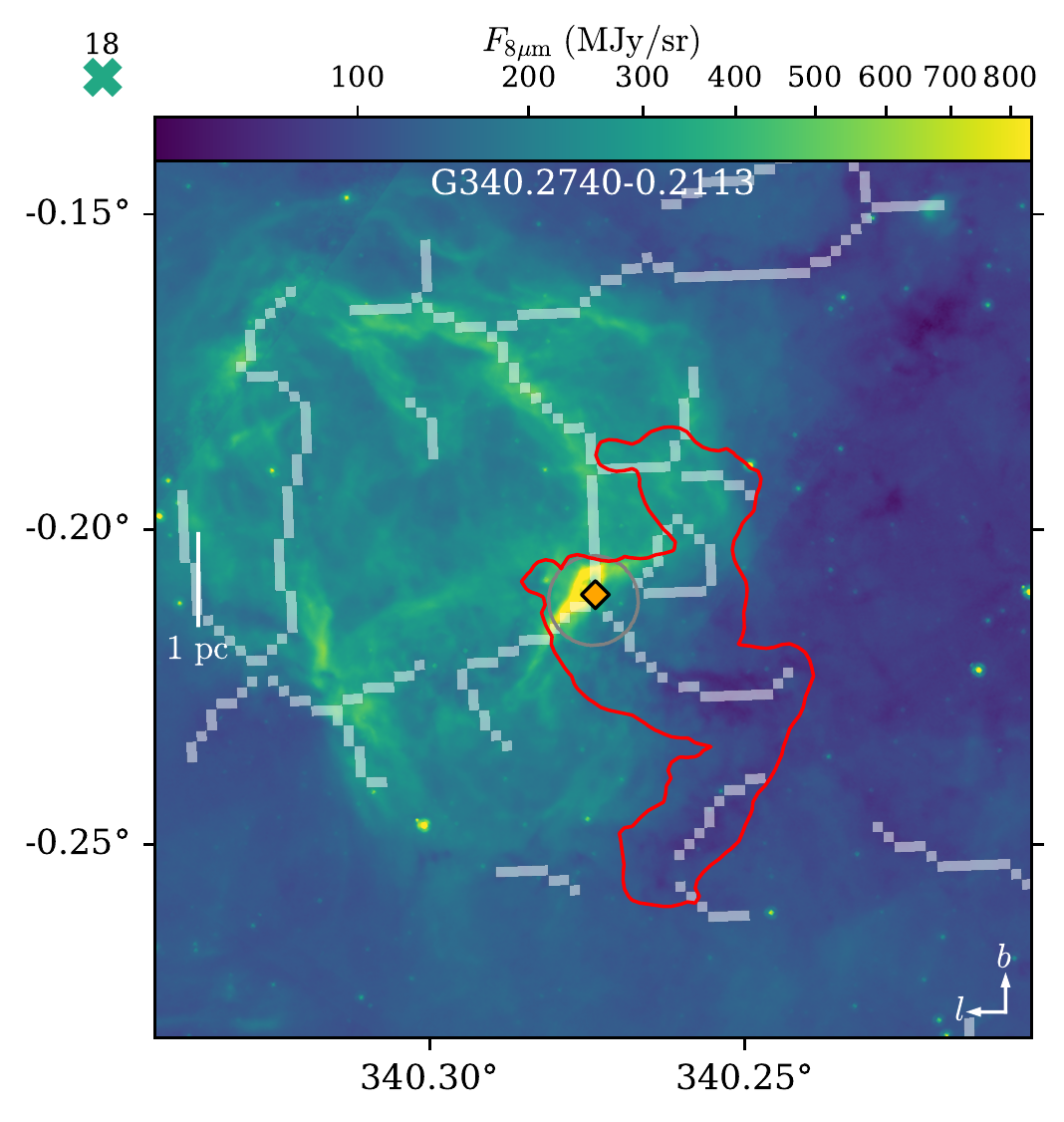}
	}
	\subfloat{
	\includegraphics[width=43mm]{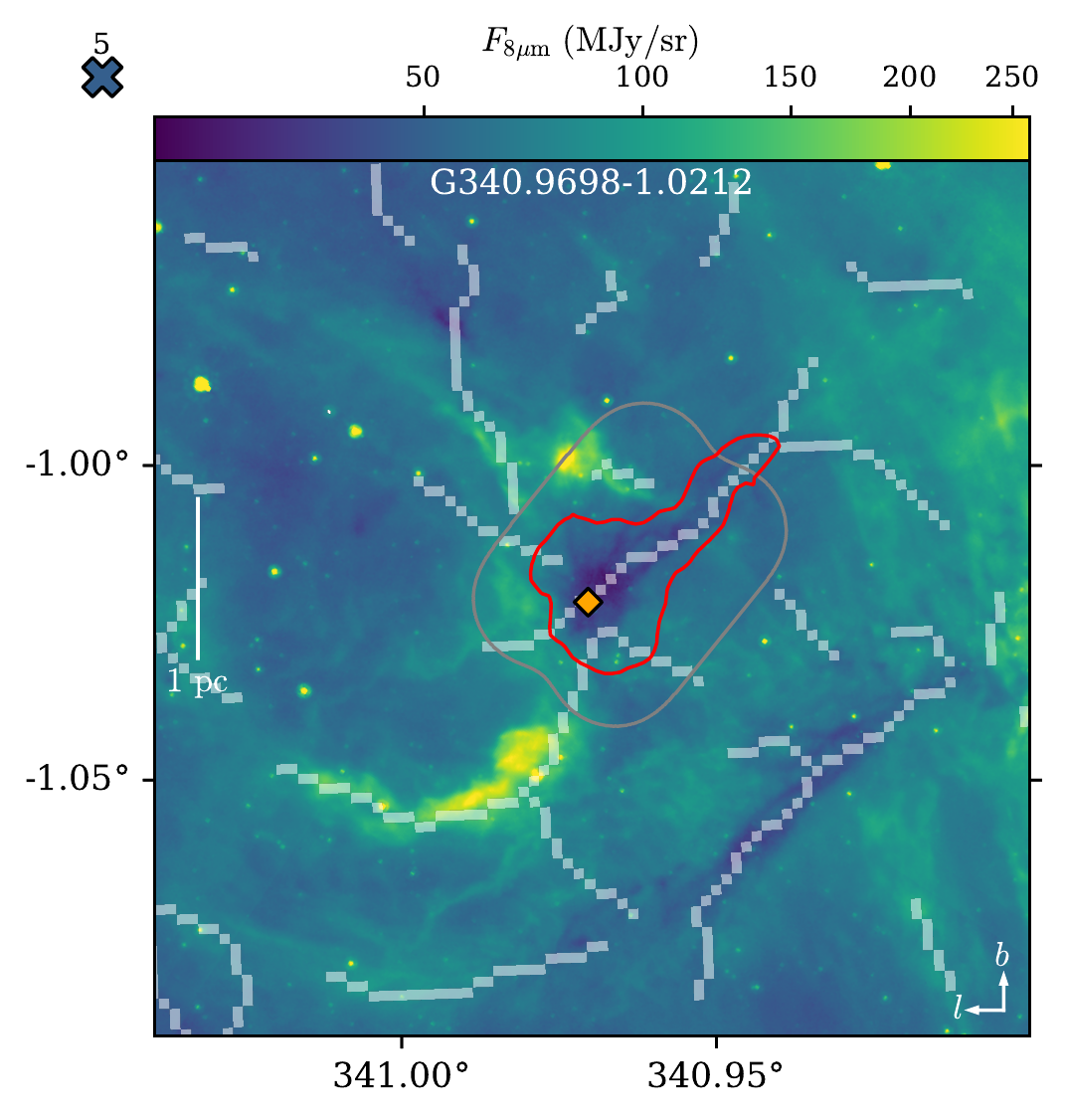}
	}
	\subfloat{
	\includegraphics[width=43mm]{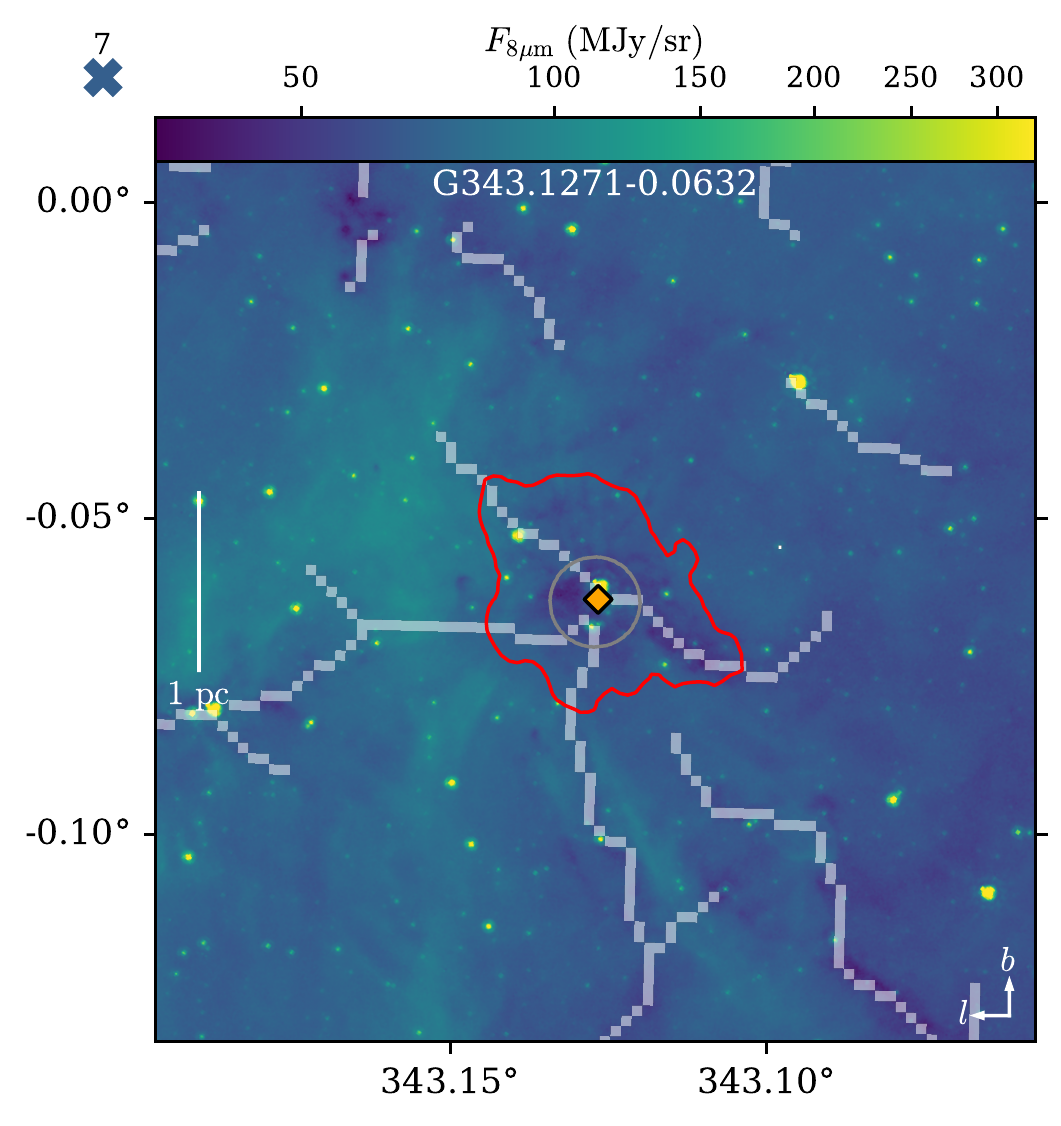}
	}
	\caption{\emph{(continued)}
	}
\end{figure*}
\begin{figure*}\ContinuedFloat
	\centering
	\subfloat{
	\includegraphics[width=43mm]{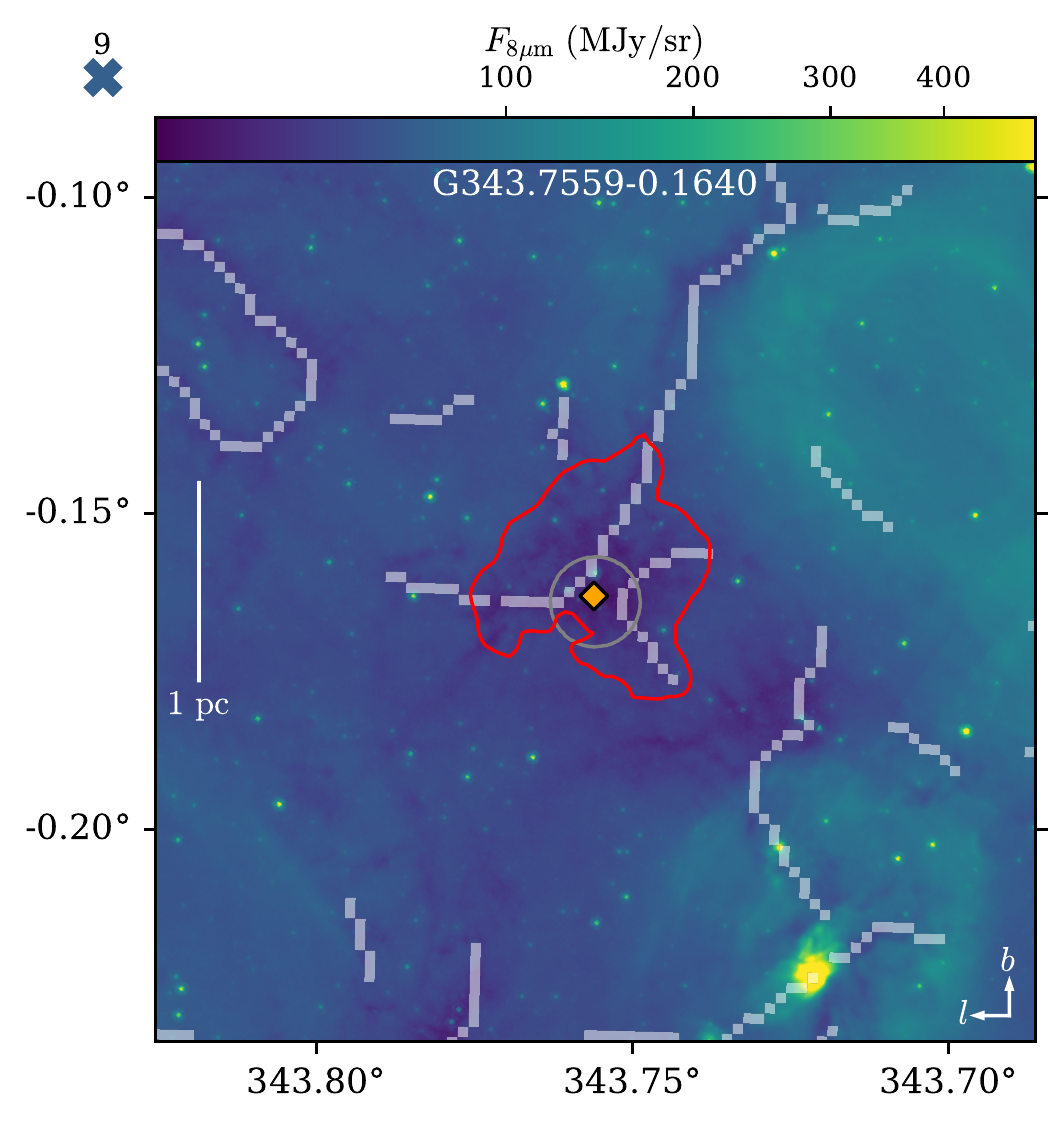}
	}
	\subfloat{
	\includegraphics[width=43mm]{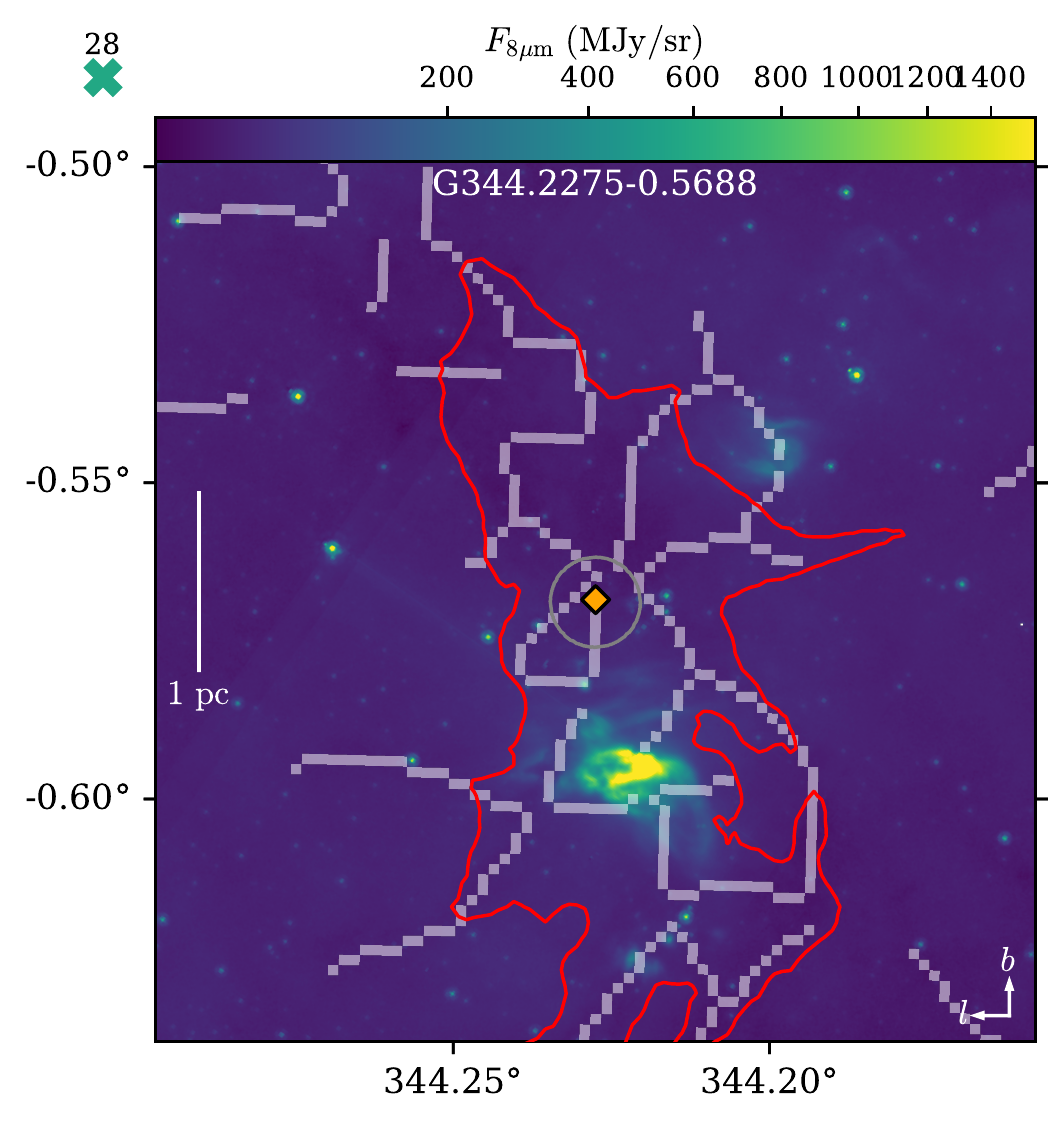}
	}
	\subfloat{
	\includegraphics[width=43mm]{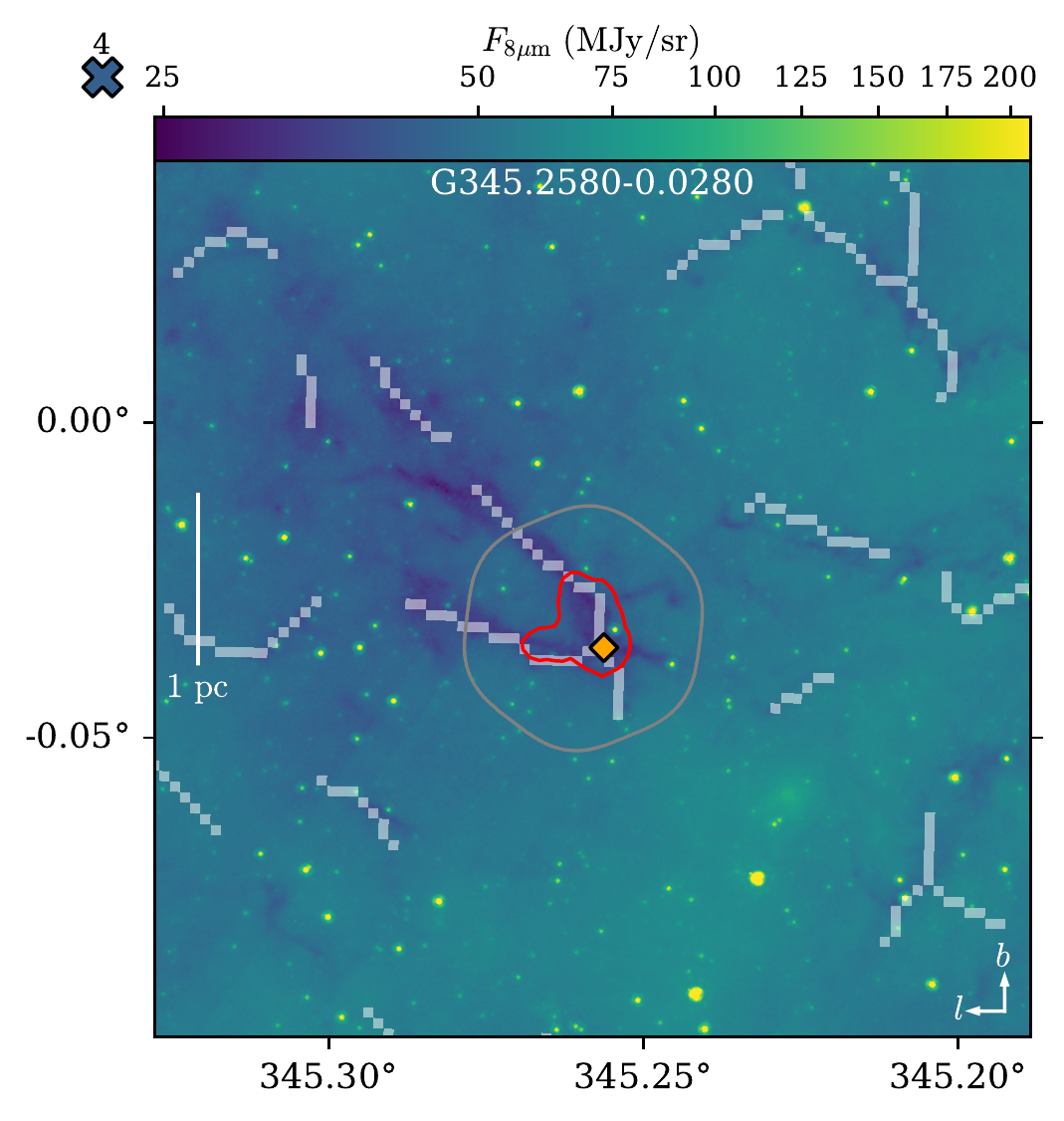}
	}\\[-5mm]
	\hspace{0mm}
	\subfloat{
	\includegraphics[width=43mm]{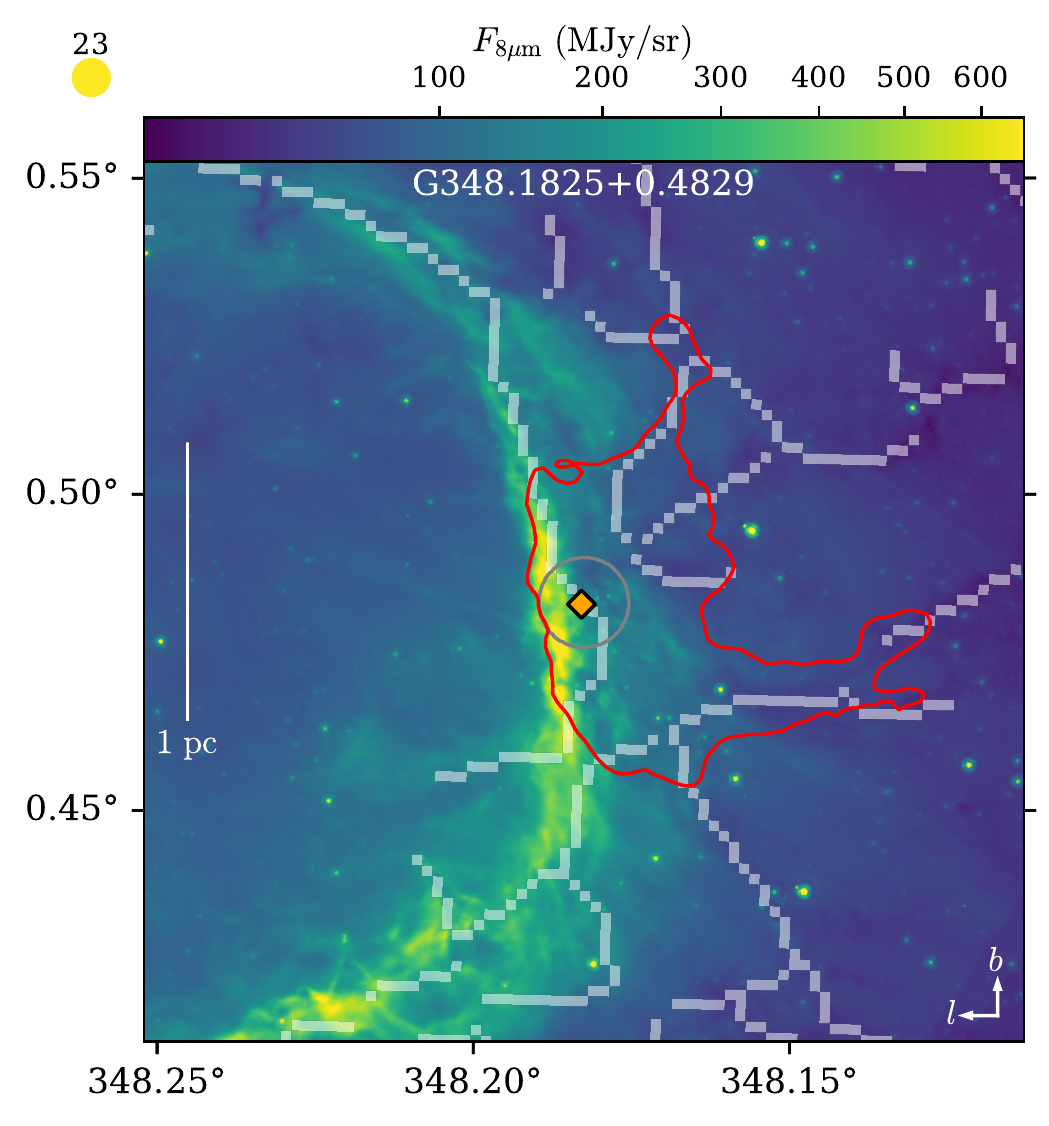}
	}
	\subfloat{
	\includegraphics[width=43mm]{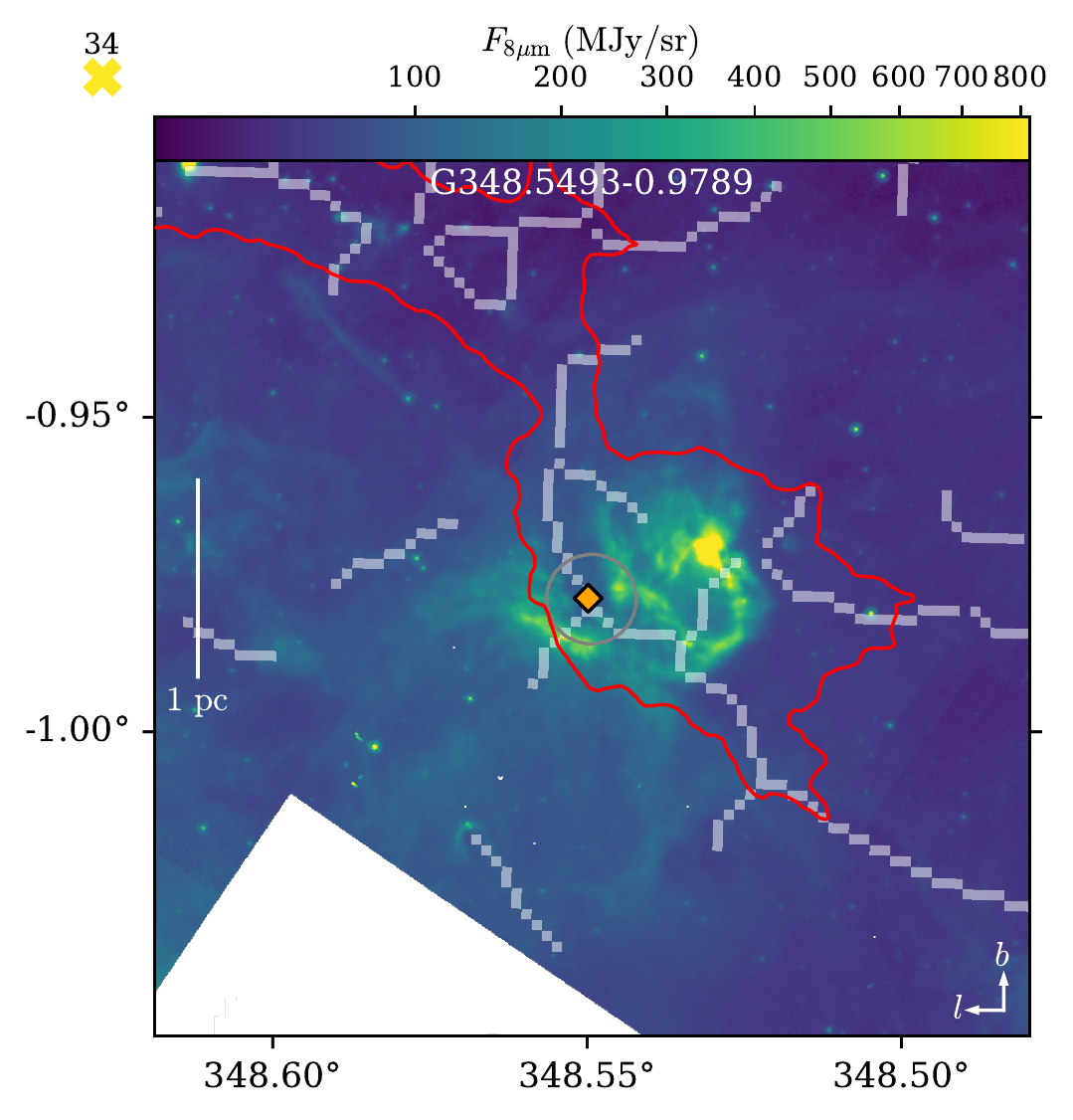}
	}
	\subfloat{
	\includegraphics[width=43mm]{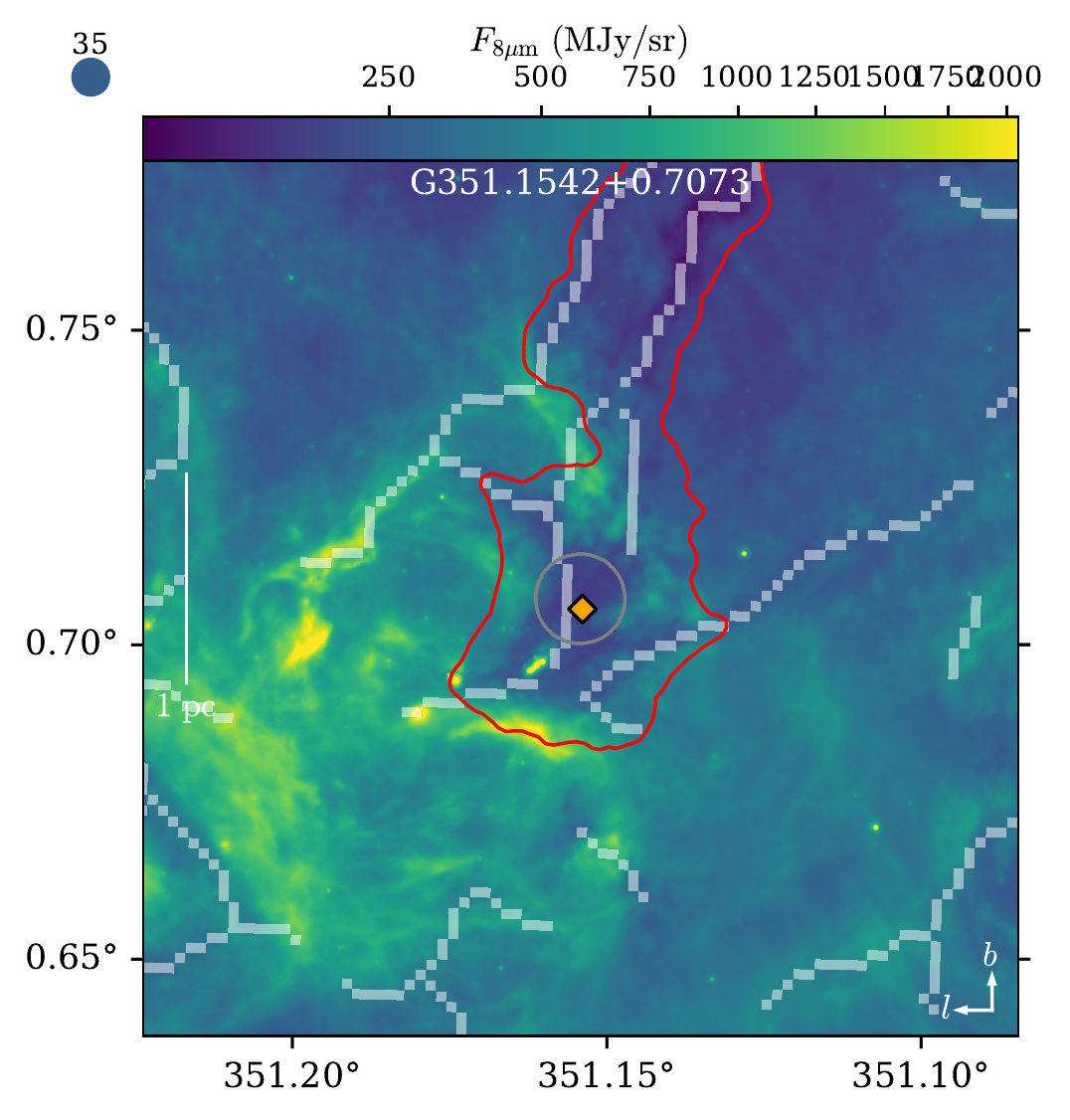}
	}\\[-5mm]
	\hspace{0mm}
	\subfloat{
	\includegraphics[width=43mm]{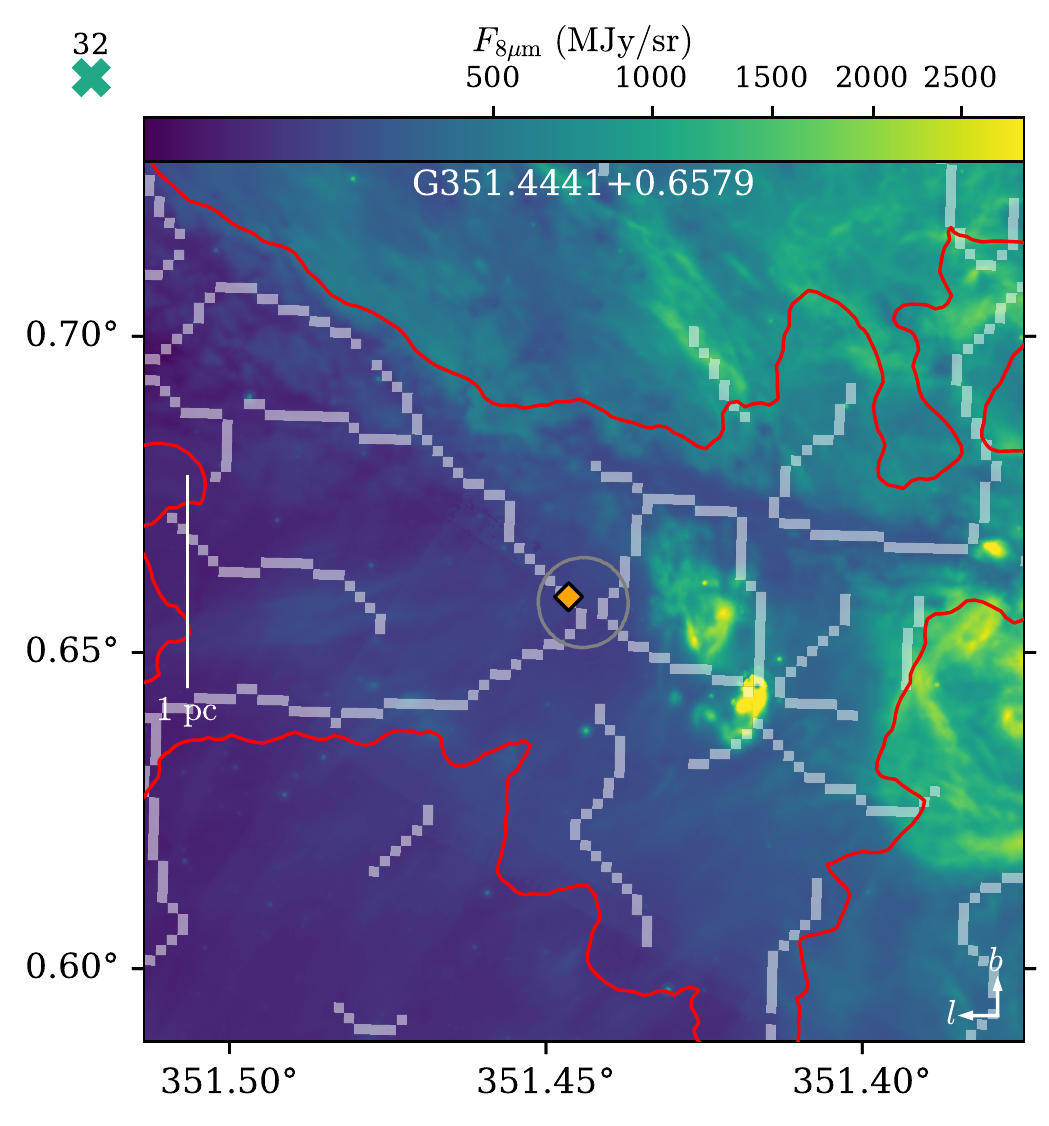}
	}
	\subfloat{
	\includegraphics[width=43mm]{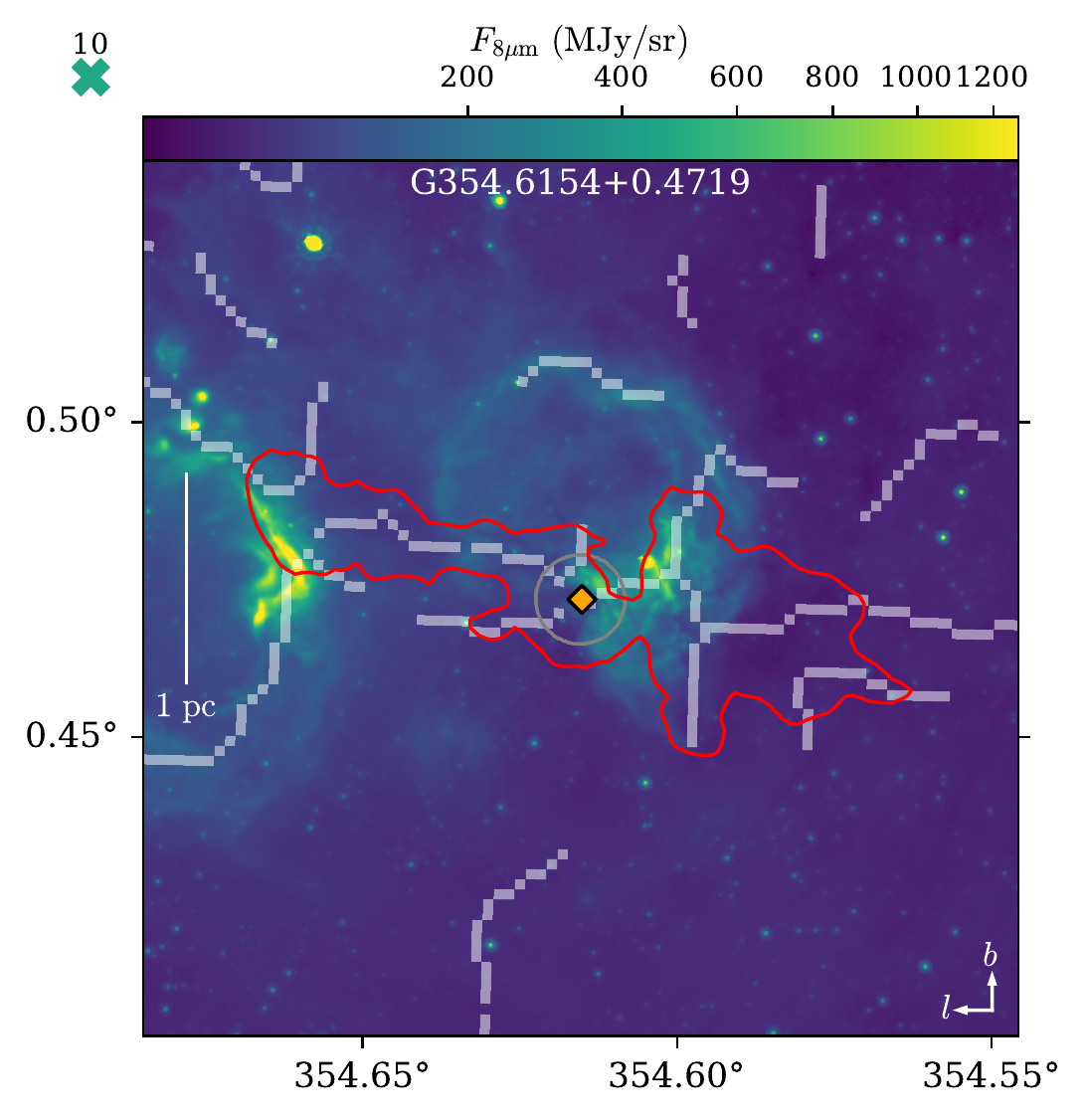}
	}
	\subfloat{
	\includegraphics[width=43mm]{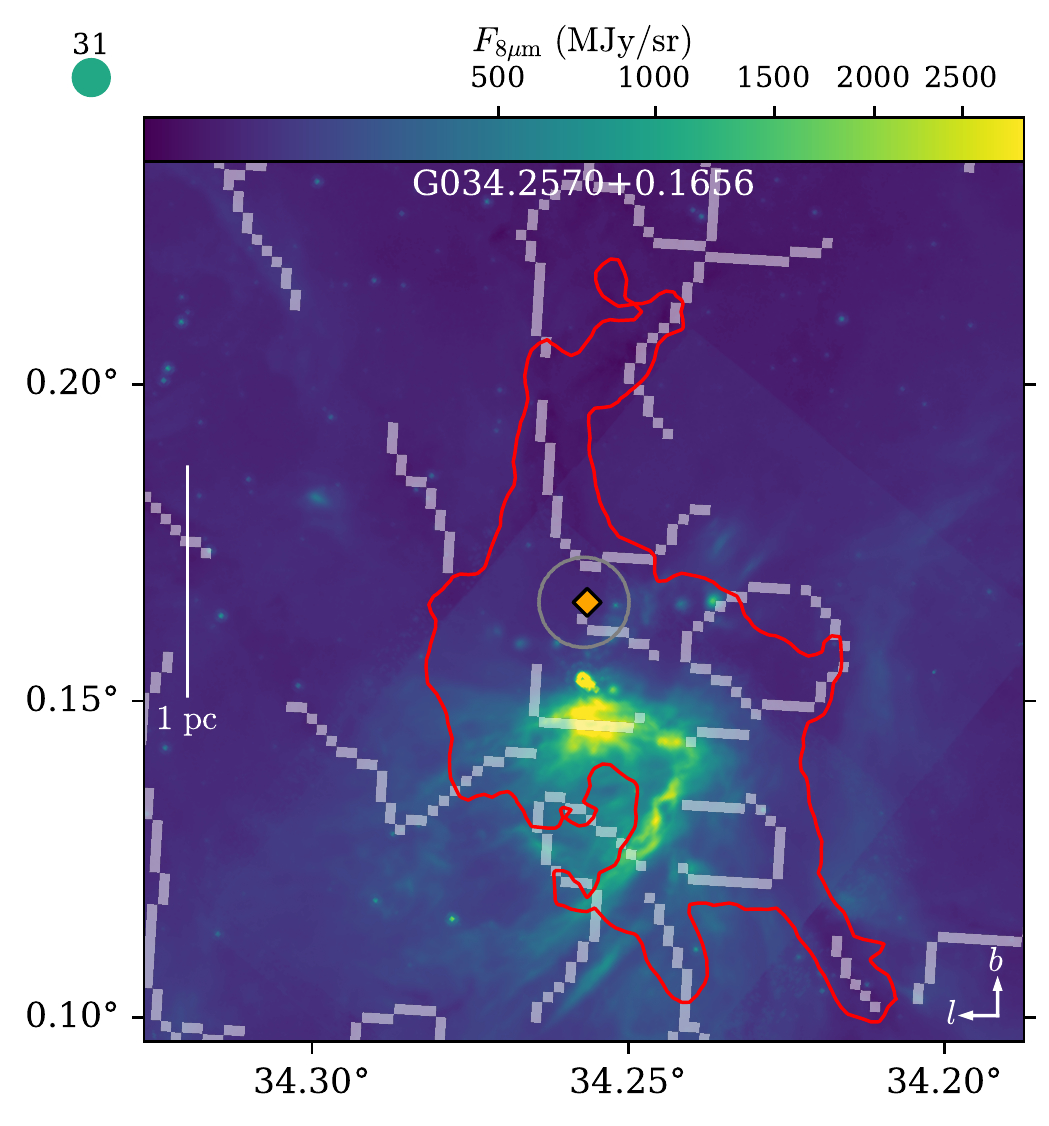}
	}\\[-5mm]
	\hspace{0mm}
	\subfloat{
	\includegraphics[width=43mm]{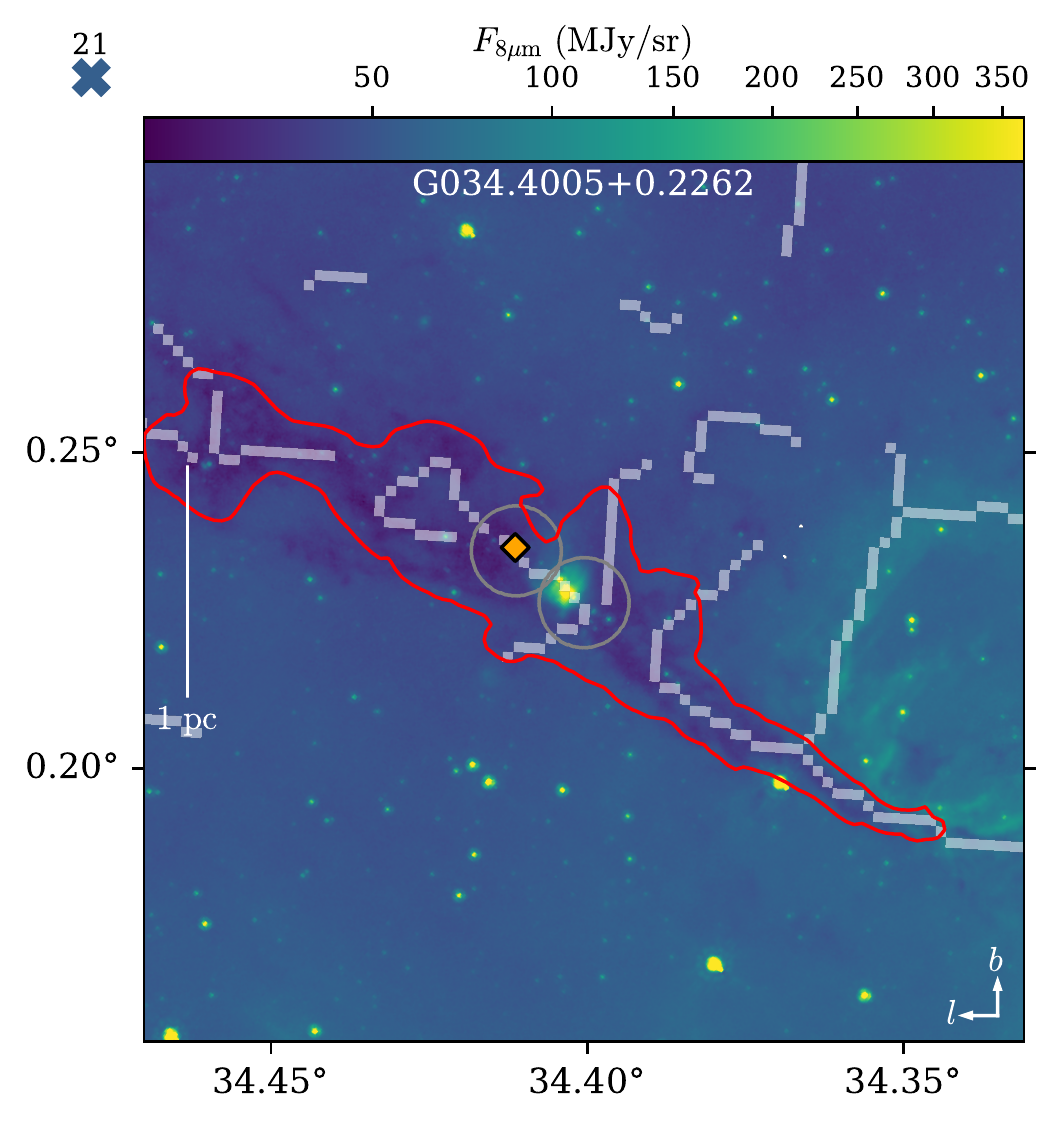}
	}
	\subfloat{
	\includegraphics[width=43mm]{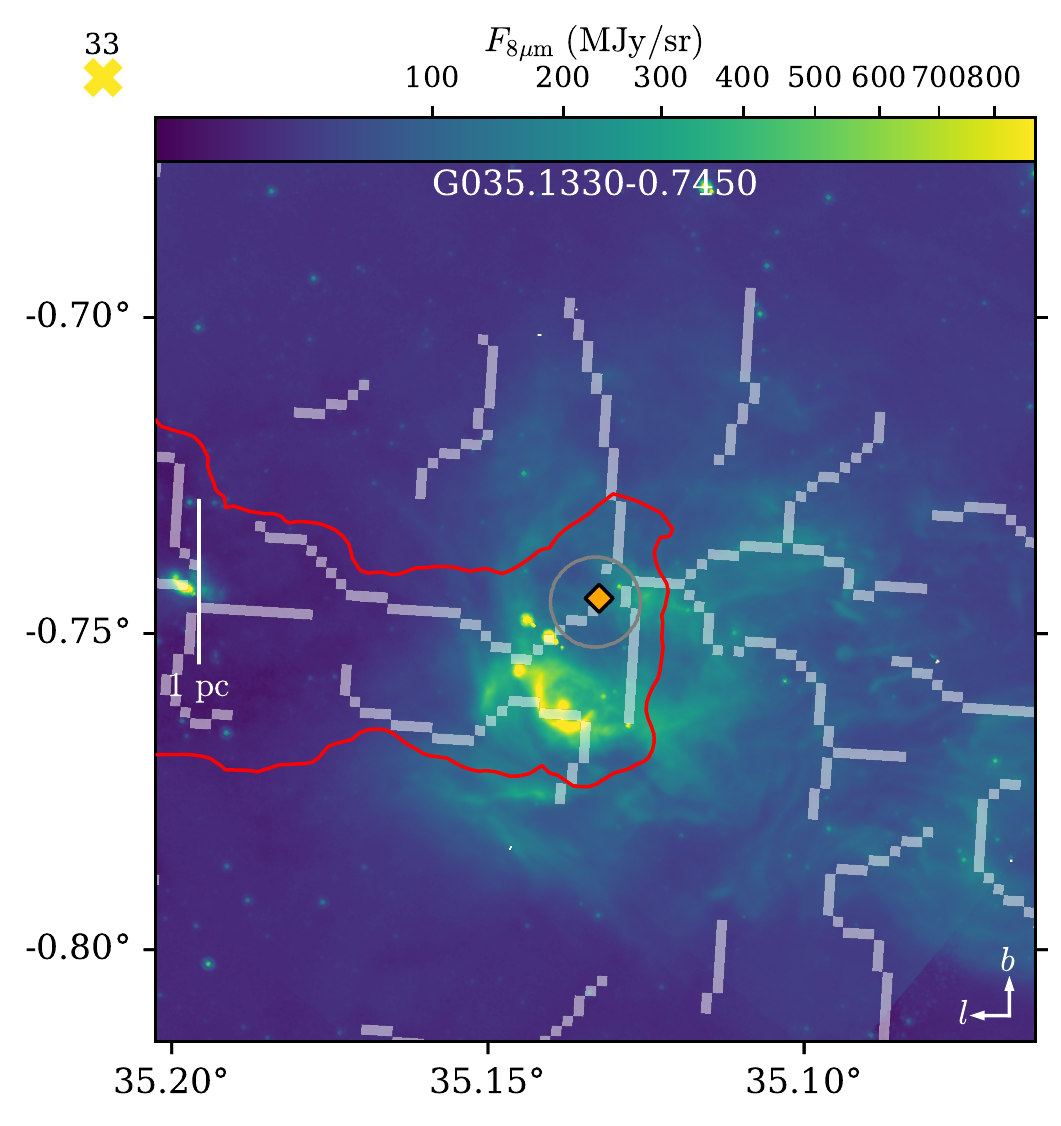}
	}
	\caption{\emph{(continued)}
	}
\end{figure*}

\twocolumn
\section{Core detection quality checking}
\label{sec:core_detection_quality_checking}

\begin{figure}
	\centering
	\includegraphics[width=\columnwidth]{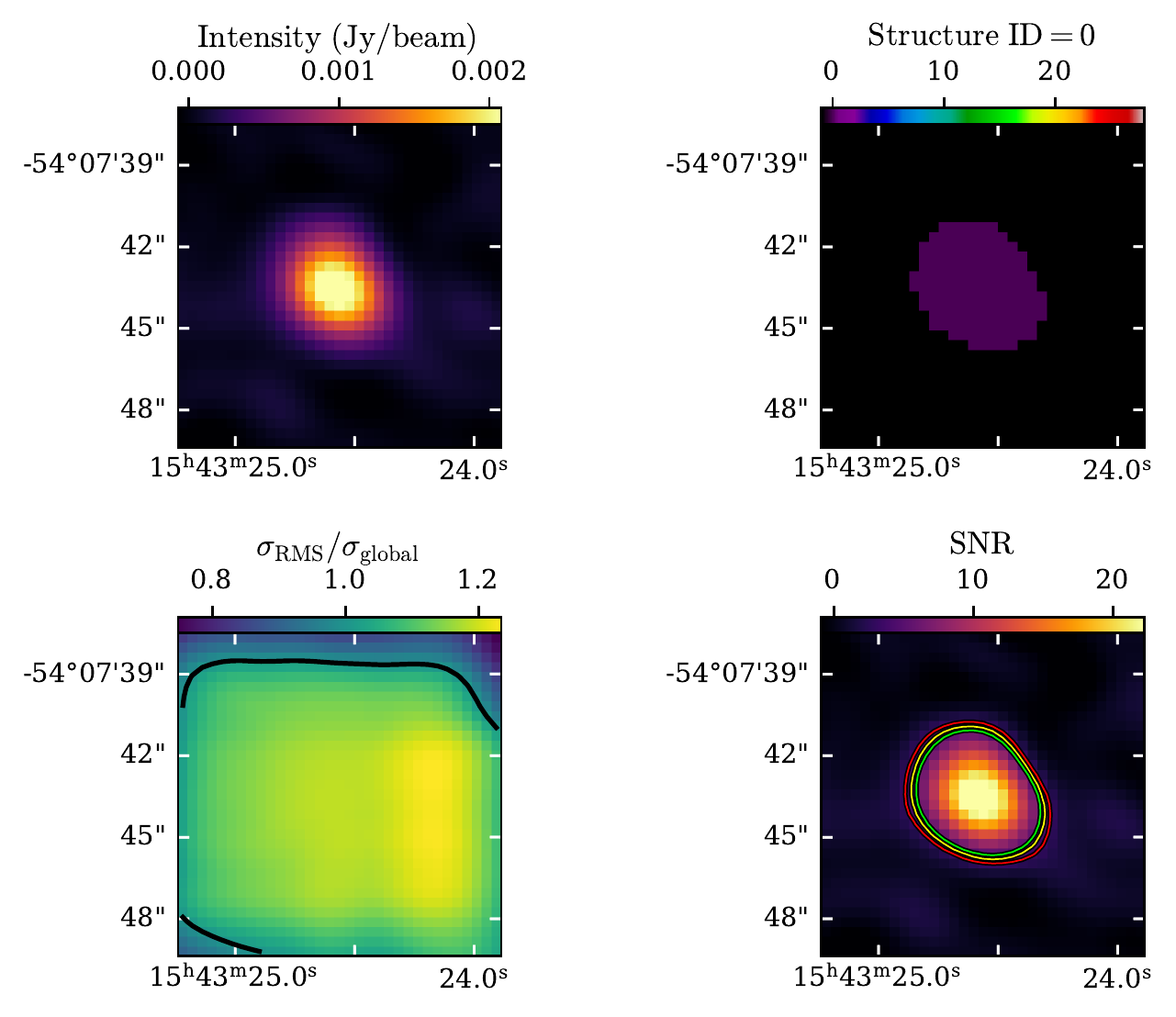}
		\caption{Example of core checking plot for Structure ID 0 in SDC326 (G326.4745+0.7027-MM4), which was flagged as a detection. \emph{(top left)} Continuum image. \emph{(top right)} Dendrogram leaf of structure extracted from image. \emph{(bottom left)} RMS noise map divided by global RMS value used for dendrogram construction. The black contour represents where $\sigma_\mathrm{RMS}/\sigma_\mathrm{global}=1$. \emph{(bottom right)} SNR map, with contours for SNR levels of 3, 4 and 5 shown in red, yellow and green, respectively.
		}
	\label{fig:core_checking}
\end{figure}

\begin{figure}
	\centering
	\includegraphics[width=\columnwidth]{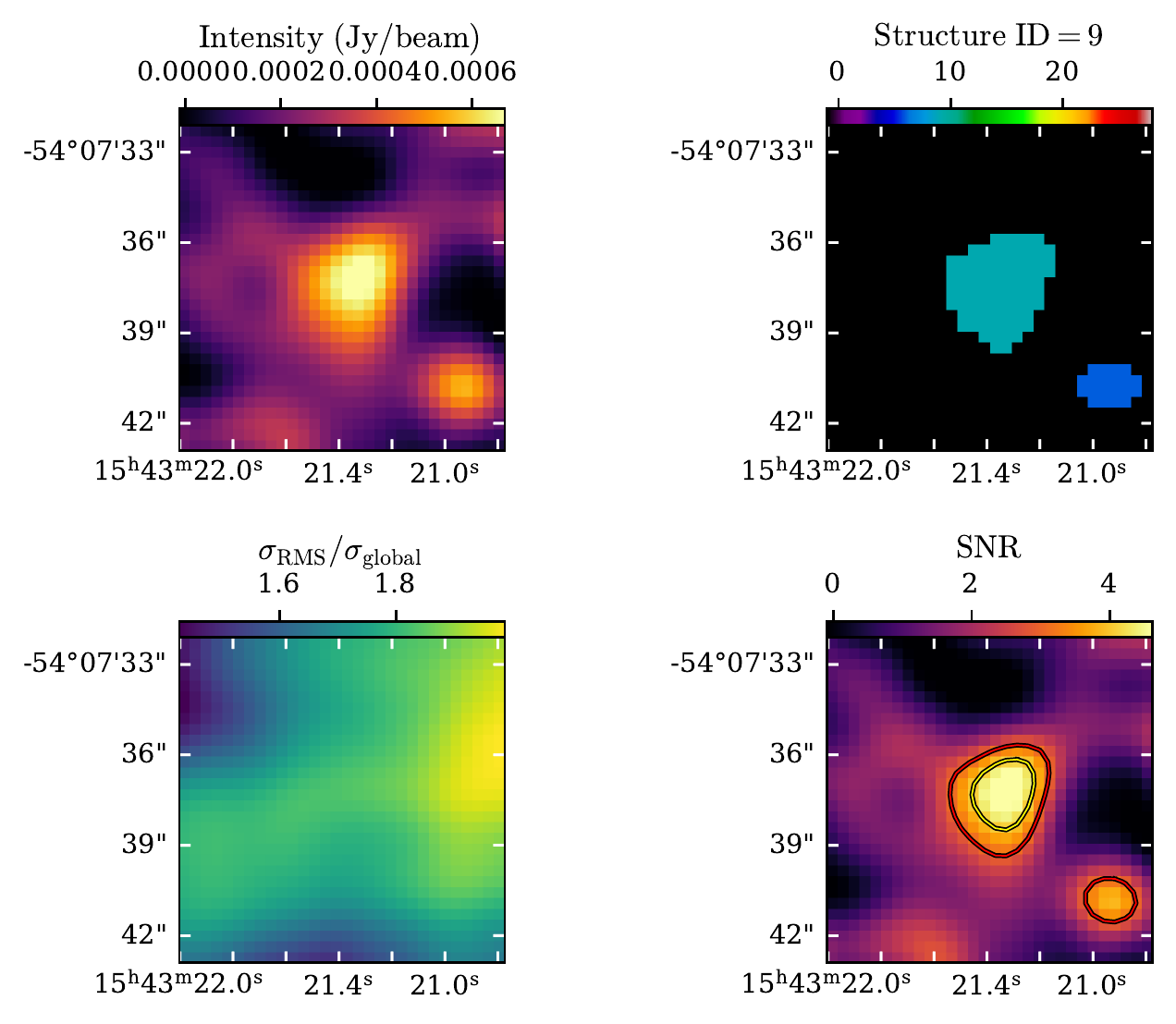}
		\caption{Another example of a core checking plot (as above) for Structure ID 9 in SDC326 (G326.4745+0.7027-MM11), which was also flagged as a detection. Structure ID 6 is shown to the lower right, which was discarded due to its small size and low signal-to-noise.
		}
	\label{fig:core_checking_weak}
\end{figure}

\bsp	
\label{lastpage}
\end{document}